\documentclass[fleqn,usenatbib]{mnras}

\usepackage[T1]{fontenc}
\usepackage{ae,aecompl}

\usepackage{graphicx}	
\usepackage{amsmath}	
\usepackage{amssymb}	

\newcommand{\vwindow}{400}	

\title[Quasar Sightline and Galaxy Evolution Survey I]{Quasar Sightline and Galaxy Evolution (QSAGE) Survey - I. The Galaxy Environment of O~{\sc vi} Absorbers up to $z=1.4$ around PKS~0232-04}

\author[Bielby et al.]{R. M. Bielby$^{1}$\thanks{E-mail: richard.bielby@durham.ac.uk (RMB)},
J. P. Stott$^{2}$,
F. Cullen$^{3}$,
T. M. Tripp$^{4}$,
J. Burchett$^{5}$,
M. Fumagalli$^{1,6}$,
\newauthor
S. L. Morris$^{1}$,
N. Tejos$^{7}$,
R. A. Crain$^{8}$,
R. G. Bower$^{6}$,
J. X. Prochaska$^{5}$
\\
$^{1}$Centre for Extra-galactic Astronomy (CEA), Durham University, South Road, Durham, DH1 3LE, UK\\
$^{2}$Department of Physics, Lancaster University, Lancaster, LA1 4YB, UK\\
$^{3}$Institute for Astronomy, University of Edinburgh, Royal Observatory, Edinburgh EH9 3HJ, UK\\
$^{4}$University of Massachusetts - Amherst, Amherst, MA 01003, USA\\
$^{5}$UCO/Lick Observatory, University of California, Santa Cruz, CA, USA\\
$^{6}$Institute for Computational Cosmology (ICC), Durham University, South Road, Durham, DH1 3LE, UK\\
$^{7}$Instituto de F\'{i}sica, Pontificia Universidad Cat\'{o}lica de Valpara\'{i}so, Casilla 4059, Valpara\'{i}so, Chile\\
$^{8}$Astrophysics Research Institute, Liverpool John Moores University, 146 Brownlow Hill, Liverpool L3 5RF, UK\\
}

\date{Accepted XXX. Received YYY; in original form ZZZ}

\pubyear{2018}

\begin{document}
\label{firstpage}
\pagerange{\pageref{firstpage}--\pageref{lastpage}}
\maketitle

\begin{abstract}
We present the first results from a study of O~{\sc vi} absorption around galaxies at $z<1.44$ using data from a near-infrared grism spectroscopic Hubble Space Telescope Large Program, the Quasar Sightline and Galaxy Evolution (QSAGE) survey. QSAGE is the first grism galaxy survey to focus on the circumgalactic medium at $z\sim1$, providing a blind survey of the galaxy population. Using the first of 12 fields, we provide details of the reduction methods, in particular the handling of the deep grism data which uses multiple position angles to minimise the effects of contamination from overlapping traces. The resulting galaxy sample is H$\alpha$ flux limited ($f({\rm H\alpha}) > 2\times10^{-17}$~erg~s$^{-1}$~cm$^{-2}$) at $0.68<z<1.44$, corresponding to $\gtrsim0.2-0.8$~M$_\odot$~yr$^{-1}$. We combine the galaxy data with high-resolution STIS and COS spectroscopy of the background quasar to study O~{\sc vi} in the circumgalactic medium. At $z>0.68$, we find 5 O~{\sc vi} absorption systems along the line of sight with identified galaxies lying at impact parameters of $b\approx100-350$~kpc (proper), whilst we find a further 13 galaxies with no significant associated O~{\sc vi} absorption (i.e. $N({\rm OVI})<10^{13.5-14}$~cm$^{-2}$) in the same impact parameter and redshift range. We find a large scatter in the stellar mass and star-formation rates of the closest galaxies with associated O~{\sc vi}. Whilst one of the O~{\sc vi} absorber systems is found to be associated with a low mass galaxy group at $z\approx1.08$, we infer that the detected O~{\sc vi} absorbers typically lie in the proximity of dark matter halos of masses $10^{11.5}~{\rm M_\odot}\lesssim M_{\rm halo}\lesssim10^{12}~{\rm M_\odot}$.
\end{abstract}




\section{Introduction}

The peak epoch in the volume-averaged star formation rate (SFR) for galaxies is at $z=1-2$ with the SFR in typical galaxies being an order of magnitude higher than in the local Universe \citep[e.g.][]{1996ApJ...460L...1L, 1996MNRAS.283.1388M, 2013MNRAS.428.1128S}. A picture is emerging in which star formation at this epoch is very different to that at the present day. Rather than the subdued formation of stars that is the norm in today's Universe, violent  episodes of star formation are driven by the formation of super-star clusters within unstable gas rich discs \citep[e.g.][]{2010Natur.464..733S, 2010Natur.463..781T}.

Theory suggests that the peak in the star-formation rate density (SFRD) is the result of a higher rate of gas accretion in the early Universe \citep[e.g.][]{2009Natur.457..451D, 2011MNRAS.416.1566L, 2011MNRAS.415.2782V}. Star formation at high redshift should be critically dependent on the inflow rate of cold gas, while the present day galaxy stellar mass function is most readily explained through an efficient feedback mechanism such as strong supernovae driven outflows. In order to fully understand the interplay and balance of these factors \citep[e.g.][]{2010ApJ...718.1001B, 2012MNRAS.421...98D}, it is crucial that we test these theories by detecting diffuse baryons inside and around the halos of galaxies, charting their abundance and physical properties as a function of redshift \citep[e.g.][]{2011MNRAS.418.1796F, 2012MNRAS.421...98D, 2013MNRAS.430.1548H, 2013MNRAS.430.2427R, 2015MNRAS.452.2034R, 2017arXiv170907577O, 2018MNRAS.473..538C}. The approximate area of influence, within which a galaxy deposits and receives baryonic material, is commonly referred to as the circumgalactic medium \citep[CGM;][]{1996ApJ...462..651L,2010ApJ...717..289S,2017ARA&A..55..389T}.

Extensive studies have been performed using the powerful combination of galaxy surveys with high resolution quasar sightline data, to analyse the nature of gas around galaxies at $z\lesssim0.5$ (e.g. \citealt{1993ApJ...419..524M, 2010MNRAS.402.1273C, 2011Sci...334..948T, 2011ApJ...740...91P, 2014MNRAS.437.2017T, 2016MNRAS.460..590F, 2017ApJ...837..169P}) and at $2\lesssim z\lesssim4$ (primarily using the Lyman Break method; e.g. \citealt{2003ApJ...584...45A, 2005ApJ...629..636A, 2006ApJ...637..648S, 2011MNRAS.414...28C,2012ApJ...751...94R,2013ApJ...776..136P,2014ApJ...796..140P,2014MNRAS.442.2094T, 2015MNRAS.450.2067T, 2016MNRAS.462.1978F, 2017MNRAS.471.3686F, 2017MNRAS.471..690T, 2017MNRAS.471.2174B}). These have provided important insights for understanding the flow of baryons around galaxies and how this impacts galaxy evolution. However, the available data \emph{between} these two epochs (i.e. $z\approx1-2$) is relatively sparse, and yet this is the critical epoch at which most stars form and the Hubble sequence changes. There are an increasing number of studies focusing on the relevance of large scale structure (i.e. voids, filaments, groups and clusters) with respect to observations of the CGM \citep[e.g.][]{2002ApJ...565..720P, 2006MNRAS.367..139A, 2012MNRAS.425..245T, 2016MNRAS.455.2662T, 2014ApJ...791..128S, 2016ApJ...832..124B, 2018MNRAS.475.2067B, 2016MNRAS.462.1978F, 2017MNRAS.471.3686F,2017MNRAS.464.2053P, 2017MNRAS.468.1373B, 2018MNRAS.tmp..710P,2018arXiv180601715F}. Such studies are adding important context to for interpreting CGM observations, but are again primarily limited to $z\lesssim0.5$ or $z>2$.

The O~{\sc vi} doublet absorption feature offers important insights into galaxy evolution \citep{2000ApJ...534L...1T}. The nature of the material traced by O~{\sc vi} absorption in quasar sightlines potentially traces a broad range of physical manifestations: as outflowing material, potentially indicative of ongoing or recent star-formation \citep[e.g.][]{2011Sci...334..948T,2013ApJ...767...49M, 2016ApJ...833...54W}; diffuse halo gas at the ambient halo temperature, potentially indicative of relic star-formation \citep[e.g.][]{2016ApJ...833...54W,2016MNRAS.460.2157O,2018MNRAS.474.4740O}; intra-group material \citep[e.g.][]{2014ApJ...791..128S,2017ApJ...838...37S}; or more widely spread warm-hot intergalactic medium (WHIM, e.g. \citealt{2012ApJ...759...23S}).

In this paper, we present results from the first field analysed as part of the Quasar Sightline and Galaxy Evolution (QSAGE) survey. We have performed a grism spectroscopic survey around 12 bright quasars at $z>1.2$, all with existing archival HST/STIS, and in some cases HST/COS, spectroscopy. By selecting quasar sightlines as target fields, we place the scientific focus on studying the CGM, whilst providing 12 effectively independent pencil beam surveys suitable for a wide range of galaxy evolution science. Whilst the HST/STIS and HST/COS data probe gas in absorption along the central sightline at $z\lesssim1.44$, our new HST WFC3 grism observations survey the galaxy population `blindly' (i.e. without any preselection aside from flux limits) around these sightlines across a comparable redshift range (predominantly via H$\alpha$ emission). In an earlier paper \citep{2017MNRAS.468.1373B}, we presented an analysis of Mg~{\sc ii} absorption associated with a group environment detected in MUSE observations on one of the fields, prior to having acquired WFC3/grism data. Here we focus on a second field, this time incorporating both MUSE IFU and WFC3 grism data, whilst also presenting the survey strategy, data-processing methods, and an analysis of the relationship between O~{\sc vi} absorption and galaxies at $0.68<z<1.4$ data. The quasar that forms the basis for this study is PKS~0232-04 (\citealt{1966AuJPh..19..837S}, originally designated PHL~1377; \citealt{1962BOTT....3...37H}) at a redshift of $z=1.44$.

In section~\ref{sec:observations} we present the observations and data reduction methods (including optical and NIR imaging, grism spectroscopy, IFU spectroscopy with MUSE, and high resolution sightline spectroscopy); in section~\ref{sec:analysis}, we present an analysis of the distribution of galaxies around O~{\sc vi} absorbers as a function of galaxy properties and environment; we discuss our results in section~\ref{sec:discussion}; and in section~\ref{sec:conclusions} we give our conclusions and a summary. Throughout this paper, we use the AB magnitude system and the Planck 2015 cosmology \citep{2016A&A...594A..13P}. Unless otherwise stated, all distances are given in proper coordinates.

\section{Observations}
\label{sec:observations}

The QSAGE survey targets 12 $z\gtrsim1.2$ quasars with the aim of mapping and characterising the galaxy populations in their foregrounds (HST Large Program 14594; PIs: R. Bielby, J. P. Stott). The full survey consists of 96 orbits with the WFC3 in imaging and grism mode (i.e. 8 hours per quasar sightline). In this paper we focus on the first field to be fully analysed, centred on the $z=1.44$ quasar PKS~0232-04, chosen due to the extensive data available to complement the WFC3 grism and STIS sightline data: Canada-France-Hawaii Telescope Legacy Survey (CFHTLS) optical imaging, MUSE IFU data and HST/COS sightline spectroscopy. In the following section we describe in detail the methods used in taking and processing the various observations on this first target field, and which going forward will be used across the entirety of the 12 survey fields.


\begin{figure*}
	\includegraphics[width=0.8\textwidth]{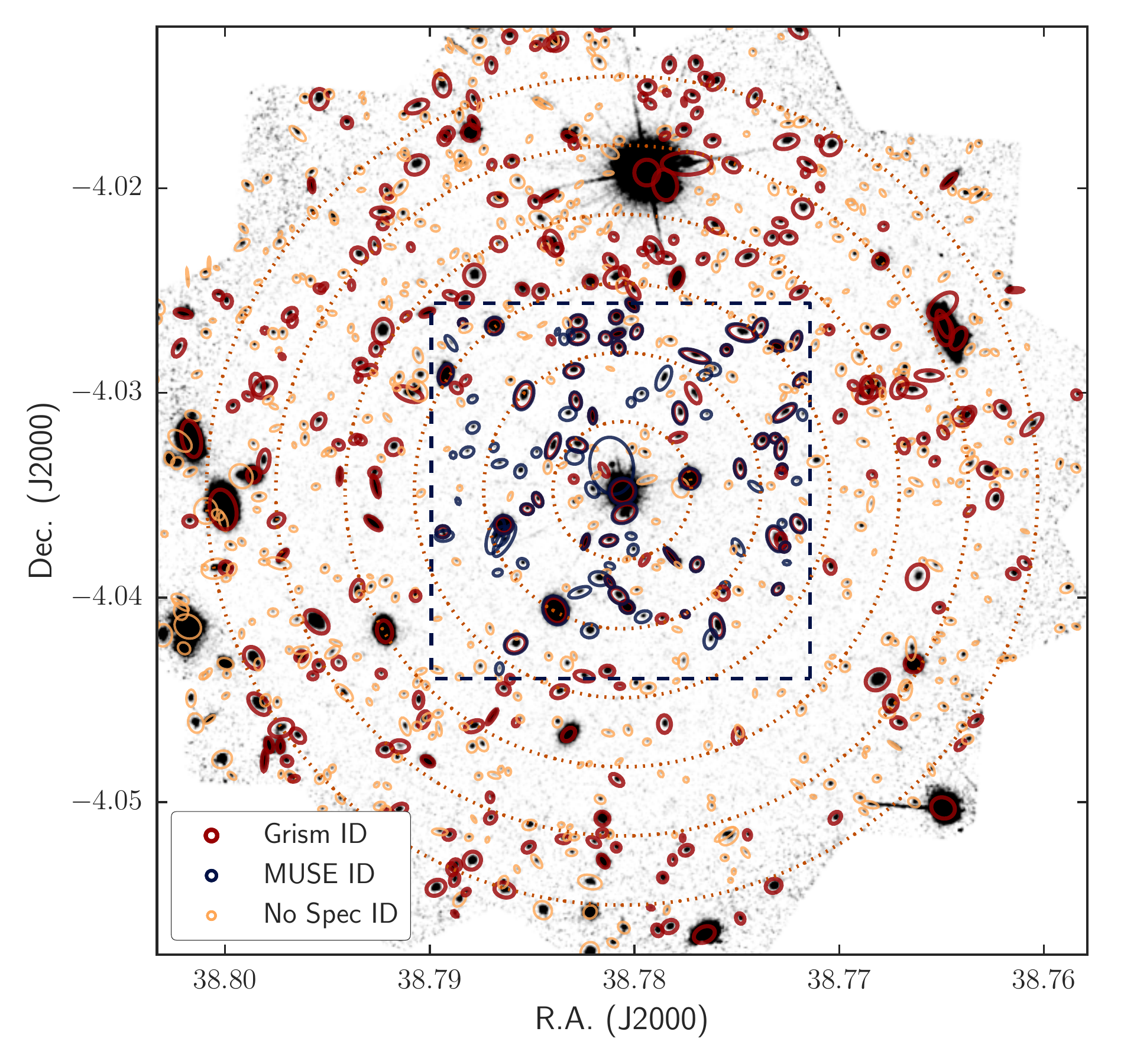}
	\caption{Stacked HST/WFC3 broad band F140W image from the 4 separate orbits observing PKS~0232-04. Galaxies with a successfully identified redshift from either VLT/MUSE (blue ellipses) or HST/WFC3 grism observations (dark red ellipses) are marked. Objects detected in the imaging, but without a reliably identified redshift are marked by the pale orange ellipses. Dotted rings indicate scales of 100kpc up to 600kpc in steps of 100kpc at $z=1$. The dashed box denotes the field of view of the VLT/MUSE observations.}
	\label{fig:wfc3f140}
\end{figure*}

\subsection{HST WFC3 data}
\label{sec:WFC3} 

For a single field, the QSAGE HST observations consist of 8 grism exposures alongside supporting NIR imaging. The imaging is required for the identification of accurate source coordinates, which act as the basis for the extraction of object spectra from the grism data. The observations for a given field were split into four `visits', with each visit providing observations at a different position angle. Each visit consisted of 2 orbits, with each orbit scheduled to acquire a single F140W image of $\approx250$s, 2 exposures of $\approx1,000$s using the G141 grism, and a single $\approx250$s F160W exposure. These exposures were taken using the `SPARS25' sampling sequence in the case of the basic imaging exposures and the `SPARS100' sampling sequence in the case of the grism exposures. The corresponding exposure times were tweaked to optimally fill the time afforded by a given orbit and all exposures were taken with the `grism1024' aperture. The total exposure times across the 4 visits for this field were 2123s with F140W; 2123s with F160W; and 16,047s with the G141 grism. These data represent some of the deepest observations using the WFC3 G141 grism thus far. As such, the different position-angle visits are necessary for the optimal extraction of spectra with the removal of contamination from overlapping sources. 

\subsubsection{WFC3 NIR imaging}

Both the direct imaging and grism data were reduced using \textsc{Grizli} (Brammer, in prep.\footnote{https://github.com/gbrammer/grizli}), a custom software package dedicated to the reduction and analysis of slitless spectroscopic datasets which builds on previous software packages such as \textsc{aXe} \citep{2009PASP..121...59K} and \textsc{threedhst} \citep{2012ApJS..200...13B, 2016ApJS..225...27M}.
For the direct images, individual exposures were corrected for small astronomic offsets using \textsc{TweakReg} before \textsc{AstroDrizzle} was used to produce background-subtracted drizzled images in both the F140W and F160W filters. 
The F140W and F160W sets of images were each stacked using {\sc SWarp} \citep{swarp} using a median combination. The resulting F140W stacked image is shown in Fig.~\ref{fig:wfc3f140}. Photometric zeropoints for the F140W and F160W imaging were taken directly from the STScI guidelines ($m_{\rm ZP}({\rm F140W})=26.45$ and $m_{\rm ZP}({\rm F160W})=25.95$).


Extraction of the grism spectra requires a catalogue of sources with accurate positions from the associated imaging data. We produced this catalogue from the F140W stacked image using {\sc SExtractor} \citep{sextractor}. We used a detection threshold of $1.5\sigma$, a minimum area for detection of 5 contiguous pixels, and we adjusted the deblending parameter to optimally extract any objects visually identified as being blended with the quasar point spread function (${\rm DEBLEND\_MINCONT}=0.005$).

\begin{figure}
	\includegraphics[width=\columnwidth]{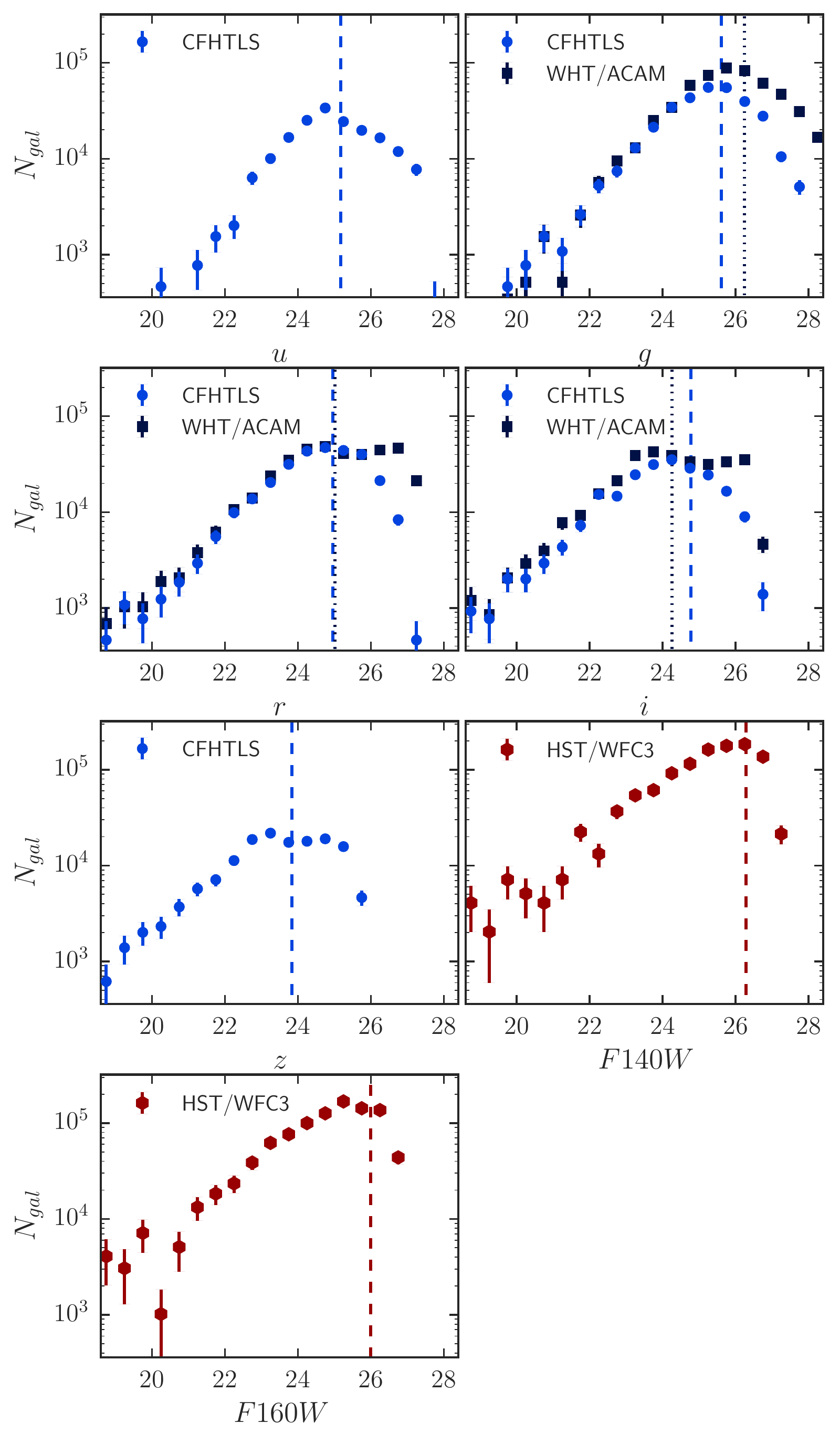}
	\caption{Galaxy number counts estimated for all of the broad band imaging used in this work, obtained from single image source extraction using {\sc SExtractor}. Vertical lines show the estimated 80\% completeness limit for point sources. Where both CFHTLS and WHT/ACAM data are available, the dashed line denotes the CFHTLS limit and the dotted line shows the WHT/ACAM limit.}
	\label{fig:numcount}
\end{figure}

The number counts of sources detected in the WFC3 F140W and F160W imaging are given in the final two panels of Fig.~\ref{fig:numcount}, whilst the corresponding 80\% completeness depths are given in Table~\ref{tab:imagingprops}. The completeness estimates were made by iteratively placing model sources (Gaussian profiles with the instrument full-width half-maximum - FWHM) directly into the imaging data, and re-running {\sc SExtractor} with our given parameters on these modified images.

\subsection{WFC3 G141 grism data}

The catalogues and segmentation images generated with \textsc{SExtractor} were then used to extract the spectrum of each object using \textsc{Grizli}. First, each G141 exposure was divided through by the F140W flat field, neglecting any wavelength-dependent flat field effects which are at most a few percent \citep{2016ApJS..225...27M}. An initial background was subtracted using the `master sky' images from \citet{2015wfc..rept...17B}.
These master images are necessary to account for the variation in background structure due to variations in zodiacal continuum, scattered light, and He emission across the sky \citep{2016ApJS..225...27M}.
The remaining residuals (typically $0.5-1 \%$ of the initial background levels) were then removed by subtracting average values of the sky pixels in each column. Finally 1D and 2D spectra were extracted from the background-subtracted grism images at the individual exposure level, which, for the \textsc{QSAGE} observations, resulted in 16 individual spectra per object.

An unavoidable feature of slitless spectroscopy is contamination from nearby sources. Despite the fact that this effect is substantially mitigated by the four independent grism orientations employed in the \textsc{QSAGE} survey, we nevertheless estimated the quantitative contamination for each individual spectrum. \textsc{Grizli} initially generates models for each object in the catalogue assuming a linear continuum, based on their observed magnitudes. These models are then refined using a second-order polynomial fit directly to the observed spectra after subtracting off the initial contamination estimate. The contamination model of each spectrum can then be used to mask severely contaminated pixels when producing the final stacked spectrum across all orientations as discussed below.

\begin{figure*}
	\includegraphics[width=\columnwidth]{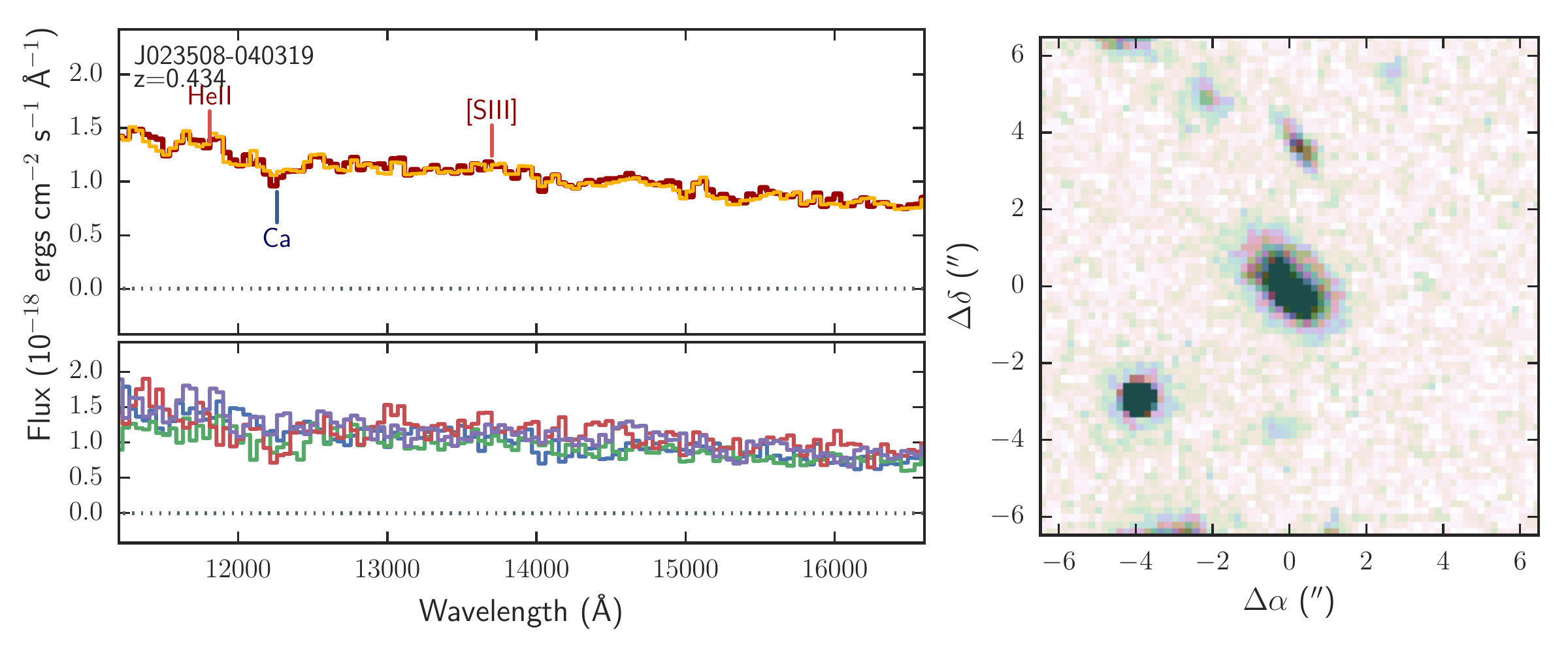}
	\includegraphics[width=\columnwidth]{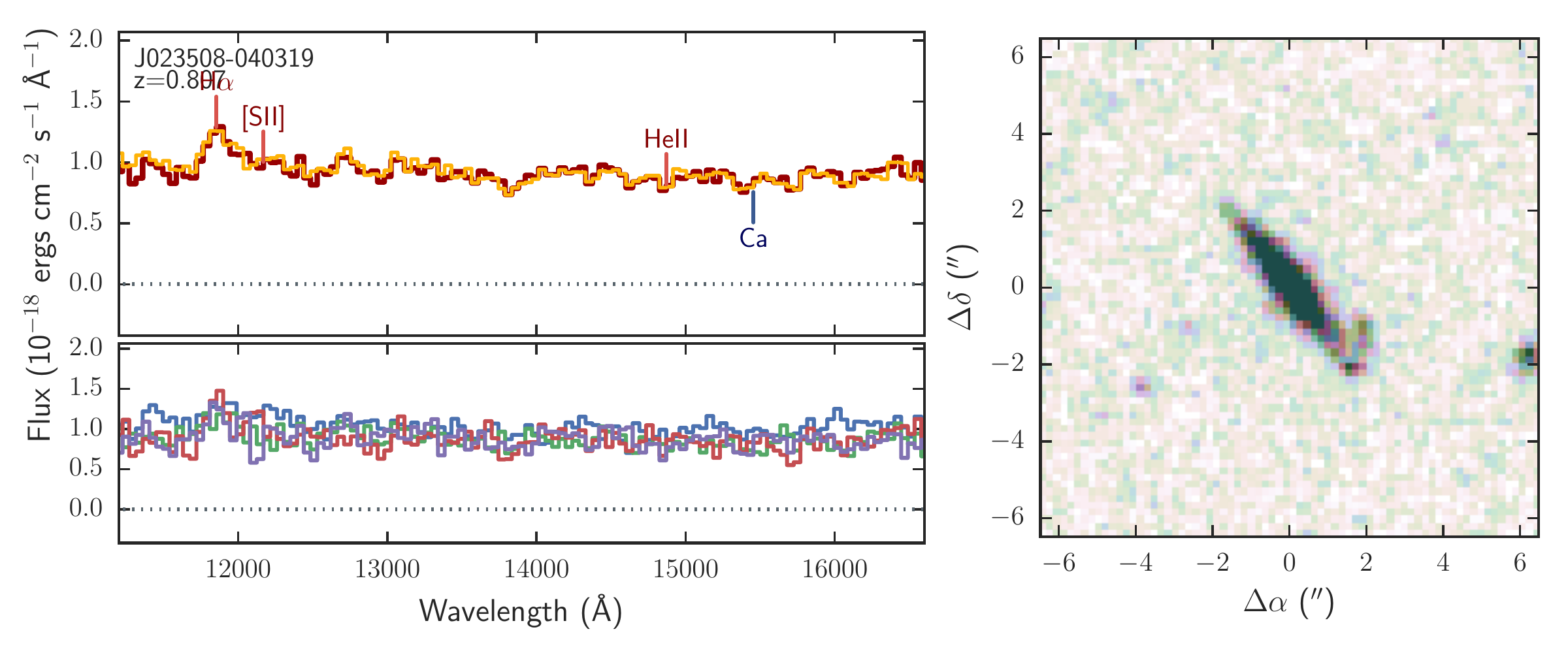}
	\includegraphics[width=\columnwidth]{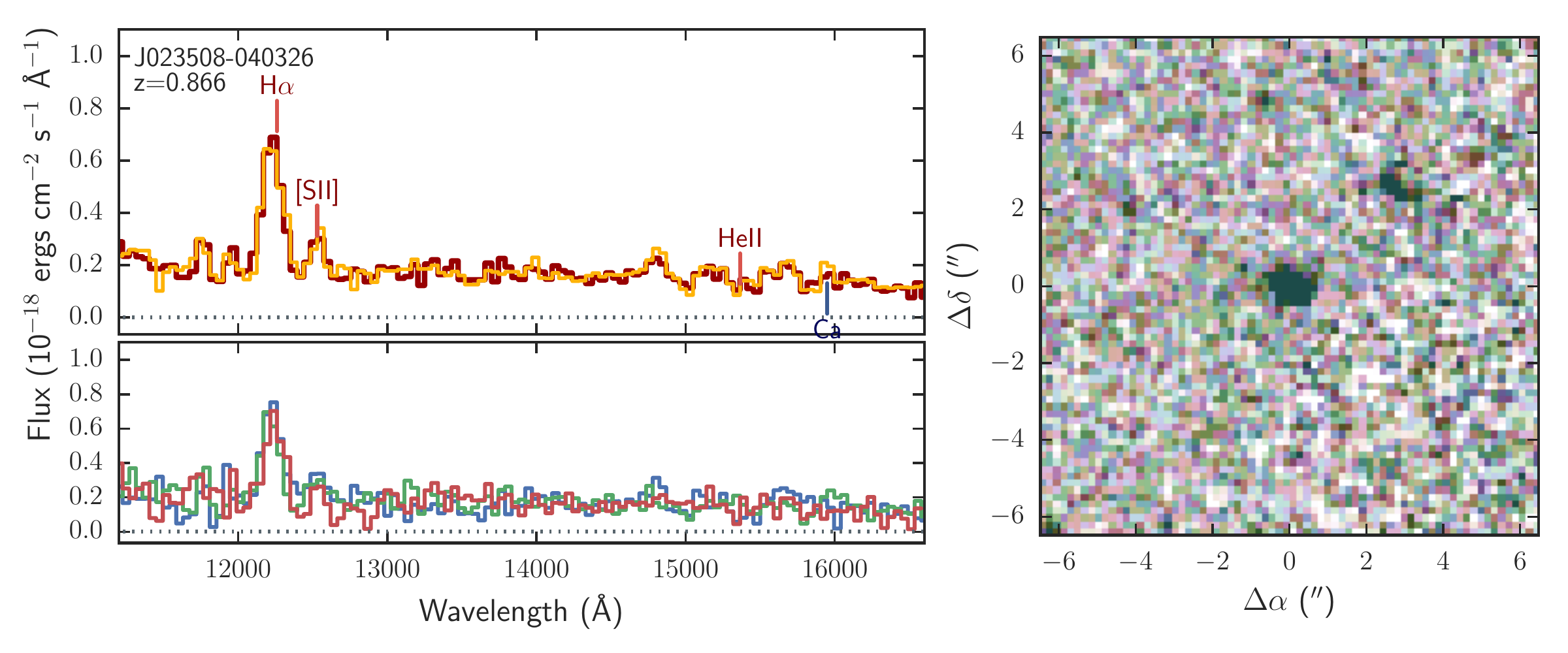}
	\includegraphics[width=\columnwidth]{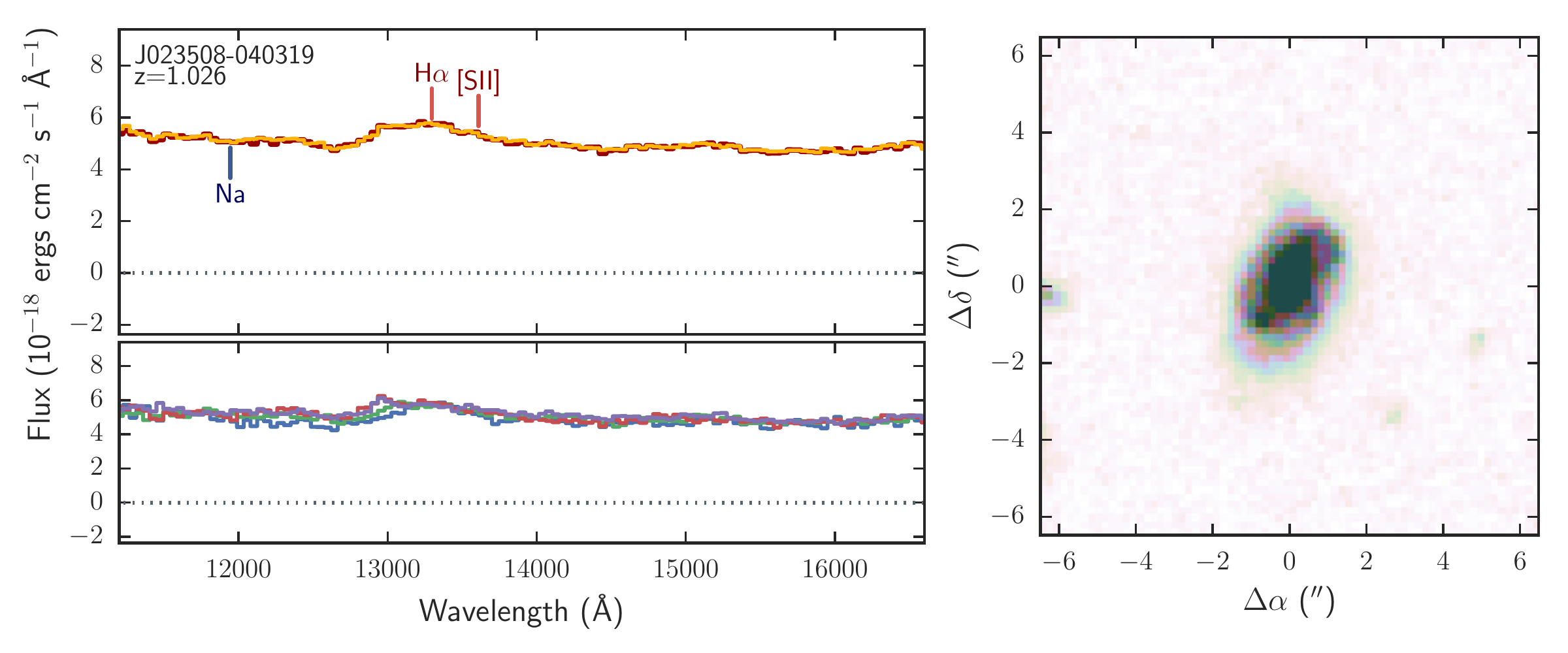}
	\includegraphics[width=\columnwidth]{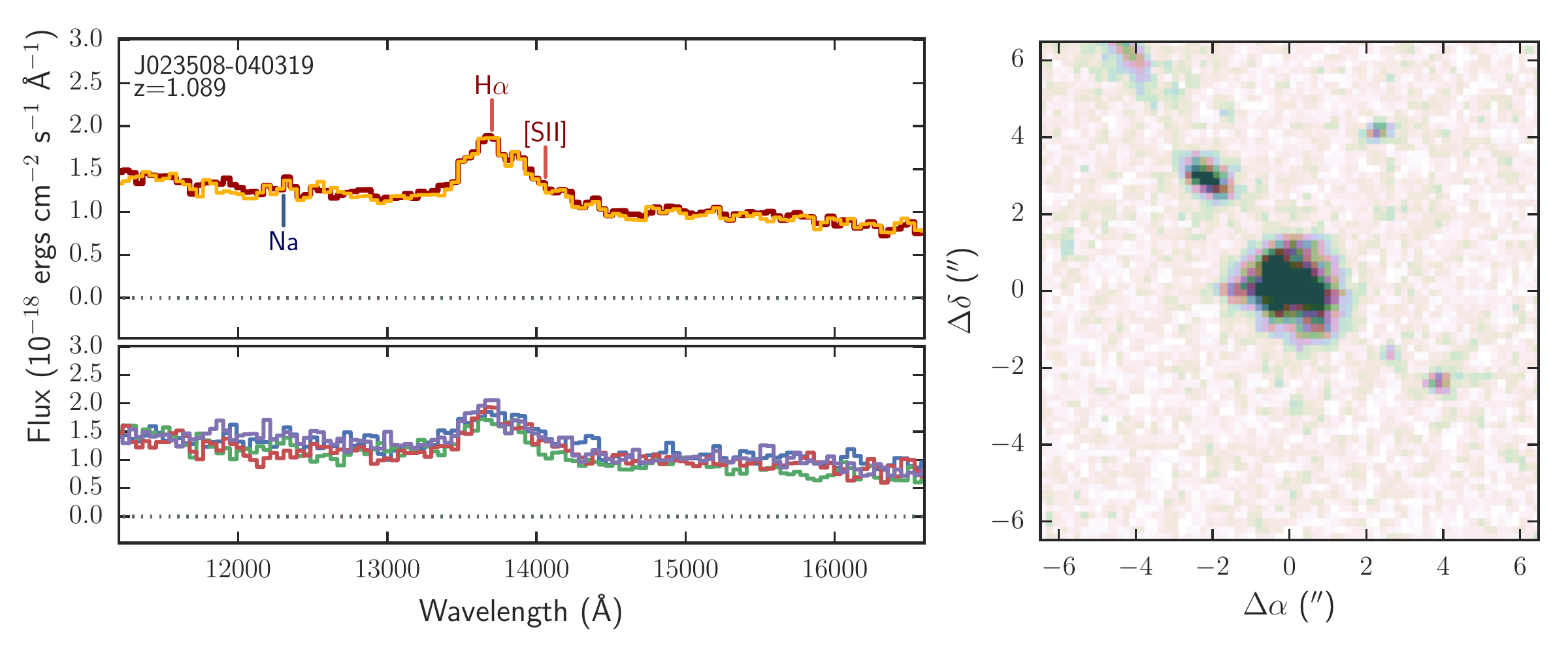}
	\includegraphics[width=\columnwidth]{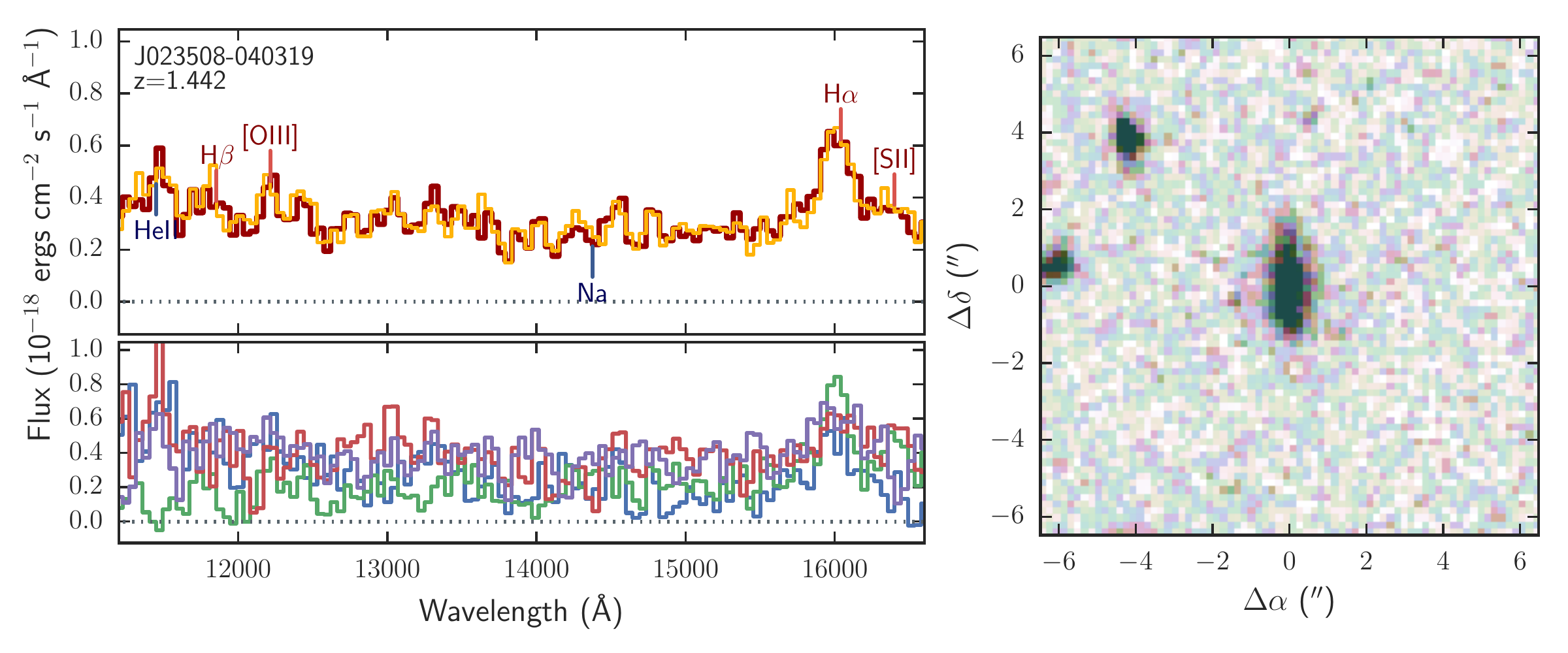}
	\includegraphics[width=\columnwidth]{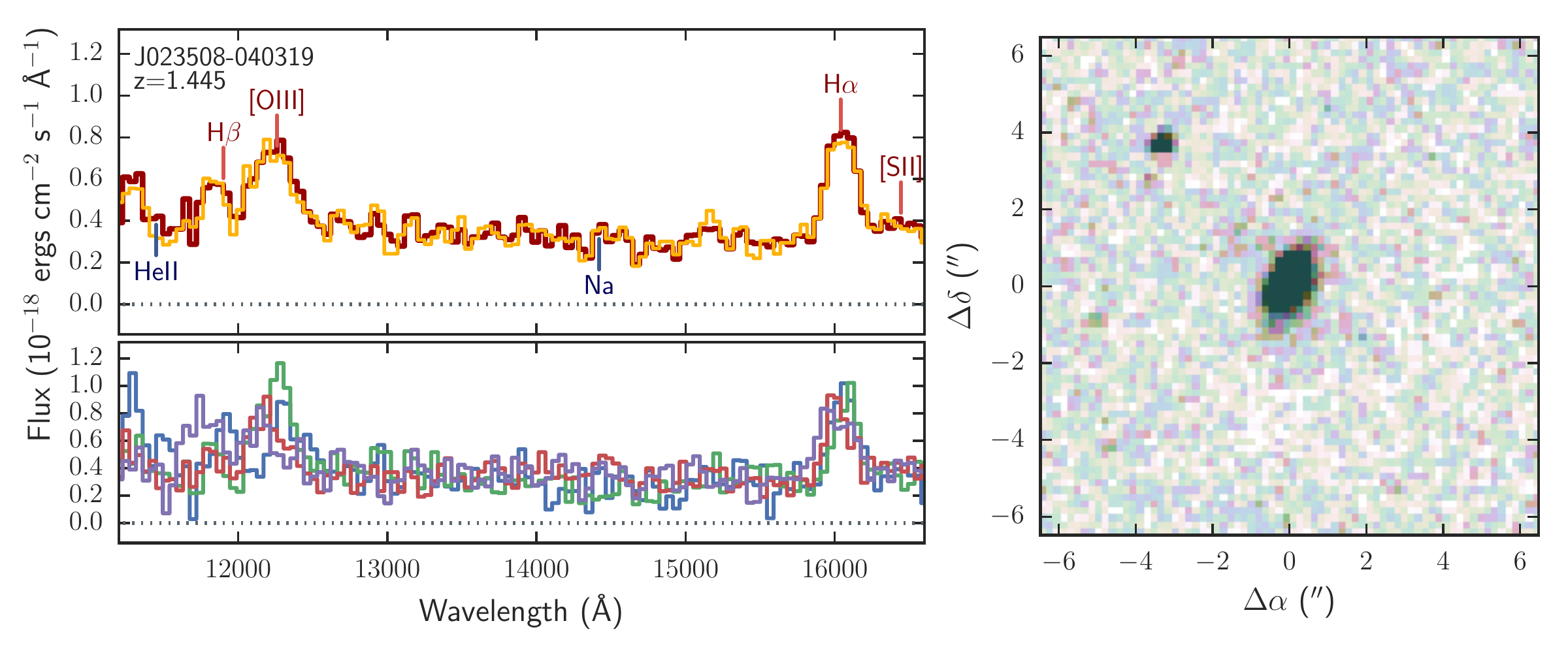}
	\includegraphics[width=\columnwidth]{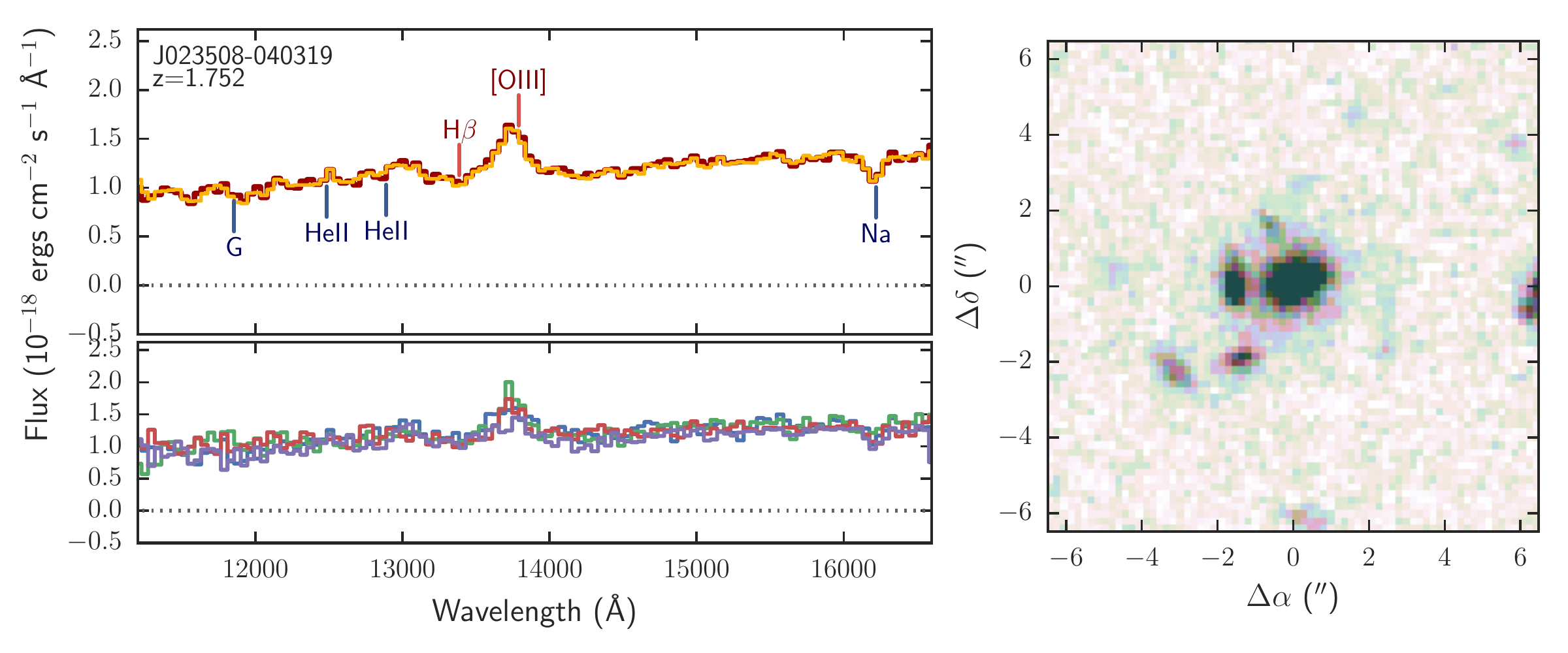}
	\includegraphics[width=\columnwidth]{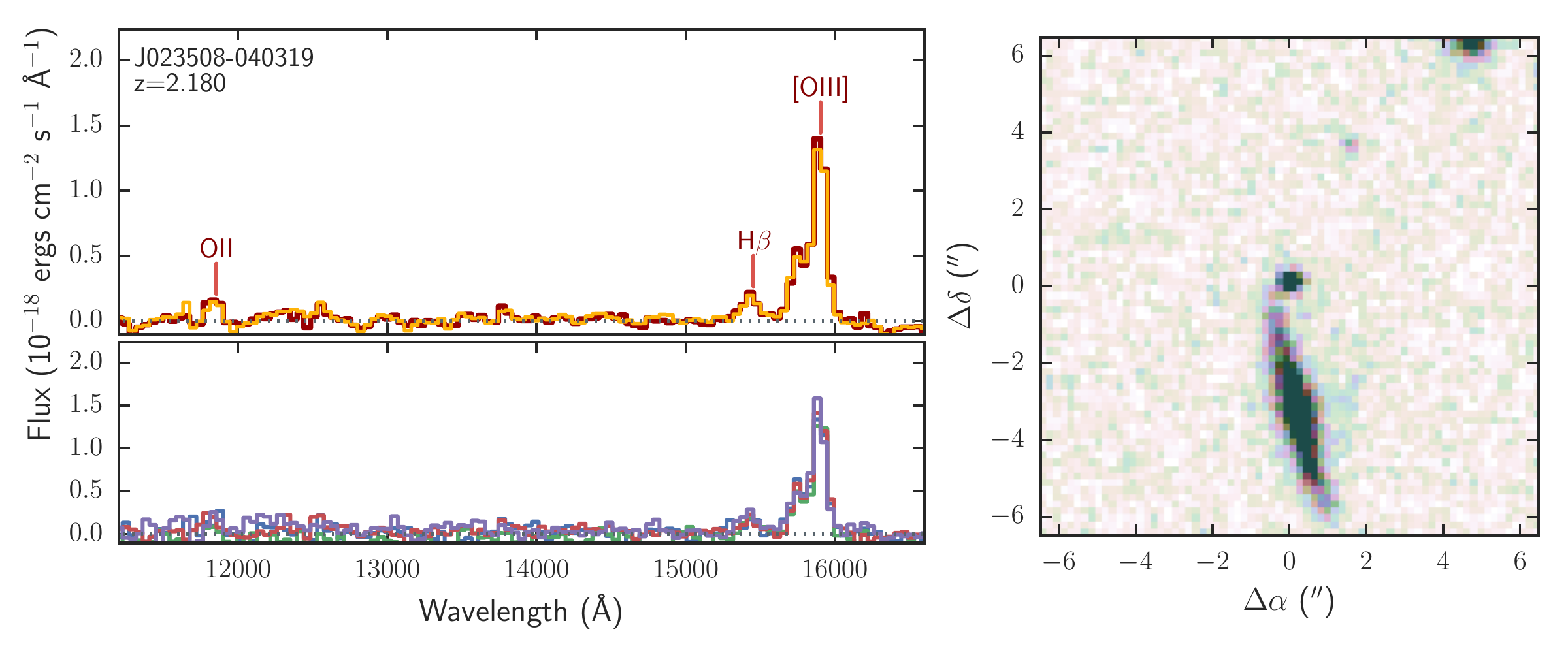}
	\includegraphics[width=\columnwidth]{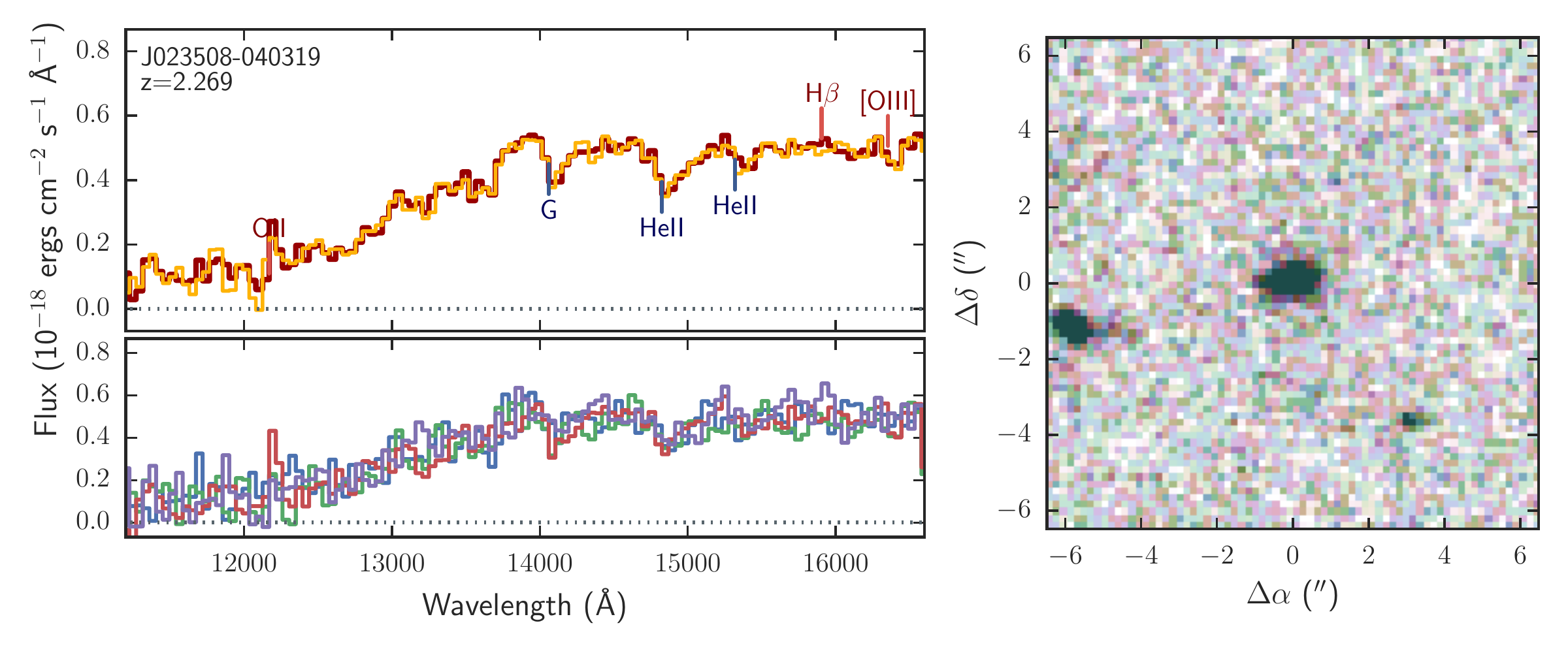}
	\caption{Example objects from the HST WFC3 grism data. In each case the top-left panel shows the median spectrum (orange) and contamination-masked mean spectrum (dark red). The lower-left panel in each case shows the corresponding spectra taken at each roll angle, whilst the right hand panel in each case shows a thumbnail taken from the WFC3 F140W stacked image.}
	\label{fig:wfc3-exspec}
\end{figure*}

Redshifts for sources in the grism spectra were primarily identified from the 1D spectra using a combination of H$\alpha$, H$\beta$, [O~{\sc ii}], and [O~{\sc iii}] emission features. Given the low-resolution nature of the grism spectra, significant blending of lines can occur, e.g. [O~{\sc iii}] 4953\AA\ with [O~{\sc iii}] 5007\AA\, and  H$\alpha$ 6562\AA\ with [N~{\sc ii}] 6583\AA. In the higher signal to noise spectra these blends exhibit an asymmetry in the overall line profile, allowing (alongside the presence of H$\beta$ emission) reliable differentiation between these lines. This asymmetry (and any H$\beta$ emission) is often lost in lower signal-to-noise spectra and as such it can be difficult to differentiate between H$\alpha$ at $z\sim0.9$ and [O~{\sc iii}] at $z\sim1.6$. We therefore perform the line identification in 2 stages. Firstly, lines are identified using an automatic fitting algorithm which then determines matches in wavelength ratios to common emission lines. We then visually inspect each spectrum, identify reliable detections, and attribute each detection a quality rating on a scale of $Q_w=1$ to $Q_w=4$ (where 4 represents a high S/N, multiple emission line detection and 1 represents a low-S/N less secure redshift). We mitigate the effect of contamination between nearby sources by simultaneously inspecting: a median stacked spectrum of all 16 exposures; the 4 separate median spectral stacks from the different roll-angle positions; and a mean-combination stack where contaminated pixels in individual exposures are rejected based on the contamination model described above (see Fig.~\ref{fig:wfc3-exspec}). Whilst the observations are most sensitive to detecting emission line objects at $z>0.68$, in the absence of such emission lines, galaxies are also identifiable via absorption features and, at $z\gtrsim2$, the 4000\AA\ break (examples of which are shown in Fig.~\ref{fig:wfc3-exspec}).

We produce a catalogue of secure sources which require at least one of the following criteria to be met: multiple lines clearly observed in the grism spectrum; clearly asymmetric lines indicating blended [O~{\sc iii}] emission or H$\alpha$/[N~{\sc ii}] emission; or confirmed redshift from optical spectroscopy (i.e. VLT/MUSE data described below). The secure redshifts are used as a calibration/training set in our photometric redshift fitting analysis described below. With the photometric redshifts in hand, we then constrain the single blended emission line objects to the redshift most closely matched by the available photometric data. These objects are assigned a quality flag of $Q_w=2$. For the subset of targets falling within the VLT/MUSE data field of view, we also use the optical spectra as a guide - both using the existence or the lack-of expected emission line features in the MUSE data to guide the identification of the grism data in this region.

We show the on-sky distribution of spectroscopically identified sources in Fig.~\ref{fig:wfc3f140}, whilst example spectra from the grism data are shown in Fig.~\ref{fig:wfc3-exspec}. The magnitude distribution of sources identified in the WFC3 G141 grism spectroscopy is given in the bottom panel of Fig.~\ref{fig:specnm} and the redshift distribution is shown in Fig.~\ref{fig:nz}.

\subsection{VLT MUSE IFU Spectroscopy}

\subsubsection{Overview and Data Reduction}

The VLT/MUSE data for this field were taken as part of the Guaranteed Time MUSE Program (096.A-0222, PI: Schaye). MUSE provides IFU data across a $1'\times1'$ field of view, with a pixel scale of $0.2''$/pixel \citep{2010SPIE.7735E..08B}. All the observations were taken in the normal spectral mode with a wavelength coverage of 480-930~nm and resolution ranging from $R=1770$ at the lower end of the wavelength range to $R=3590$ at the higher end. A total of 12 exposures were taken, each of 900s, totalling 3hrs of integration on source. We have reduced the entirety of these data using the ESO MUSE pipeline \citep{2014ASPC..485..451W} with custom python scripts\footnote{\url{https://github.com/mifumagalli/mypython}} \citep{2014MNRAS.445.4335F, 2017MNRAS.467.4802F} to optimize the illumination correction across different CCDs and using {\sc zap} \citep[Z\"{u}rich Atmosphere Purge;][]{2016MNRAS.458.3210S} to optimise the sky subtraction.

\subsubsection{Source extraction and identification}

Given the deep nature of the WFC3 NIR imaging, we use this as the basis for the source identification in the MUSE data. We first match the astrometry between the MUSE cube and the WFC3 F140W image (again using {\sc scamp}, \citealt{scamp}) and then run {\sc SExtractor} on the F140W image.  From the resulting catalogue, we extract 1D spectra from the MUSE cube along with the variance and sky background measured for each spectrum. We use {\sc marz} \citep{2016A&C....15...61H} to measure redshifts from the spectra based on template fitting and visual inspection. For this we use A. Griffiths (Sept. 2016) fork of the {\sc marz} tool\footnote{\url{https://a-griffiths.github.io/Marz/}}, which includes templates covering redshifts up to $z\sim6$ (for our wavelength coverage). Each identification is allocated a confidence flag, $Q_m$ based on the following categorization:

\begin{itemize}
	\item 1 - Low confidence/low S/N (can be single or multiple possible features);
	\item 2 - Single line emitter, low S/N continuum or continuum fit with weak or no emission lines;
	\item 3 - Single line emitter, some hint of absorption lines in continuum or multiple emission lines with some at low signal-to-noise;
	\item 4 - multiple high S/N emission and/or absorption lines;
	\item 6 - Star.
\end{itemize}
The single line emitters are predominantly [O~{\sc ii}] emission at $0.88\lesssim z\lesssim1.5$ (which is partially resolved in the MUSE data) or Ly$\alpha$ at $z\gtrsim2.9$ (which can be identified as such via asymmetries and continuum shape).

\begin{figure}
	\includegraphics[width=\columnwidth]{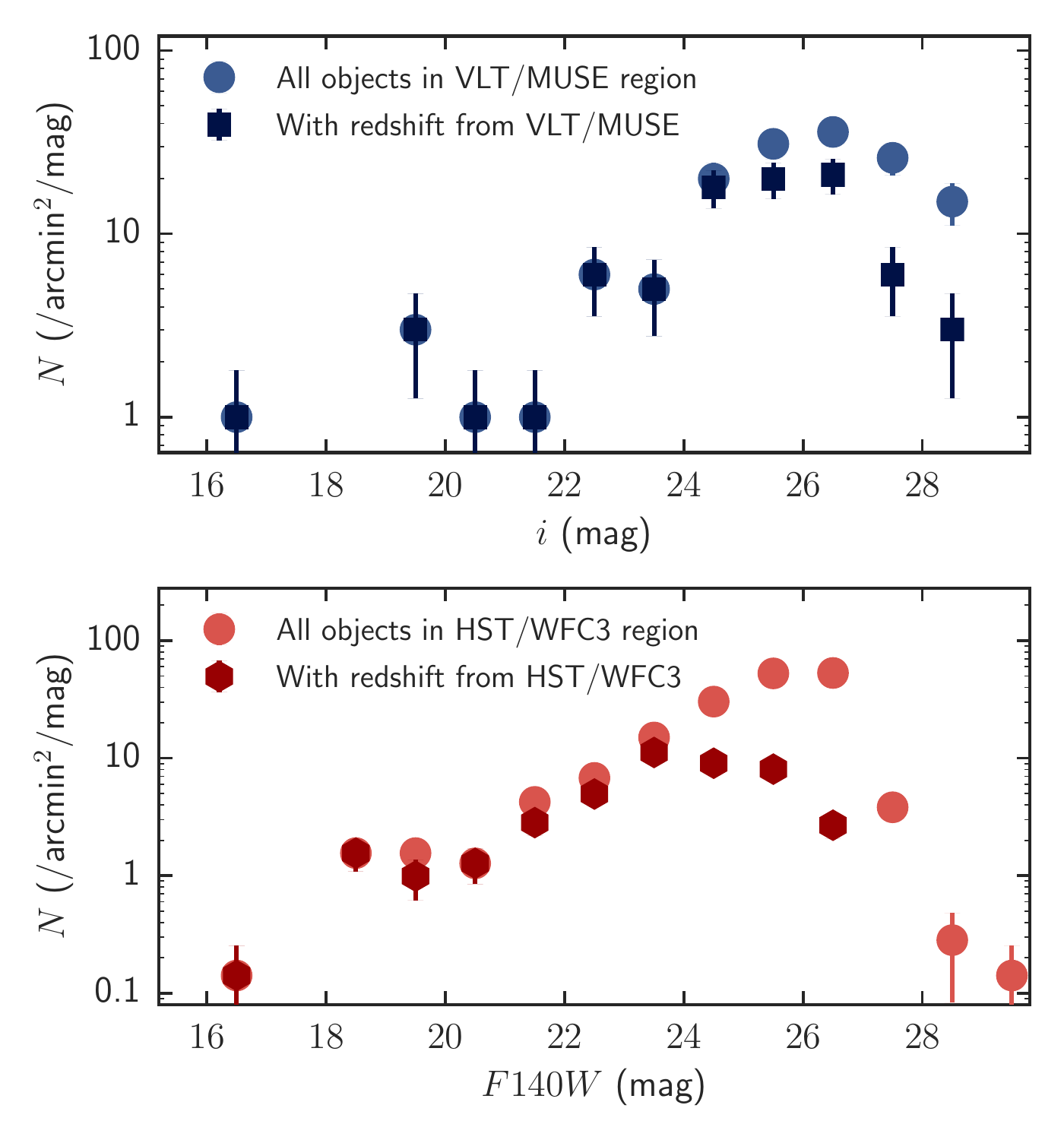}
	\caption{Magnitude distribution of sources in the MUSE (top panel) and WFC3 (lower panel) field of views. In both panels, the fainter circles denote number counts for all detected sources, whilst the darker circles show those with successfully identified redshifts from the MUSE and WFC3 grism spectroscopic data.}
	\label{fig:specnm}
\end{figure}

\begin{figure}
	\includegraphics[width=\columnwidth]{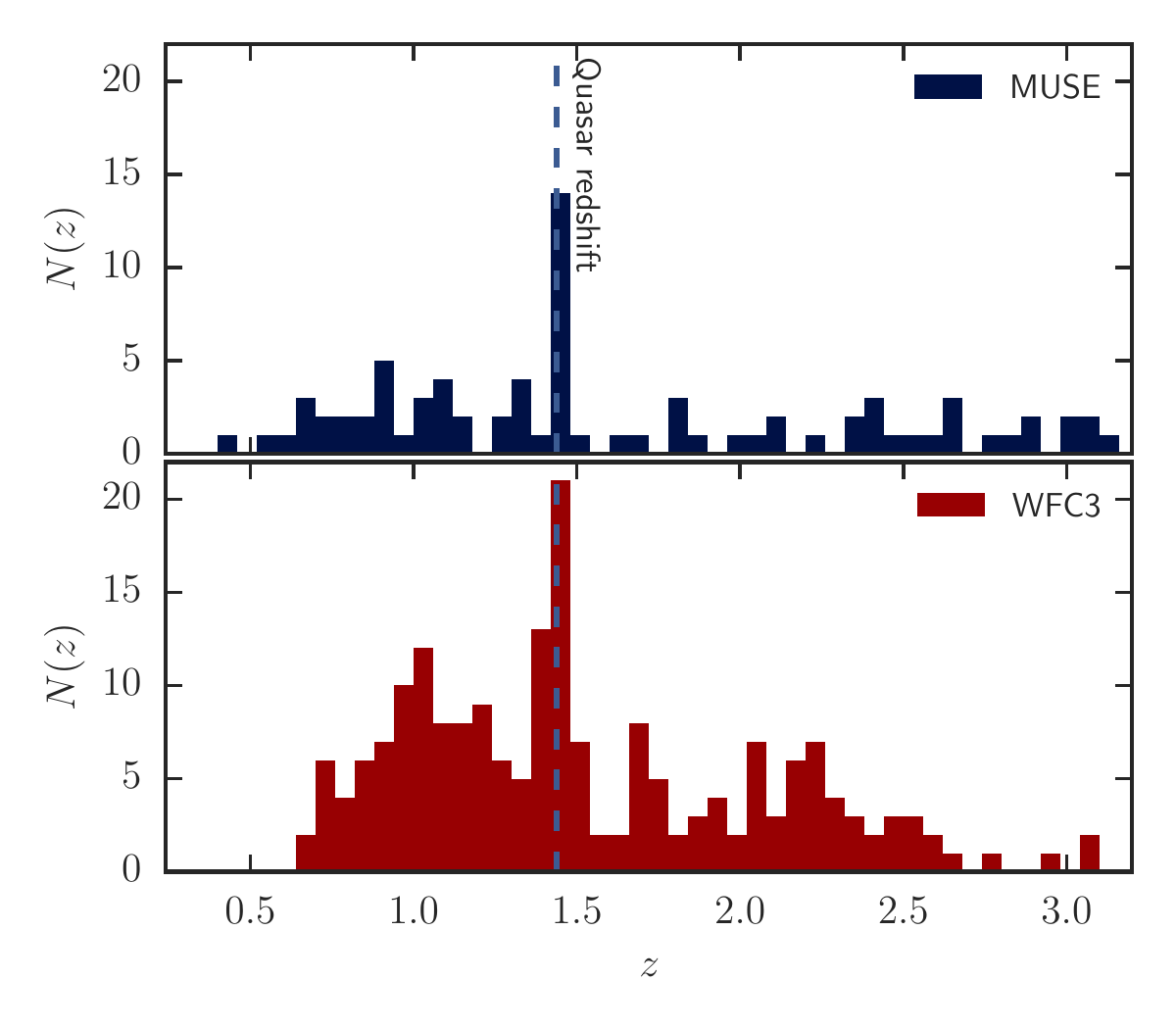}
	\caption{Redshift distribution of the MUSE spectroscopic redshifts (top panel) and the HST/WFC3 grism redshifts (lower panel). The vertical dashed line marks the redshift of the central quasar, PKS~0232-04.}
	\label{fig:nz}
\end{figure}

The magnitude distribution of sources in the MUSE field of view are shown in Fig.~\ref{fig:specnm} (top panel). The pale blue circles show the total number of sources extracted from the WFC3 data in the MUSE field-of-view, whilst the dark blue circles show the magnitude distribution of sources identified with confidence flags of $Q_{\rm m}=2$, 3, 4, and 6. The spectroscopic identification is $>50\%$ complete up to $i<25.8$ (where the $i$ band photometry is taken from the broad band imaging described later). The redshift distribution of sources identified in the MUSE cube is given in the top panel of Fig.~\ref{fig:nz}. The quasar redshift is marked by the vertical dashed line and coincides with an over-density of [O~{\sc ii}] emitting galaxies at $z\approx1.4$. The redshift distribution appears to show some structure with a particular peak at the redshift of the QSO. For a detailed discussion of the QSO environments please see Stott et al. (in prep).

\begin{figure}
	\includegraphics[width=\columnwidth]{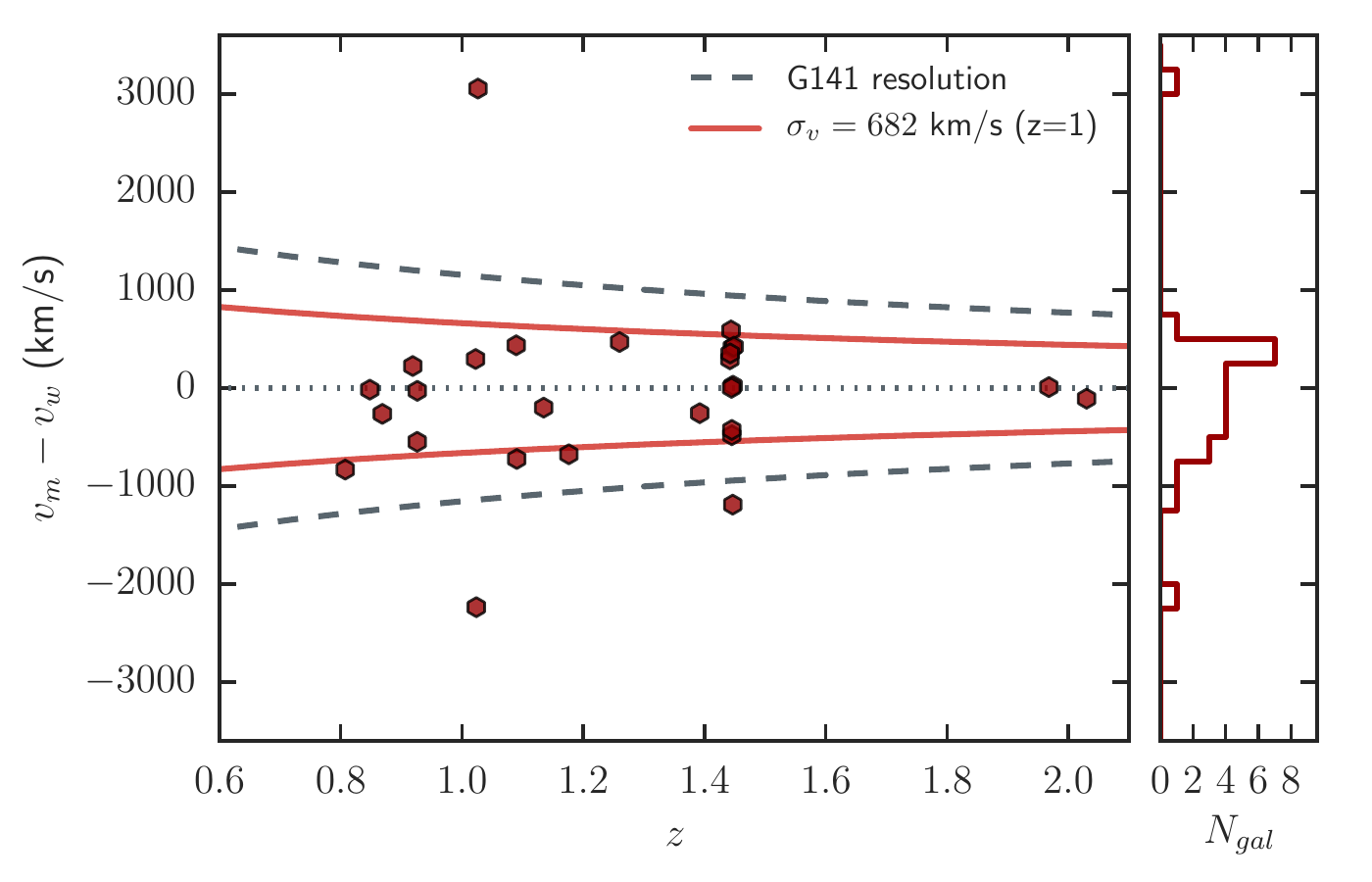}
	\caption{Rest-frame velocity offsets between redshifts for the same objects measured separately with VLT/MUSE and the HST/WFC3 G141 grism (red hexagons and histogram). The dashed curves denotes the rest-frame velocity scale given by $\pm c/R/(1+z)$, where $R=130$ is the instrumental spectral resolution of the G141 grism. The achieved velocity accuracy of $\sigma_v=682$~km~s$^{-1}$ (at $z=1$) is represented by the solid red curves.}
	\label{fig:wfc3-velacc}
\end{figure}

Using the MUSE spectroscopic data, we are able to measure the accuracy of the WFC3 grism derived redshifts. The grism data are at a low resolution - $R=130$, which equates to a rest-frame velocity uncertainty of $\approx1000$~~km~s$^{-1}$ at $z\approx1$. With a resolution of $R\approx1800-3500$, MUSE has greater redshift accuracy (i.e. $\approx50$~~km~s$^{-1}$ at $z\approx1$), allowing a robust quantification of the grism line-fitting process's accuracy. A comparison of the redshifts determined using WFC3 and MUSE as a function of redshift is shown in Fig.~\ref{fig:wfc3-velacc}, with $v_m-v_w$ giving the velocity offset between the MUSE and WFC3 redshifts in km~s$^{-1}$. The dashed curves show the velocity uncertainty derived simply taking the G141 resolution limit of $R=130$. From the 19 objects matched between the two samples, we find a velocity uncertainty on the G141 data of $\sigma_{v}=680$~km~s$^{-1}$ (at $z=1$). A closer inspection of the 2 outliers ($|\Delta v|>2,000$~km~s$^{-1}$) shows that one is a very extended bright object, where the emission line is heavily `smeared' in the grism spectrum due to the internal kinematics of the galaxy; whilst the second appears to be a statistical outlier due to spectral noise.

\subsection{Optical imaging data}

Supporting imaging is an important element in the analysis of the grism data, in particular in providing complementary photometric redshifts and galaxy properties. As described above, the photometric redshifts are needed for robust identification of objects detected with only single line features in the grism data. Below we describe the imaging datasets available on this field: 1) targeted William Herschel Telescope (WHT) auxiliary-port camera (ACAM) observations; and 2) coverage from the CFHTLS Wide-1 field. 

\subsubsection{WHT ACAM}

The WHT ACAM imaging provides $i\sim25$~mag depth imaging, with the SDSS $g$, $r$, and $i$ filters. ACAM is mounted at the folded-Cassegrain of the WHT and covers a field of view with diameter $\diameter\approx8'$ at a pixel scale of $0.25''$/pixel. All observations were taken in observing runs W13BN5 (24-25th November 2013, PI: R. Crain), W14AN16 (6-7th April 2014, PI: R. Crain), and W17AP6 (3-5th March 2017, PI: R. Bielby). Seeing conditions were generally good with an image quality of $\approx0.8-0.9''$.

The ACAM data were reduced using standard image reduction methods. Master bias and flat-field frames were produced by stacking individual calibration frames, which were then applied to the science images. Individual weight maps for use with {\sc Astromatic}\footnote{\url{https://www.astromatic.net/}} software were produced from the flat field images combined with a bad pixel mask. Astrometric solutions were derived for the images in each field using {\sc Scamp} and these were then used in constructing stacked images using {\sc Swarp}.

Photometric calibration was performed by matching to the available (shallower) SDSS photometry in the region. Number counts for each of the filters are shown in Fig.~\ref{fig:numcount} (dark blue squares), based on source catalogues extracted using {\sc SExtractor} on the individual images. The corresponding depths, based on 80\% completeness of artificial point sources placed in the images, are given in Table~\ref{tab:imagingprops}.

\begin{table}
\caption{Overview of the imaging data used in this paper. Completeness estimates are based on an 80\% limit for point sources (with a $2\sigma$ detection threshold). Values for the CFHTLS data are reproduced directly from the CFHTLS synoptic table.}
\label{tab:imagingprops}
\centering
\begin{tabular}{lccc}
\hline
Instrument   & Filter         & IQ       & Completeness\\
             &                &          & (80\%, Mag)\\
\hline
WHT/ACAM     & $g_{\rm sdss}$ & $0.87''$ & 26.24       \\
WHT/ACAM     & $r_{\rm sdss}$ & $0.88''$ & 25.01       \\
WHT/ACAM     & $i_{\rm sdss}$ & $0.88''$ & 24.26       \\
CFHT/MegaCAM & $u_{\rm cfht}$ & $0.90''$ & 25.17 \\
CFHT/MegaCAM & $g_{\rm cfht}$ & $0.74''$ & 25.61 \\
CFHT/MegaCAM & $r_{\rm cfht}$ & $0.72''$ & 24.96 \\
CFHT/MegaCAM & $i_{\rm cfht}$ & $0.68''$ & 24.78 \\
CFHT/MegaCAM & $z_{\rm cfht}$ & $0.72''$ & 23.84 \\
HST/WFC3     & F140W          & $0.26''$ & 26.28 \\
HST/WFC3     & F160W          & $0.29''$ & 26.00 \\
\hline
\end{tabular}
\end{table}

\subsubsection{CFHTLS Wide}

The central quasar studied here falls $\approx3'$ from the Eastern edge of the CFHTLS \citep{2012SPIE.8448E..0MC} W1 (wide) field (specifically in the `W1.+4+3' region). The CFHTLS Wide survey provides relatively uniform imaging across a large area with the CFHLT $ugriz$ filter set. Image quality is constrained to $\approx0.9''$ across the survey, with the imaging reaching depths of $u,g,r,i\approx25$ and $z\approx24$ (AB), i.e. well suited to our needs. We note for reference that almost all of the W1 field has been surveyed spectroscopically to $i\approx22.5$ as part of the VIPERS survey \citep{2018A&A...609A..84S}, however the central quasar studied here lies marginally outside the extent of that survey ($12'$ from the nearest VIPERS spectroscopic data point).

We downloaded the publicly available median stacked images for each of the 5 bands in W1.+4+3 from CADC (Canadian Astronomy Data Centre\footnote{\url{http://www.cadc-ccda.hia-iha.nrc-cnrc.gc.ca/}}). Using {\sc scamp}, these were then matched to the WFC3 imaging astrometry and then resampled and cropped using SWarp to match the WFC3 stack field-of-view. The image seeing and 80\% point-source completeness levels are reproduced in Table~\ref{tab:imagingprops} (taken directly from the CFHTLS synoptic table\footnote{\url{http://terapix.iap.fr/cplt/T0007/table_syn_T0007.html}}).  We show the number counts for objects detected in each of the bands across the WFC3-stack field of view in Fig.~\ref{fig:numcount}, based on catalogues extracted from the images using {\sc SExtractor}.

\subsection{Galaxy Properties}

\subsubsection{Collated Photometric Catalogue}

In addition to the individual F140W catalogue for grism extraction, we also produce a collated catalogue of all 10 ACAM, CFHTLS and WFC3 filters. To do so, all images were astrometrically matched and resampled to a common grid using a combination of the {\sc scamp} and {\sc swarp} software. {\sc SExtractor} was then run in dual image mode using the F140W image as the detection image.

\subsubsection{Photometric Redshifts}

Photometric redshifts are determined using the {\sc Le Phare} photometric fitting code \citep{1999MNRAS.310..540A,ilbert06}, using the `COSMOS' set of SED templates (as used by \citealt{2010ApJ...709..644I}). The template SED fitting was performed on all the available imaging data and using all secure spectroscopic galaxy redshifts from the MUSE IFU and WFC3 grism data (i.e. $Q_{w,m}=3$ and $Q_{w,m}=4$) as a training set. As discussed above, the photometric redshifts for the single line emitters with non-secure redshifts were then incorporated into optimising the redshift identification for these objects. 

\begin{figure}
	\includegraphics[width=\columnwidth]{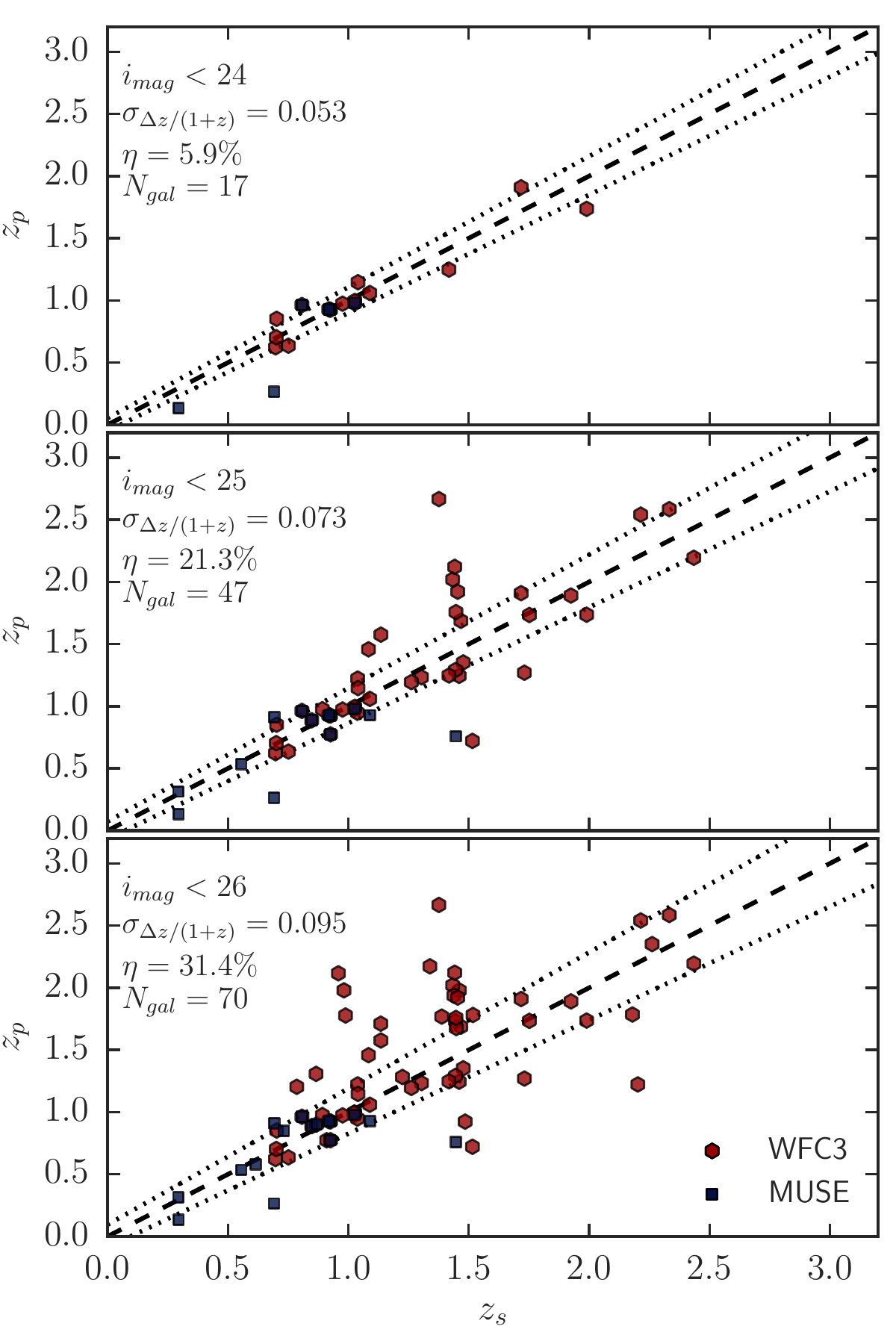}
	\caption{Photometric ($z_{\rm p}$) versus spectroscopic ($z_{\rm s}$) redshifts for all galaxies identified with confidence $Q_{op}=3$ or $Q_{op}=4$ in the HST/WFC3 grism (red hexagon points) or the VLT/MUSE (blue square points) data. The dashed line shows $z_{p}=z_{s}$ and the dotted lines show the $1\sigma$ on the distribution of $(z_{p}-z_{s})/(1+z_s)$.}
	\label{fig:photozacc}
\end{figure}

A comparison between the {\sc Le Phare} photometric redshifts and the available spectroscopic data from WFC3 and MUSE is shown in Fig.~\ref{fig:photozacc} for magnitude limits of $i_{mag}<24$, $i_{mag}<25$ and $i_{mag}<26$ (top, middle and bottom panels respectively). The red hexagons denote spectroscopic redshifts from WFC3, whilst the blue squares denote those from MUSE. In both cases, we only use those with confidence flags of 3 or 4 (given that objects with flags of 2 have partially been assigned redshifts with the photometric redshifts as a prior). Where an object is detected with both WFC3 and MUSE, we take the more accurate MUSE redshift. For each sample we quote the resultant photometric redshift accuracy, $\sigma_{z/(1+z)}$, and outlier percentage, $\eta$.

The photometric fitting shows reliable results at $i<24$, with an accuracy of $\sigma_{z/(1+z)}=0.053$ and an outlier rate of $\eta=5.9\%$, whilst the results degrade at $i>24$ as the photometric uncertainties increase. We note that the varying image quality across the imaging bands could cause issues for the photometric fitting, this is countered to some degree by the adaptive fitting incorporating the spectroscopic training set, which determines and applies constant offsets (of $\Delta m\lesssim0.1$) to the input photometry.

\subsubsection{Star Formation Rates}

Over the main redshift range of interest ($0.68<z<1.44$), we primarily use the H$\alpha$ fluxes measured from the grism data to estimate star-formation rates for the individual galaxies in our sample. Using the relation given in \citet{1998ARA&A..36..189K}, divided by a factor of 1.8 to convert from Salpeter to Chabrier IMF. The galaxy intrinsic H$\alpha$ luminosities are estimated using an extinction correction of $A_{H\alpha}=0.818A_V$ \citep{1989ApJ...345..245C}, where we use a value of $A_V=1$ (consistent with the mean absorption estimated from the template fitting with {\sc Le Phare} and with \citealt{2010MNRAS.409..421G} for galaxies of mass $M_{\star} \approx 10^{9.5-10}$~M$_\odot$). We remove the estimated [N~{\sc ii}] contribution to the blended H$\alpha$-[N~{\sc ii}] emission assuming log([N{\sc ii}]/H$\alpha)=-1.0$ \citep{2018ApJ...855..132F}.

We supplement the H$\alpha$ derived SFRs with [O~{\sc ii}] SFR estimates where [O~{\sc ii}] emission is observed either in the grism data or the MUSE data. SFRs are estimated from the [O~{\sc ii}] fluxes based on the relationship given by \citet{1998ARA&A..36..189K}, using $A_V=1$ and $A_{[OII]}=1.54A_V$ based on \citet{1989ApJ...345..245C}. We note that for the objects where both [O~{\sc ii}] and H$\alpha$ are detected (i.e. a number of the galaxies observed with both MUSE and WFC3), we find the two indicators give on average consistent measures of the SFR, albeit with significant scatter. In these instances, where we have both [O~{\sc ii}] and H$\alpha$ measures of the SFR, we favour the H$\alpha$ for the analyses presented here given that it will be less affected by extinction and have a lesser dependence on metallicity and ionization parameter. The resulting galaxy SFRs are given as a function of redshift in the upper-left panel of Fig.~\ref{fig:photprops}. We find a H$\alpha$-derived SFR systematic lower-limit in the primary redshift range of interest in this study (i.e. $0.68<z<1.44$) of $\approx0.2-0.8~{\rm M}_\odot{\rm yr}^{-1}$. For galaxies with upper limits on the H$\alpha$ and/or [O~{\sc ii}] fluxes, we propagate these upper limits to determine upper limits on the SFR. 

\subsubsection{Galaxy Stellar Masses}

Given the catalogue of spectroscopic and photometric redshifts for our sample, we next derive galaxy stellar masses also using {\sc Le Phare} (via the {\sc GazPar} online interface\footnote{https://gazpar.lam.fr}). Following \citet{2010ApJ...709..644I}, we use a set of SED templates calculated using the stellar population synthesis models of \citet{bruzualcharlot03}. We assume a \citet{2003PASP..115..763C} IMF and an exponentially declining star-formation history. Dust extinction is then applied to the templates using the \citet{2000ApJ...533..682C} law,  with $E(B-V)$ in the range $0-0.7$.

\begin{figure*}
	\includegraphics[width=0.96\textwidth]{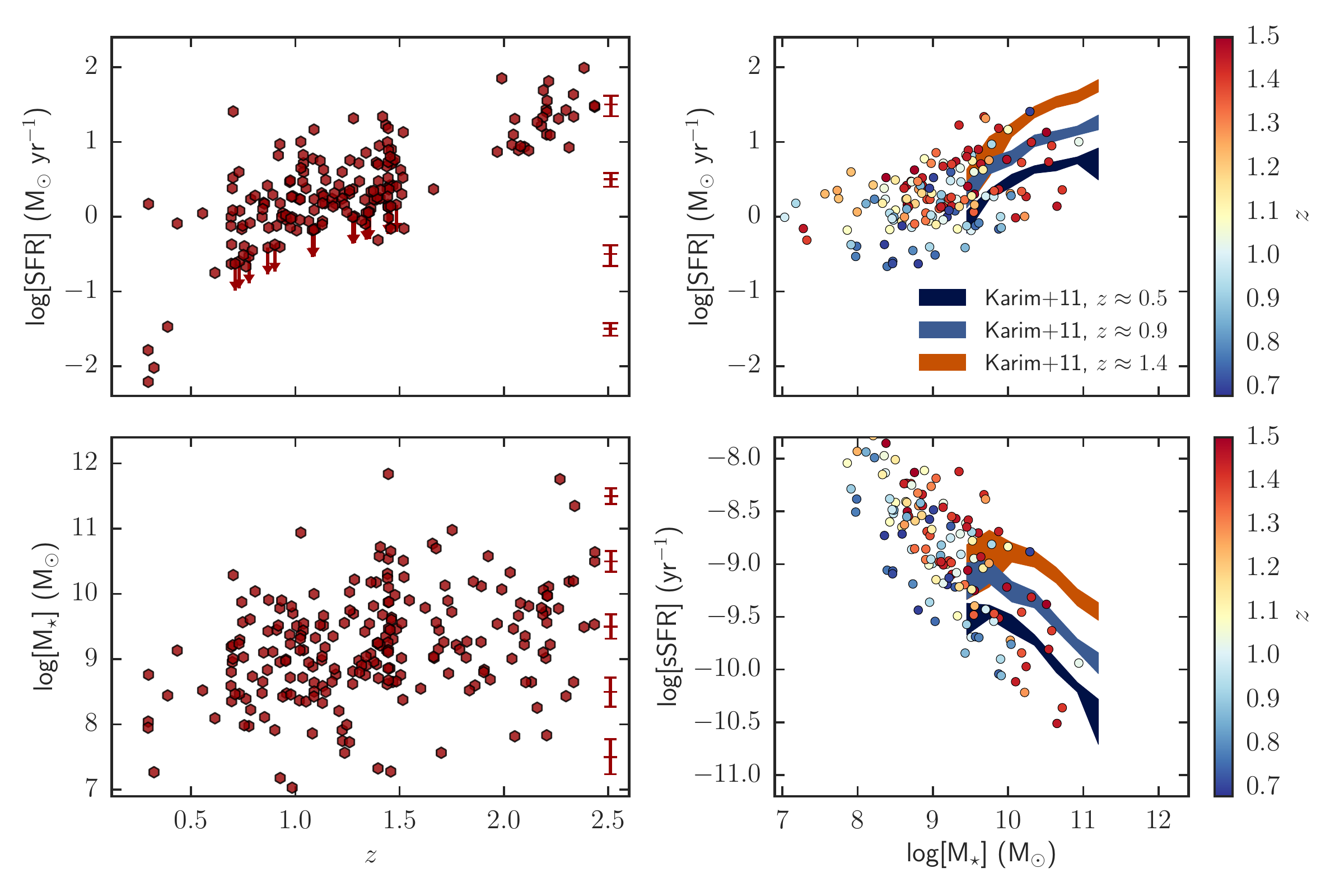}
	\caption{Derived photometric properties of the spectroscopically identified galaxy sample. SFRs were estimated primarily from measured H$\alpha$ emission where available and [O~{\sc ii}] where H$\alpha$ was not covered by the spectral range of the data. Galaxy masses were estimated using photometric fitting to the available photometric data. The error bars along the right hand axes of the left hand panels show the mean estimated uncertainty on the points in bins of 1~dex. In the right hand panels we limit the data to the prime redshift range of interest ($0.68<z<1.4$) and colour code points by redshift (as defined by the associated colour-bars). For reference, we also show the $M_\star-{\rm SFR}$ relations derived at $z\approx0.5$, $z\approx0.9$ and $z\approx1.4$ given by \citet{2011ApJ...730...61K} (solid curves color-coded by redshift, with the line-widths indicating the uncertainties on the mean SFR measurements).}
	\label{fig:photprops}
\end{figure*}

Stellar masses for the spectroscopically identified galaxy sample are given in the lower-left panel of Fig.~\ref{fig:photprops}. The depths of our NIR imaging correspond to approximate limiting masses of $M_\star\approx10^{7.5-8}$~M$_\odot$ across the redshift range $0.68<z<1.44$. In the right hand panels of Fig.~\ref{fig:photprops} we show the SFR and sSFR versus stellar mass for the galaxy sample in the redshift range $0.68<z<1.44$. Points are colour-coded by each galaxy's redshift. For comparison, the plotted line show the $z\approx1$ star-formation main-sequence trend reported by \citet{2011ApJ...730...61K}. From this, we can see that the galaxies lie on or around the star-forming main sequence.

\subsubsection{Inferred Galaxy Halo Properties}

Based on the derived stellar masses, we infer galaxy halo masses ($M_{\rm halo}$), virial radii ($R_{\rm vir}$), virial velocity dispersions ($\sigma_{\rm v}$) and virial temperatures ($T_{\rm vir}$) estimates for each of our galaxies. Clustering studies suggest a correlation between galaxy stellar mass and host halo mass over a wide range of redshift based on the $\Lambda$CDM prediction of distribution of dark matter \citep[e.g.][]{2011ApJ...728...46W, 2012A&A...542A...5C, 2014A&A...568A..24B, 2018MNRAS.475.3730C}. Here we infer halo masses based on the individual estimated stellar masses of the galaxies in our sample using the relations based on abundance matching presented by \citet{2013ApJ...770...57B}. From these halo masses, we then (with a number of simplifying assumptions), infer virial radii and virial temperatures for these structures \citep[e.g.][]{2014ApJ...784..142S, 2017ASSL..430..301V}. The virial radius for a given halo mass can be expressed as:

\begin{equation}
\centering R_{\rm vir} \approx 340~{\rm kpc} \left(\frac{M_{\rm halo}}{10^{12}{\rm M}_\odot}\right)^\frac{1}{3} \frac{1}{1+z}
\end{equation}

\noindent using a ``virial overdensity'' threshold of $18\pi^2$ \citep[e.g.][]{1998ApJ...495...80B,2017ASSL..430..301V}. The halo velocity dispersion, $\sigma_{\rm v}$ is estimated using the above halo mass and virial radius, whilst the corresponding virial temperature can be expressed as:

\begin{equation}
\centering T_{\rm vir}\approx3\times10^5~{\rm K}\left(\frac{M_{\rm halo}}{10^{12}{\rm M}_\odot}\right)^\frac{2}{3}(1+z)
\end{equation}

\noindent assuming a mean molecular mass of $\mu=0.59$ \citep[e.g.][]{2017ASSL..430..301V}. We make a note of caution that the environments in which a number of the galaxies in our survey exist are perhaps unlikely to be virialised. It is with this caveat that we use the virial radius in this work, and as such use such a metric largely to give a relative sense of scale between different galaxies within our study. For context, at $R_{\rm vir}$ (for $M_{\rm halo}\approx10^{12}$~M$_\odot$ at $z\approx1$) the escape velocity is $v_{\rm esc}\approx260$~km~s$^{-1}$, whilst at $2R_{\rm vir}$ the escape velocity is $v_{\rm esc}\approx180$~km~s$^{-1}$.

\subsection{High-resolution ultraviolet absorption spectroscopy}

The HST spectroscopic observations with COS were obtained as part of the the COS Absorption Survey of Baryon Harbors (CASBaH, HST Programs 11741 and 13846, PI: T. Tripp).  Full details about the CASBaH program design and data handling are reported by Tripp et al. (In Prep); here we briefly summarize some important aspects of the data set.  CASBaH was designed to obtain high-resolution spectra with complete coverage from observed wavelength $\lambda _{\rm ob} =$ 1150 \AA\ to the redshifted wavelength of the Ly$\alpha$ emission line of each target QSO.  This was achieved using the far-UV G130M and G160M COS gratings (resolving power $R = 18000 - 22000$ or FWHM $\approx 15$ km s$^{-1}$), the near-UV COS G185M and G225M gratings ($R = 16000 - 24000$, \citealt{2012ApJ...744...60G,2011Ap&SS.335..257O}), and the STIS E230M echelle spectrograph ($R = 30000$, \citealt{1998PASP..110.1183W, 1998ApJ...492L..83K, 2018stis.book.....R}). Table~\ref{tab:observations} provides a log of the COS and STIS observations.\footnote{For further information about the design and performance of COS, please see \citet{2012ApJ...744...60G,2011Ap&SS.335..257O}. Information about the design and performance of STIS can be found in \citet{1998PASP..110.1183W}, \citet{1998ApJ...492L..83K}, and \citet{2018stis.book.....R}.}

The FUV G130M and G160M spectra were obtained to survey very weak absorption lines such as the Ne~\textsc{viii} doublet and accordingly required higher signal-to-noise (S/N) ratios.  The NUV were obtained to record the stronger \textsc{H~i} and longer-wavelength metal species (e.g., \textsc{C~iii}, Si~\textsc{iii}, and \textsc{O~vi}) affiliated with Ne~\textsc{viii} absorbers; for the detection of these stronger lines, lower S/N was acceptable.  For PKS0232-042, the COS G130M and G160M spectra have median S/N ratios of 13 and 17, respectively, while the COS G185M and G225M spectra have median S/N = 5.   The STIS E230M spectrum has median S/N = 7.

\begin{table}
 \centering
 \caption{Ultraviolet spectroscopy of PKS0232-042 with HST\label{tab:observations}}
 \begin{tabular}{@{}llll}
 \hline \hline
Instrument/  & Observation & Exp. time & HST Program  \\
grating  & date & (ksec)          & (PI)\\
\hline
COS/G130M & 2010 Feb. & 16.019 & 11741 (Tripp)\\
COS/G160M & 2010 Jan.  & 22.841 & 11741  \\
   \         & 2010 Feb. &  \          & \\
COS/G185M & 2015 July  & 23.424 & 13846 (Tripp) \\
COS/G225M & 2015 July  & 29.060 & 13846 \\
   \                  & 2015 Aug. & \           & \\
STIS/E230M & 2001 Feb.   & 41.934 & 8673 (Jannuzi) \\ 
   \                  & 2002 Jan. & \            & \\
\hline \hline
\end{tabular}
\end{table}

\begin{table}
	\centering
	\caption{Parameters for the identified O~{\sc vi} doublet absorption systems found in the PKS~0232-04 sightline using the HST COS and STIS data.}
	\label{tab:ovi_params}
	\begin{tabular}{lcccc} 
\hline	
	\hline
	$z_{sys}$  & $\Delta v$    & EW    & log[N(O~{\sc vi}])        & $b$      \\
	           & (km~s$^{-1}$) & (\AA)  & (cm$^{-2}$) & (km~s$^{-1}$)  \\
\hline	
0.17355 &  $   +0\pm  4$ & 0.54 & $13.7\pm 0.1$ & $ 18.9\pm 6.5$ \\
0.21802 &  $  -67\pm  8$ & 0.64 & $13.8\pm 0.1$ & $ 34.8\pm11.1$ \\
        &  $   +0\pm  1$ & 1.87 & $14.5\pm 0.1$ & $ 23.9\pm 2.2$ \\
        &  $  +74\pm  2$ & 1.04 & $14.0\pm 0.1$ & $ 25.1\pm 3.1$ \\
0.32243 &  $   +0\pm  4$ & 0.59 & $13.7\pm 0.1$ & $ 33.4\pm 6.1$ \\
0.35589 &  $   +0\pm  6$ & 0.33 & $13.5\pm 0.1$ & $ 24.3\pm 9.2$ \\
0.36384 &  $  -87\pm 12$ & 0.36 & $13.5\pm 0.1$ & $ 42.4\pm20.3$ \\
        &  $   +0\pm  5$ & 0.54 & $13.7\pm 0.1$ & $ 21.5\pm 7.1$ \\
        &  $  +76\pm  2$ & 0.35 & $13.5\pm 0.1$ & $ 10.9\pm 4.8$ \\
0.43431 &  $  -81\pm  3$ & 0.77 & $13.9\pm 0.1$ & $ 30.4\pm 4.6$ \\
        &  $   +0\pm  2$ & 0.69 & $13.9\pm 0.1$ & $ 14.8\pm 3.2$ \\
0.51208 &  $  -95\pm  4$ & 0.32 & $13.5\pm 0.1$ & $ 14.2\pm 6.2$ \\
        &  $   +0\pm  6$ & 0.39 & $13.5\pm 0.1$ & $ 25.7\pm 8.6$ \\
0.73901 &  $ -189\pm 23$ & 1.07 & $14.0\pm 0.2$ & $ 63.1\pm32.9$ \\
        &  $   +0\pm  8$ & 2.54 & $14.4\pm 0.1$ & $ 73.3\pm11.7$ \\
0.80783 &  $ -167\pm  4$ & 0.88 & $14.0\pm 0.1$ & $ 16.8\pm 6.4$ \\
        &  $   +0\pm  3$ & 2.65 & $14.6\pm 0.1$ & $ 32.9\pm 4.3$ \\
0.86816 &  $  -52\pm 10$ & 0.28 & $13.4\pm 0.4$ & $  9.9\pm17.0$ \\
        &  $   +0\pm  4$ & 1.77 & $14.4\pm 0.1$ & $ 22.6\pm 6.3$ \\
1.08894 &  $   +0\pm 13$ & 2.09 & $14.4\pm 0.2$ & $ 31.6\pm13.9$ \\
        &  $  +66\pm 40$ & 0.87 & $13.9\pm 0.6$ & $ 34.2\pm45.7$ \\
1.35646 &  $  -45\pm 25$ & 0.92 & $14.0\pm 0.4$ & $ 29.2\pm28.9$ \\
        &  $   +0\pm  8$ & 1.13 & $14.1\pm 0.3$ & $ 18.4\pm 9.5$ \\
\hline	
\hline	
	\end{tabular}
\end{table}

We reduced the FUV data as described in \citet{2011ApJ...732...35M} and Tripp et al. (In Prep).  In brief, we used the CALCOS pipeline (version 3.1.7) to carry out the initial reduction steps leading to one-dimensional extractions of the spectra from the individual exposures, and we aligned the individual exposures by comparing the positions of well-detected lines with distinctive component structure. COS employs a photon-counting detector with very low backgrounds, so we next coadded the individual exposures to accumulate the total gross and background counts in each pixel, which we used to determine the uncertainty in the flux based on counting statistics.  For this absorption study, the absolute flux of the QSO is irrelevant, so no absolute flux calibrations were applied to the FUV spectra.   The COS pipeline produces highly oversampled data, so we also binned the spectra to two pixels per resolution element to optimally sample the spectra.  For the NUV COS and STIS data, we similarly used CALCOS and CALSTIS (v. 2.22) to extract 1D spectra.  However, for these data it was more convenient to flux calibrate the data (e.g., to account for the echelle blaze function of STIS), and then we aligned and coadded the individual spectra with the weighting method described by \citet{2001ApJ...563..724T}.

Absorption lines in the PKS0232-042 spectrum were identified by inspecting every single line in the spectrum in many ways with a variety of tools.  For every line, we checked whether the line would be an O~{\sc vi} candidate based on the velocity spacing and relative strength of the O~{\sc vi} doublet.  This search for O~{\sc vi} did \textit{not} require that \textsc{H~i} is detected or not; we only searched for two lines with the signature spacings of O~{\sc vi}. Interestingly, detection of O~{\sc vi} without H{\sc i} mainly occurs in absorbers with $z_{\rm abs} \approx z_{\rm quasar}$, i.e., this seems to be a signature of ejected or at least ``proximate'' absorption systems \citep[][]{2008ApJS..177...39T}. Examples of individual \textsc{O~vi} \textit{components} in intervening absorbers without \textsc{H~i} at the same velocity can be found in \citet[][]{2008ApJS..177...39T,2010ApJ...719.1526S}. However, in these examples, \textsc{H~i} is clearly detected in other components of the same absorption systems. We corroborated the line identifications by searching for \textsc{H~i} and other metal lines detected in the \textsc{O~vi} absorption systems. Affiliated metals and \textsc{H~i} lines often have similar (or identical) component structure as the \textsc{O~vi} absorption profiles \citep[][]{2006ApJ...643L..77T, 2008ApJS..177...39T, 2011Sci...334..952T}, which gives additional credence to the identifications.  Information on the other lines detected in these systems can be found in Tripp et al. (In Prep).

We fit each system using Voigt profiles to constrain the column density ($N_{\rm OVI}$), velocity width ($b$), and velocity centroid of each component, and for purposes of comparison, these Voigt-profile parameters were used to calculate the equivalent width of each component using a full curve-of-growth. We use the convention of \citet{2008ApJS..177...39T}, i.e., use the velocity centroid of the strongest O~{\sc vi} component in a system to define the systemic redshift of the absorber. The resulting measurements for each O~{\sc vi} absorption doublet are given in Table~\ref{tab:ovi_params} 

\begin{figure*}
\includegraphics[width=0.24\textwidth]{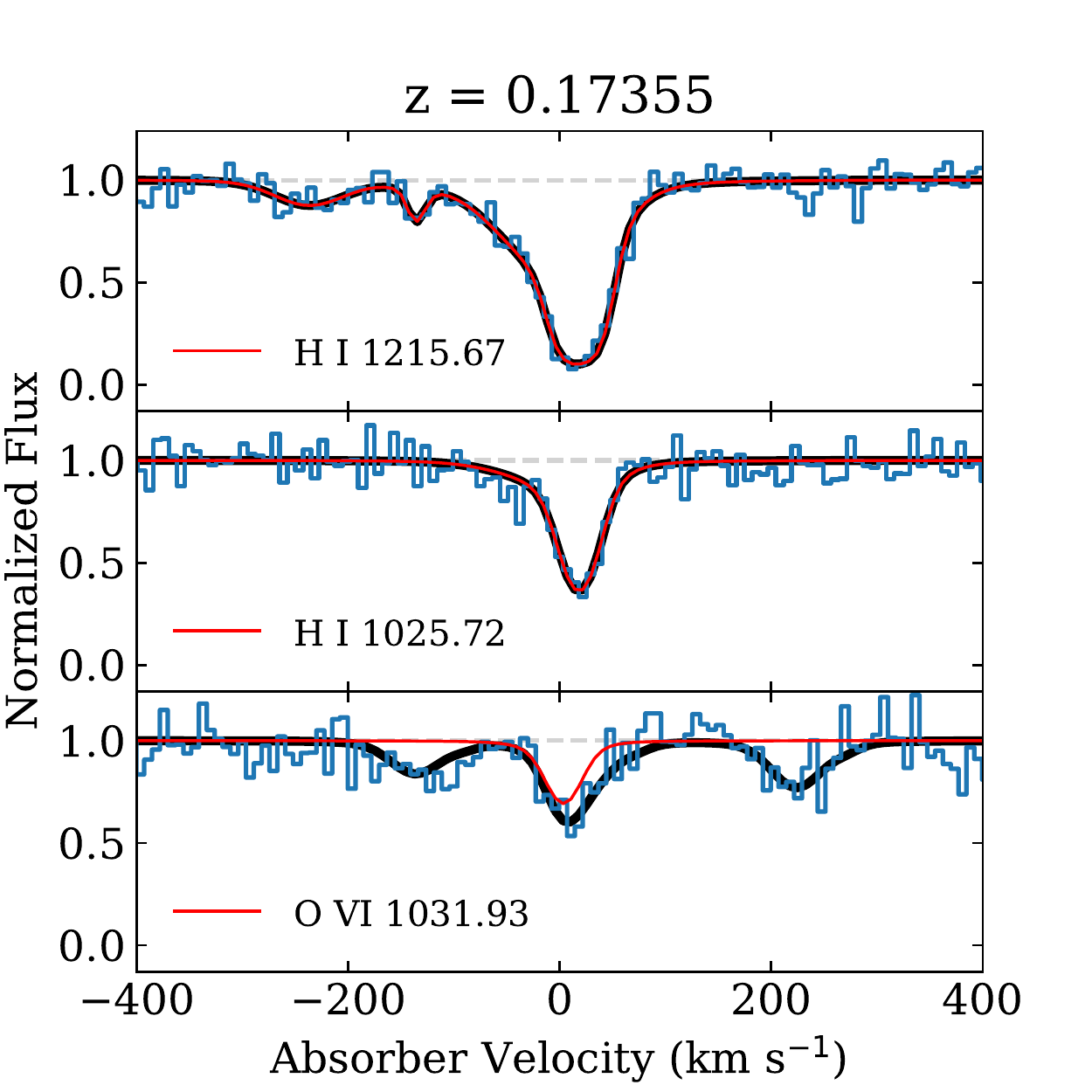}
\includegraphics[width=0.24\textwidth]{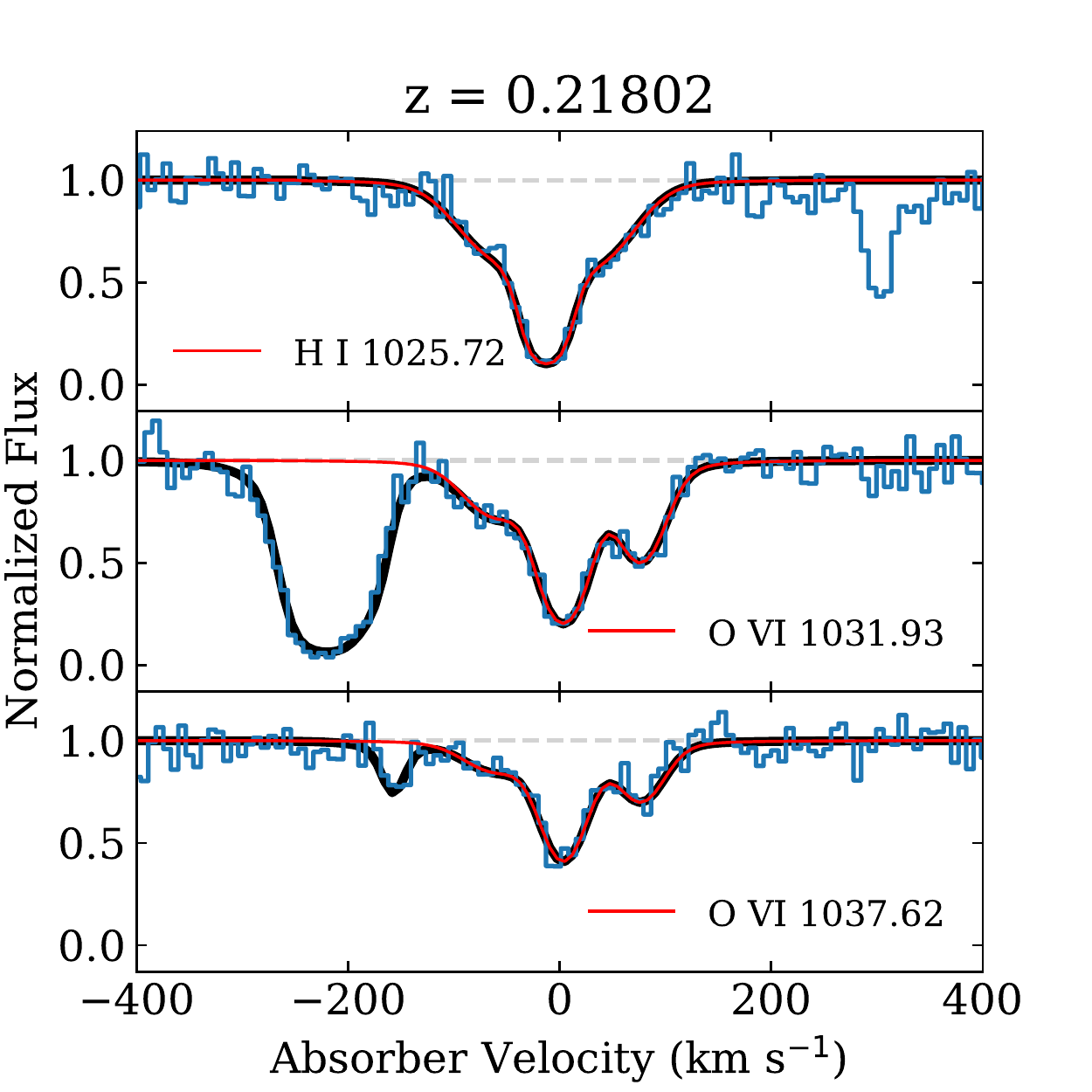}
\includegraphics[width=0.24\textwidth]{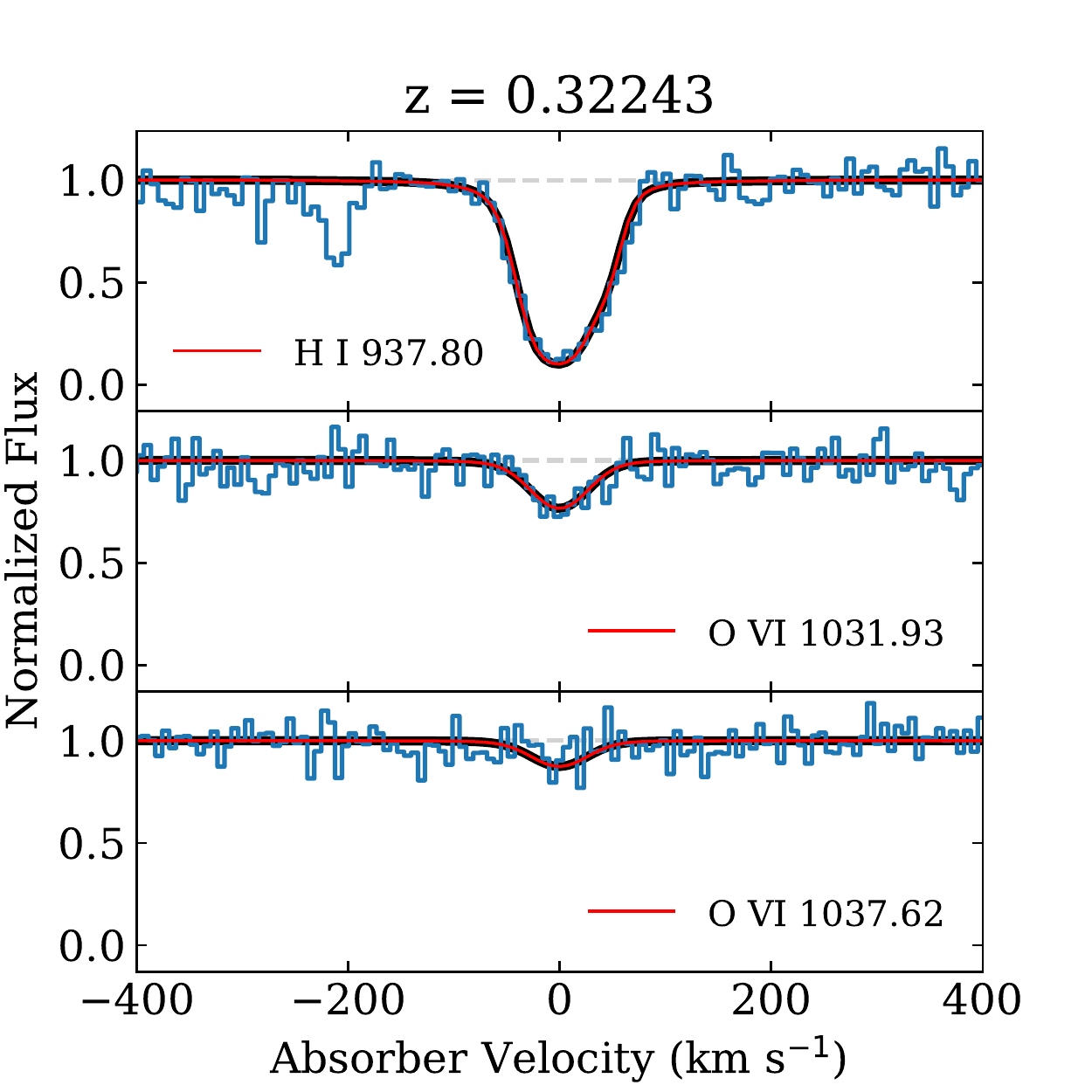}
\includegraphics[width=0.24\textwidth]{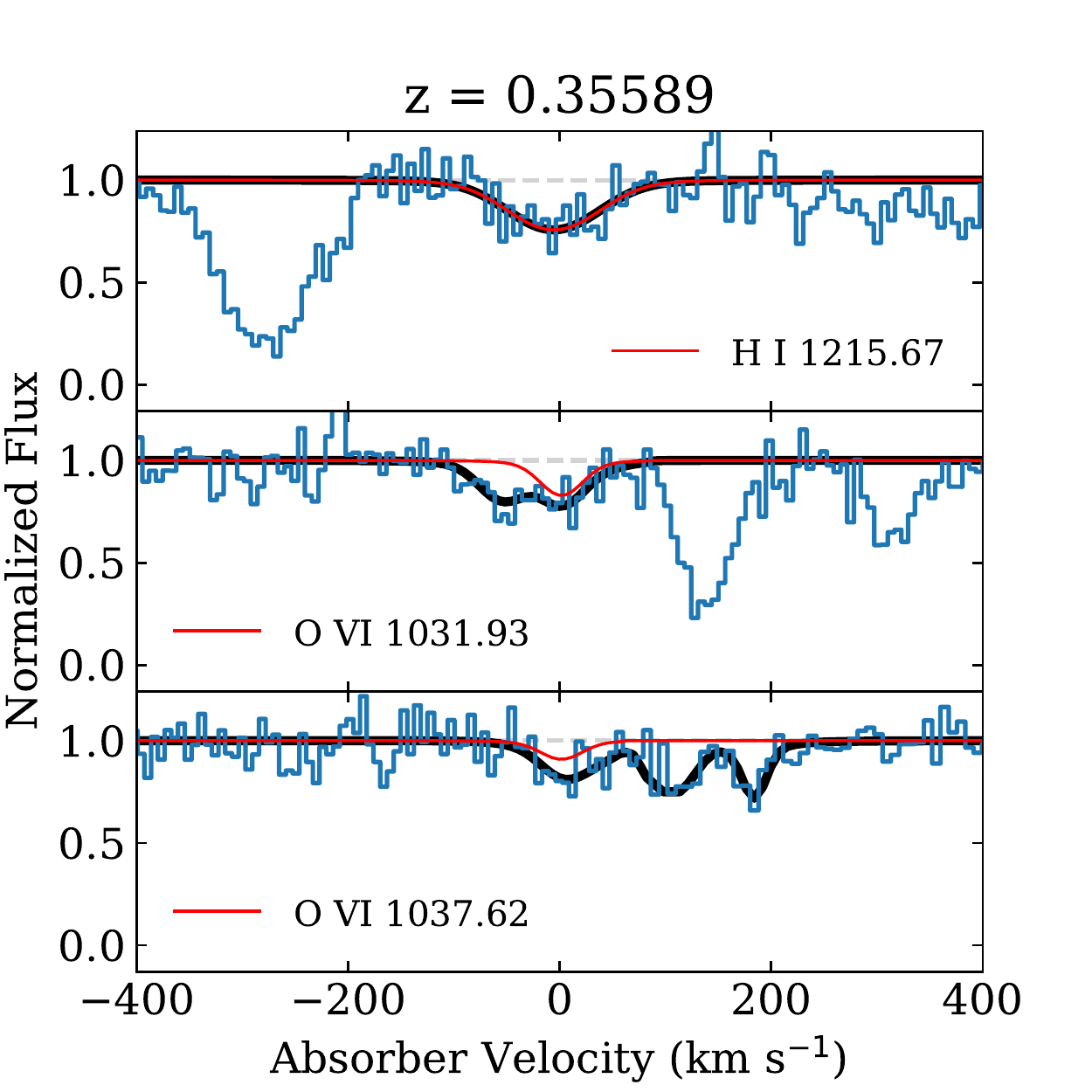}
\includegraphics[width=0.24\textwidth]{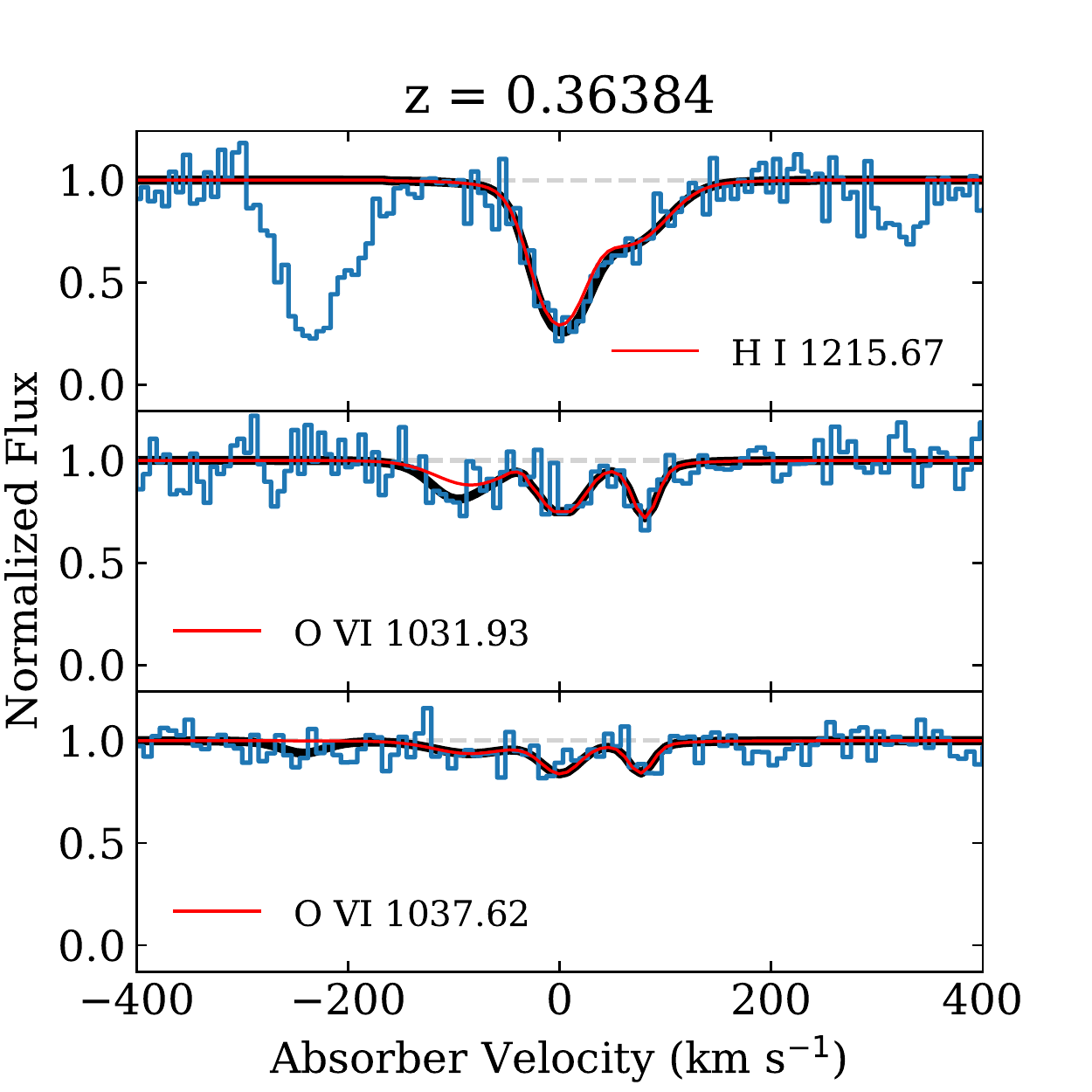}
\includegraphics[width=0.24\textwidth]{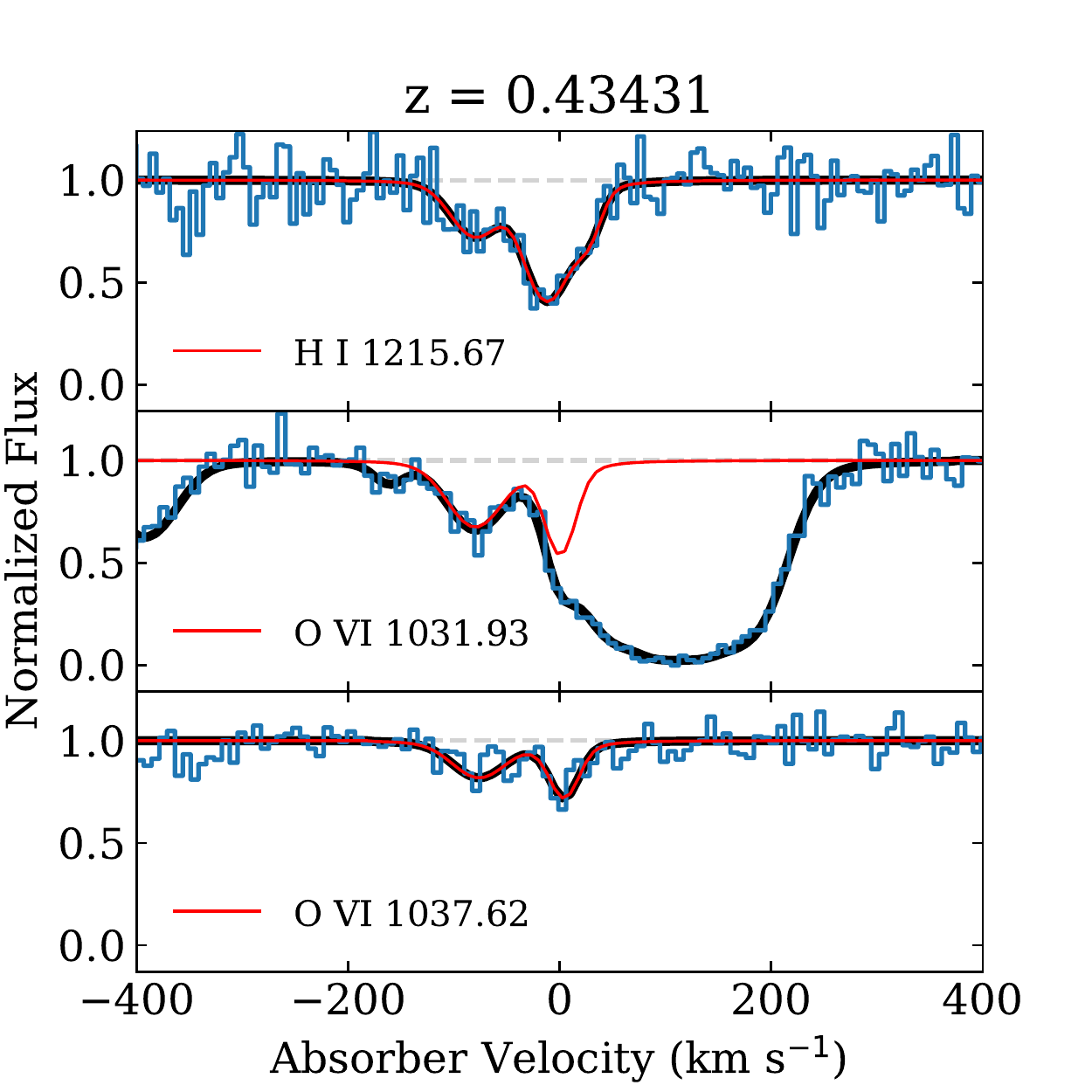}
\includegraphics[width=0.24\textwidth]{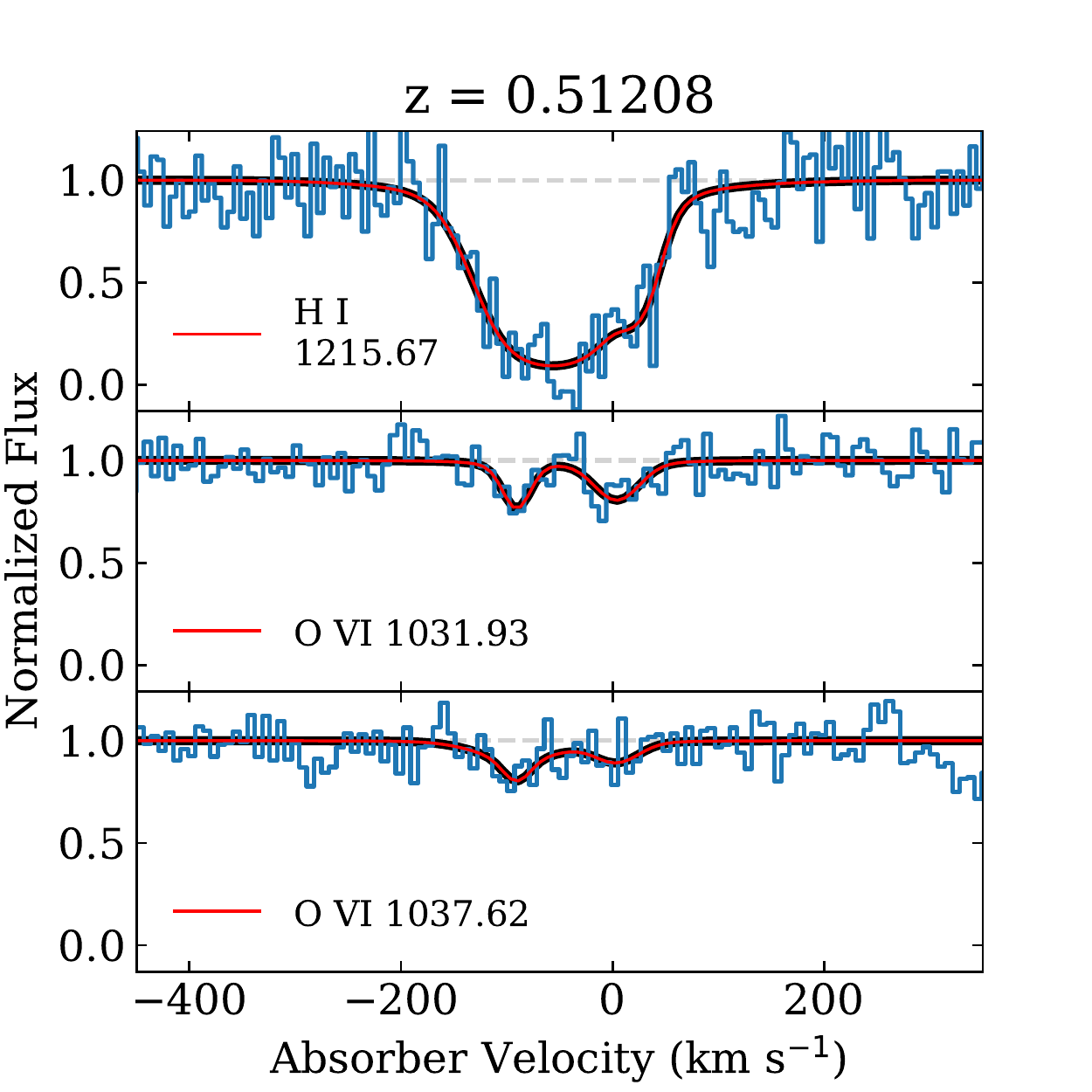}
\includegraphics[width=0.24\textwidth]{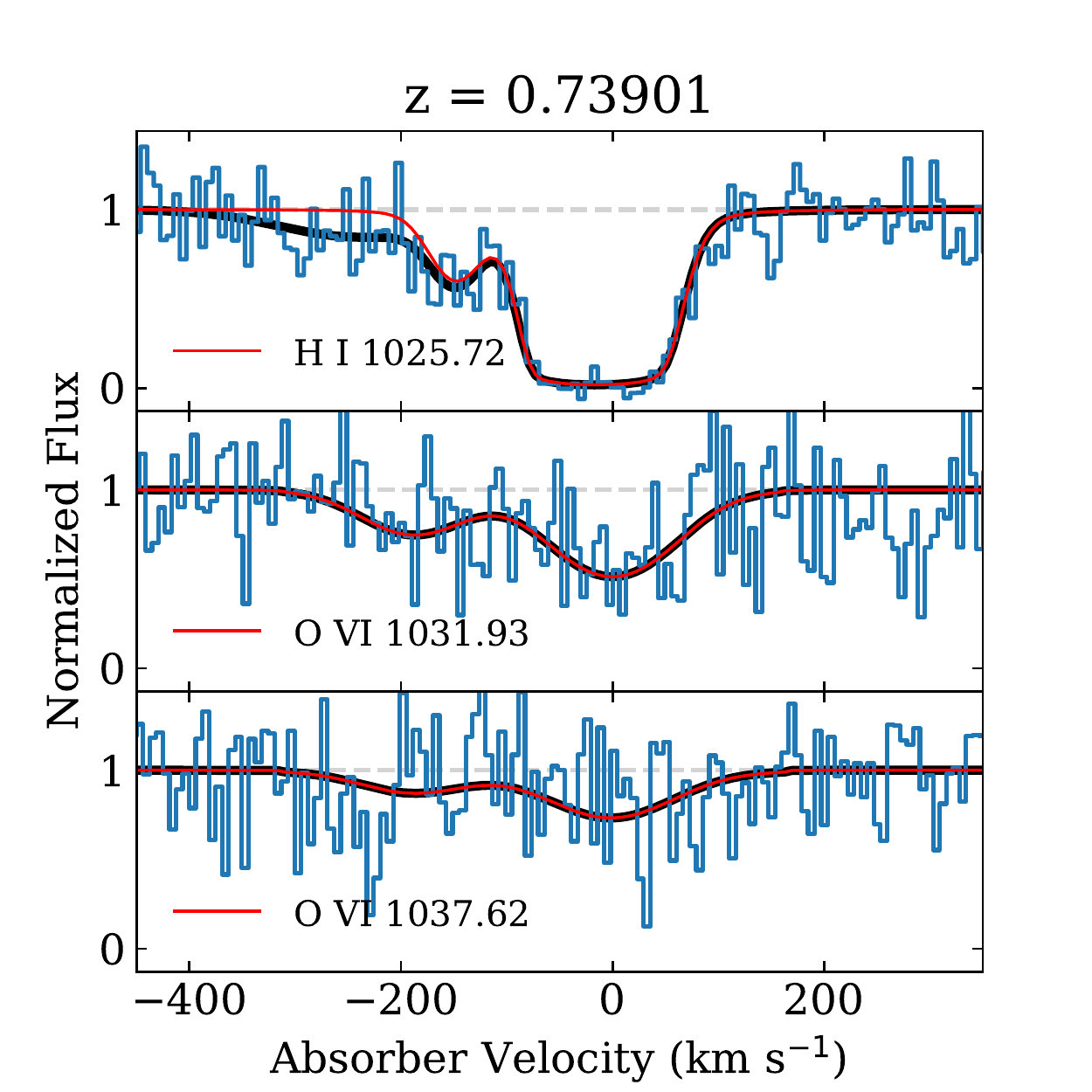}
\includegraphics[width=0.24\textwidth]{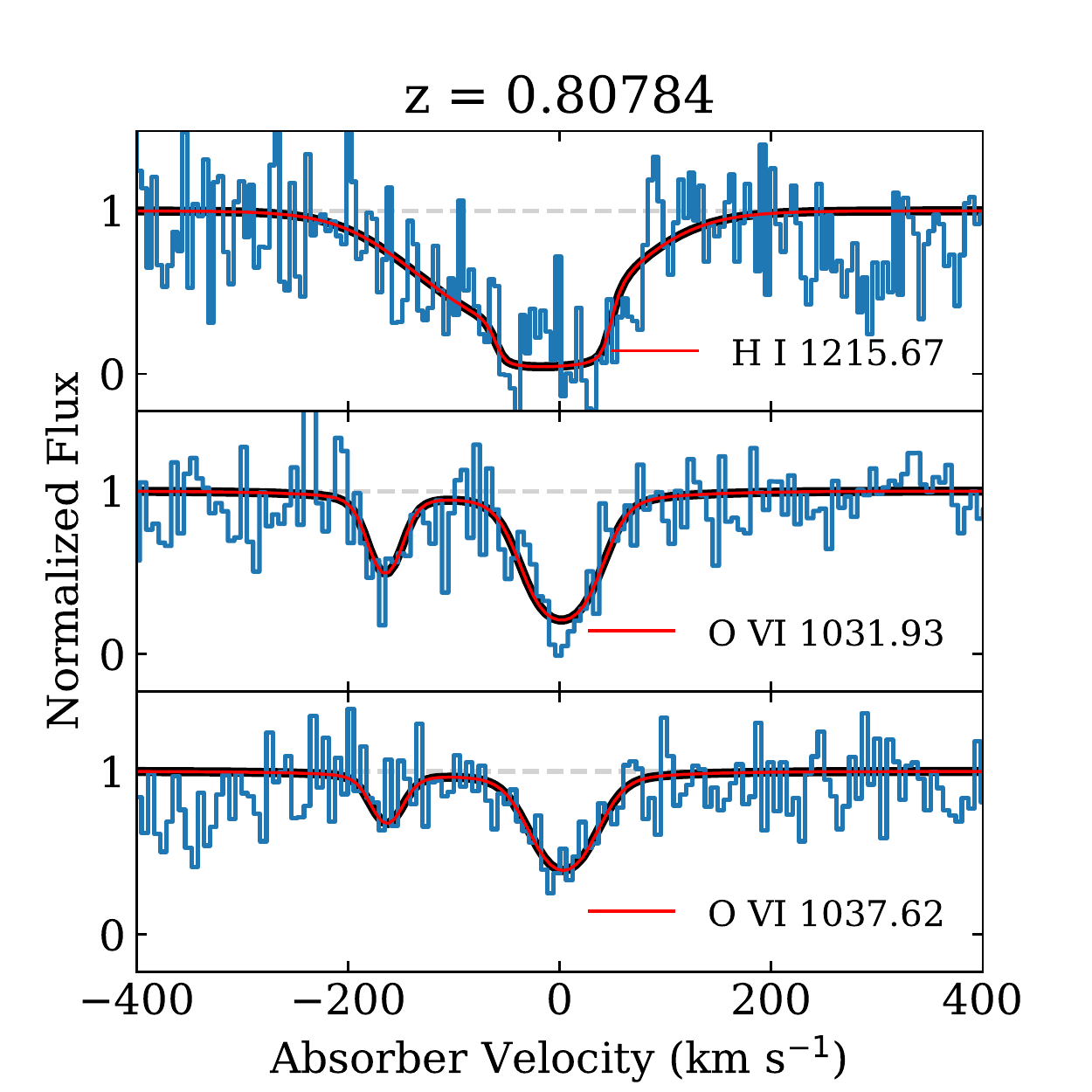}
\includegraphics[width=0.24\textwidth]{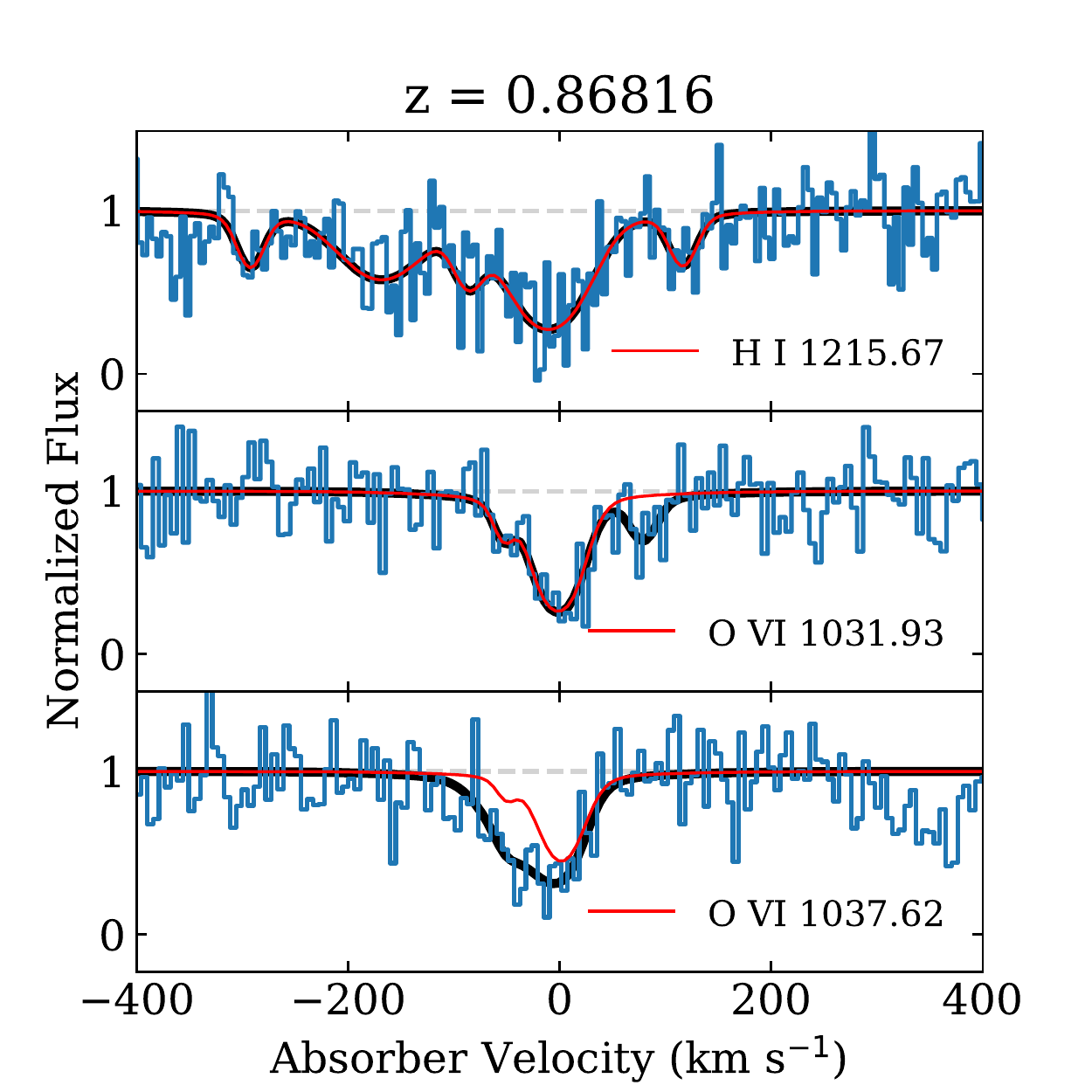}
\includegraphics[width=0.24\textwidth]{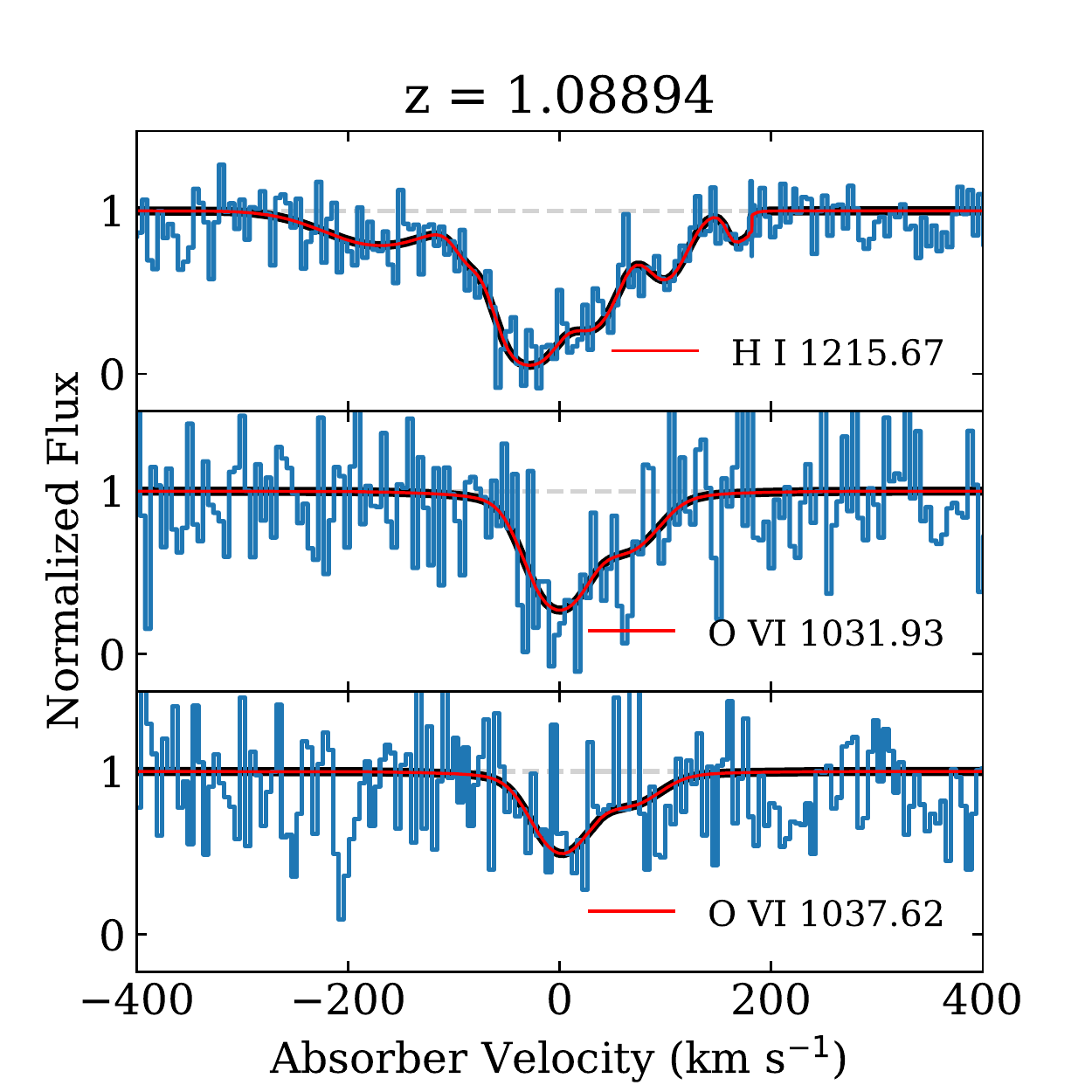}
\includegraphics[width=0.24\textwidth]{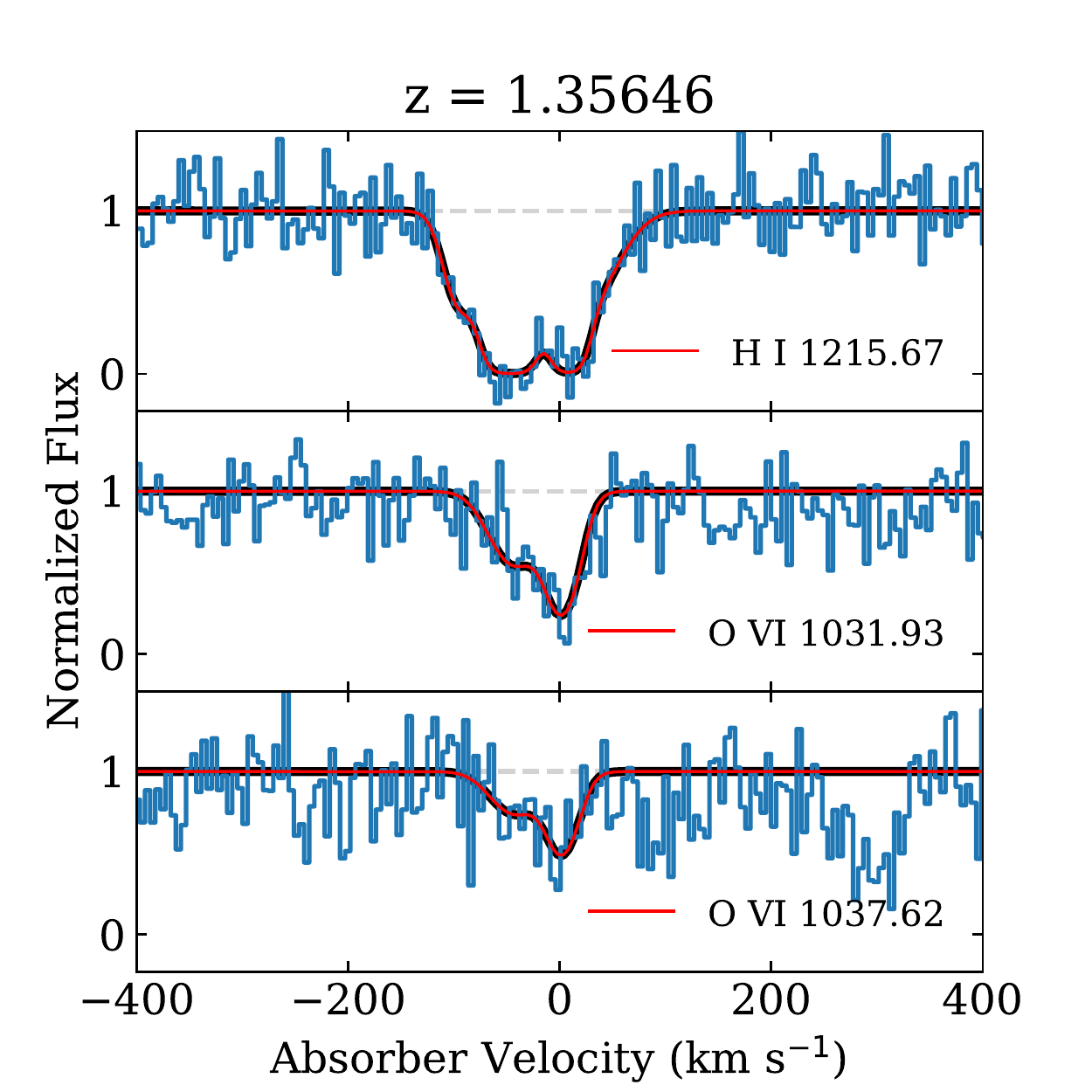}
\caption{Spectra (blue step histogram) of \textsc{O~vi} and \textsc{H~i} absorption lines (as labelled) overlayed with the absorption-profile fits.  In each panel, the black line shows the full fit (including blends from interloping lines from other redshifts), and the red line shows the Voigt-profile model for the \textsc{O~vi} or \textsc{H~i} only.}
\label{fig:absspec}
\end{figure*}


We show the absorption systems, and the associated model fits, used in this paper in Fig.~\ref{fig:absspec}. In each panel, the black line shows the entire Voigt-profile model including all blends that were included in the fit, and the red line shows the Voigt-profile model for only the species of interest (indicated in each panel). We show three panels per absorption system: one of the H I Lyman series lines (often there are many to choose from; we selected the one that was most informative, but all available Lyman series lines were used in the fit); and both of the lines of the O~{\sc vi} doublet. Whilst we are only studying O~{\sc vi} in this paper, it is helpful to see the corresponding H I to corroborate the O~{\sc vi} identification. Tripp et al. (In Prep.) gives the full suite of lines detected and measured in each of these systems. Note that for the O~{\sc vi} absorber at $z_{\rm abs} = 0.17355$, the O~{\sc vi} 1038\AA\ line is lost in the Milky Way Ly$\alpha$ + geocoronal Ly$\alpha$ complex, so we do not show the 1038\AA\ data in that stack.  Instead, we show the Ly$\alpha$, Ly$\beta$, and O~{\sc vi}  1032\AA\ lines.

\section{The galaxy environment of O{\sc vi} absorbers}
\label{sec:analysis}

The blind nature of both the HST-WFC3 grism and VLT-MUSE large field IFU data provide a large quantity of information on the galaxy environment directly surrounding the gaseous environments probed along the quasar sightline. We aim to utilise these data to probe the galaxy environments of the individual O~{\sc vi} absorption systems along the line of sight.

We consider the systems in two separate redshift ranges: i.e. a low-redshift sample (i.e. $z<0.68$), where the galaxy population is primarily probed up to $\approx0.5'$ from the sightline by the MUSE data; and a high redshift sample ($0.68<z<1.44$), where the population is revealed over a wider field of view up to $\approx1.'5$ with the grism data. For the analyses that follow, we define a velocity window within which to consider an absorber to be associated with a local galaxy. We consider a number of physical factors: 1) the rotational velocity of galaxies within the mass range covered by our sample; 2) the typical range of velocities measured for outflowing material from galaxies; 3) the velocity dispersion of galaxies within galaxy groups and clusters; and 4) the velocity uncertainties on the galaxy redshifts themselves.

Taking typical galaxy rotation curves for guidance, galaxies at $z\approx1$ with masses of $\approx10^{10}$~M$_\odot$ have median rotational velocities of $\approx120-160$~km~s$^{-1}$, although rotational velocities of up to $\approx2\times$ the median are common \citep[e.g.][]{2018MNRAS.474.5076J}. Velocity offsets of up to $\approx320$~km~s$^{-1}$ therefore seem reasonable if any O~{\sc vi} absorption is tracing co-rotating material within the galaxy halo. Considering galactic winds driven by star-formation activity, outflow velocities of $v_w\approx200-400$~km~s$^{-1}$ are typical at the redshifts we probe \citep[e.g.][]{2014ApJ...794..156R}, although wind velocities of up to $\approx600$~km~s$^{-1}$ have also been reported \citep[e.g.][]{1998ApJ...493..129S, shapley03, 2011Sci...334..952T}. In terms of large scale galaxy structures, massive galaxy clusters can have velocity dispersions up to $\approx1000$~km~s$^{-1}$ \citep[e.g.][]{1999ApJS..125...35S}. Such a large velocity window is clearly not appropriate for draw associations with the average galaxy population, however, so such a large offset will only be considered in this work where a galaxy group or cluster is detected. Allowing for the potential range in velocity offsets, we set a velocity window for associating absorbers with galaxies of $\Delta v_{\rm max}={\rm max}\{\vwindow$~km~s$^{-1},\sigma_v\}$, where $\sigma_v$ is the velocity uncertainty on any given galaxy (i.e. $\approx50-80$~km~s$^{-1}$ for galaxies with MUSE redshifts and $\approx682$~km~s$^{-1}$ for galaxies with only grism redshifts). As a reference point, \citet{2016ApJ...833...54W} find O~{\sc vi} absorber-galaxy pairs, at $z=0.2$, associated in velocity space within $|\Delta v|\lesssim\pm200$~km~s$^{-1}$.

In the following section, we provide details of individual absorber-galaxy associations, discussing system properties and environment on a case-by-case basis. We collate the data and present the statistical properties of the sample as a whole in section~\ref{sec:ovi_stats}.

\subsection{Census of O{\sc vi} systems}

\begin{figure*}
	\includegraphics[width=\textwidth]{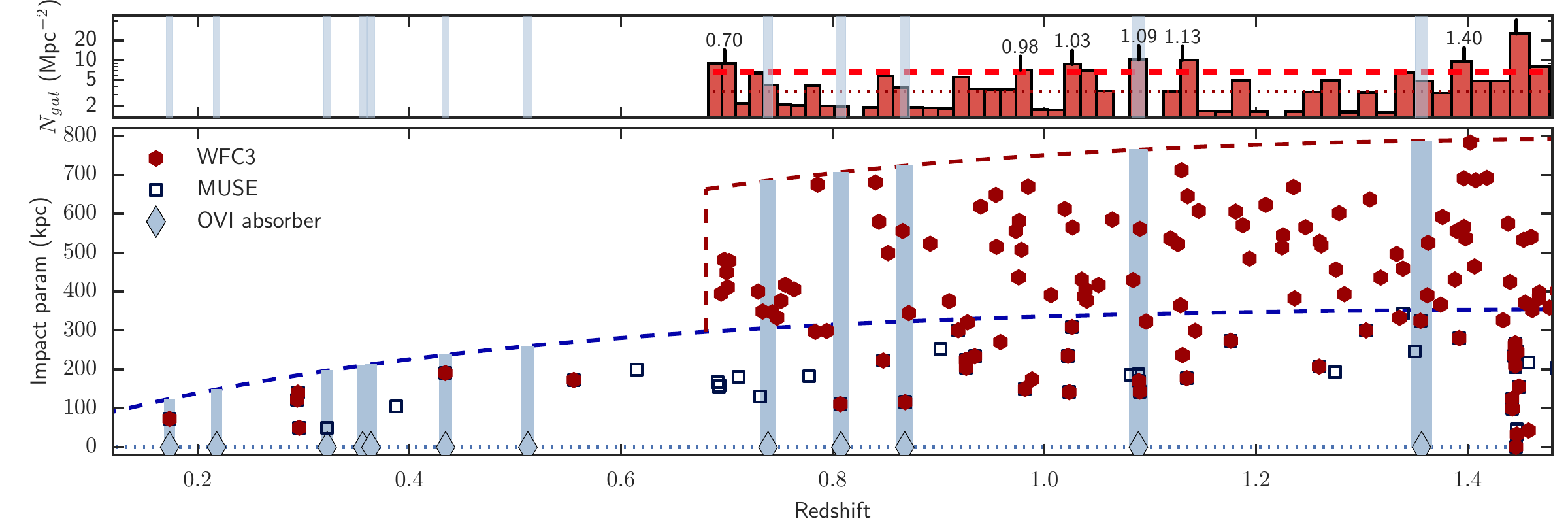}
	\caption{Redshift versus impact parameter for all identified galaxies within the MUSE (inner dashed curve) and WFC3 grism (outer dashed curve) fields of view. O~{\sc vi} absorbers along the quasar sightline (dotted line) are marked with $\pm1000$ km~s$^{-1}$ interval shown extending across the field of view to highlight any potentially associated galaxies or galaxy structures. Galaxy density within $\Delta v\pm100$~km~s$^{-1}$ bins (histogram) alongside the O{\sc vi} absorber positions. The horizontal dashed line shows the median observed galaxy density in the WFC3 grism field of view (within the redshift range 0.68<z<1.38). The O~{\sc vi} absorbers are found in a range of galaxy environments, with at least 1 out of the 5 coinciding with a potential group environment (i.e. twice the mean density as given in the text) and 2 having no associated galaxy within 200~kpc (although all have associated galaxies within $\approx350$~kpc).}
    \label{fig:zVimp}
\end{figure*}

A graphical overview of the field is presented in Fig.~\ref{fig:zVimp}, where all detected O~{\sc vi} absorbers (diamonds) and galaxies (squares and hexagons) at $z<1.5$ are shown as a function of redshift and impact parameter to the quasar sightline (lower panel). The top panel shows a histogram of the galaxy population within 600~kpc of the sightline at $0.68<z<1.44$. The vertical shaded regions highlight redshifts coincident with the O~{\sc vi} absorbers. The dotted and dashed horizontal lines in the top panel denote the median and twice the median galaxy density in the field respectively. The data reveal a diversity of environments in terms of galaxy density and potential associations between O~{\sc vi} absorption systems and the galaxy population. 

\begin{figure}
    \centering
	\includegraphics[width=\columnwidth]{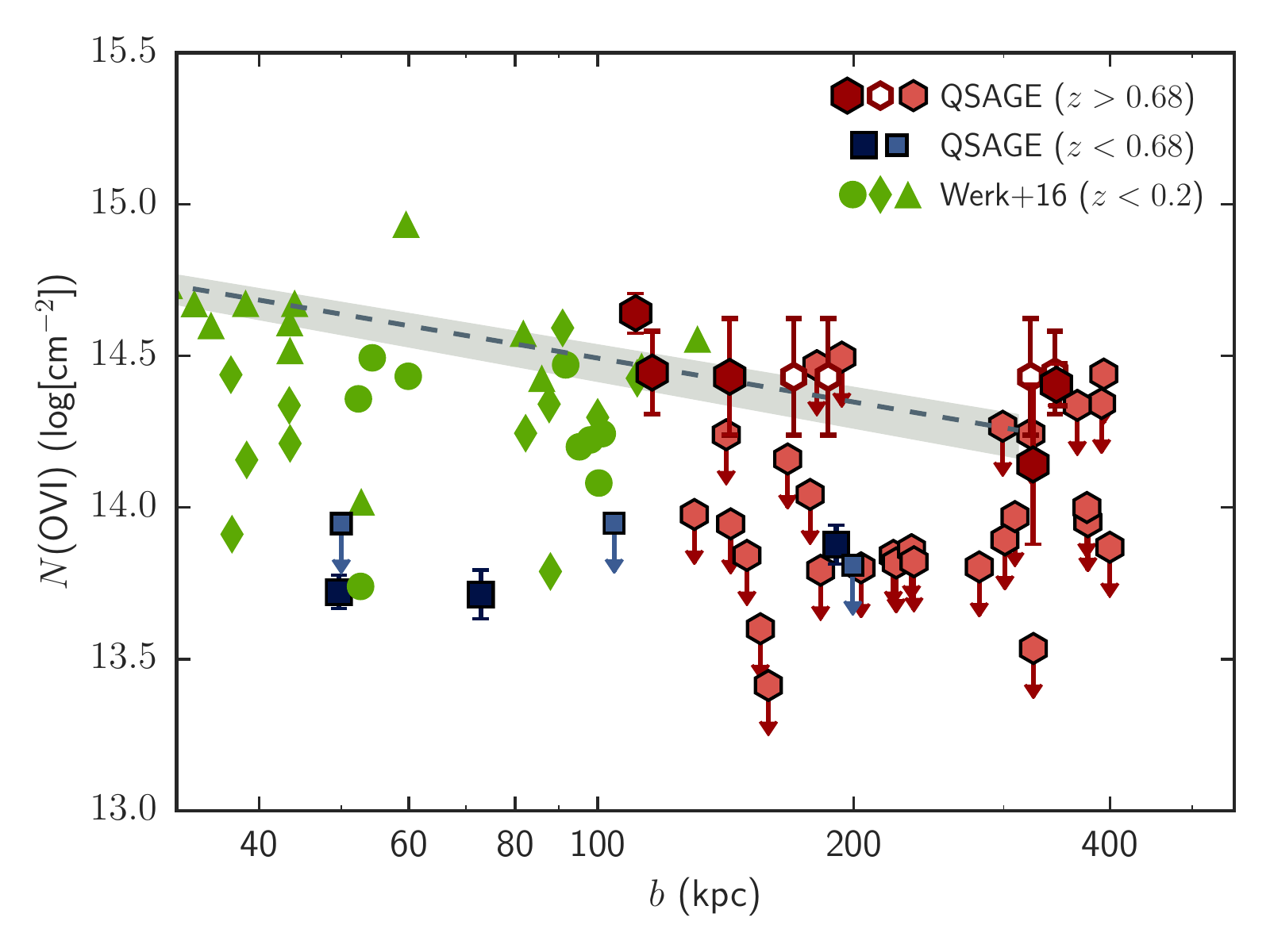}
	\caption{O~{\sc vi} column density as a function of impact parameter. The low-redshift ($z<0.68$) points from our dataset are shown by the blue squares, whilst the red hexagons denote the $0.68<z<1.44$ sample (in both cases, the paler symbols denote upper limits). The open symbols show galaxies coincident with a detected absorber, where a galaxy at smaller impact parameters is also present. The data of \citet{2016ApJ...833...54W} are shown by the green triangles, diamonds and circles (denoting broad, narrow and no low-ionization subsets). The dashed curve and shaded region show a fit to the \citet{2016ApJ...833...54W} broad absorbers (extrapolated to larger impact parameters).}
    \label{fig:impact}
\end{figure}

We now look at how the ionised oxygen is distributed around the galaxies, by analysing the column density of O~{\sc vi} absorbers as a function of impact parameter to individual detected galaxies within our given $\Delta v_{\rm max}$ constraints. This is shown in Fig.~\ref{fig:impact}, where the filled red hexagons and dark blue squares denote the QSAGE O~{\sc vi}-galaxy pairs at $z>0.68$ and $z\leq0.68$ respectively. We differentiate upper limits as pale hexagons and squares for the two samples respectively. These upper limits are calculated within $\Delta v\leq50$~km~s$^{-1}$ of detected galaxies in the survey sample. The open hexagons show the impact parameters of galaxies that are not the nearest to the sightline, but do fall within the $\Delta v_{\rm max}$ velocity window. We find 5 O~{\sc vi} absorption systems associated with galaxies in our sample at $z>0.68$ and a further 3 at $z\leq0.68$, given the $\Delta v_{\rm max}$ constraint on associations. These fall primarily at impact parameters of $\approx100$~kpc, but extend to $\approx350$~kpc in the most extreme case. Within this range in impact parameters, we found numerous non-detections with limits of $N_{\rm OVI}\lesssim10^{13.5- 14}$~cm$^{-2}$. These non-detections point to a large scatter in O~{\sc vi} column densities within $\approx300$~kpc of galaxies at $z\sim1$.

For comparison, we also show the COS-Halos measurements taken from \citet[][green points]{2016ApJ...833...54W}. The triangle, diamond, and circle points show broad, narrow and `no low-ionization' absorbers respectively, whilst the dashed curve and shaded region show the \citet{2016ApJ...833...54W} fit to the broad absorber (triangle) data-points. The distinction between the broad and narrow categories here is identified using the doppler parameter at a value of $b\sim30$~km~s$^{-1}$. \citet{2016ApJ...833...54W} exclude narrow O~{\sc vi} absorption and absorbers with no associated low-ionization lines from this fit to the data, finding that these dominate the large scatter in the observed column densities. Quantitatively, 1 out of 16 of the \citet{2016ApJ...833...54W} broad absorbers lie significantly (i.e. $\gtrsim3\sigma$) below the relation, whilst 13 of 17 narrow and `no-low' absorbers deviate significantly below the fit. Within our own data, all of the $0.68<z<1.4$ absorber-galaxy pairs are consistent with the extrapolated low-redshift fit. At $z<0.68$, all of our data lies well below the $z\sim0.2$ broad-absorber fit. In terms of absorber width, all these galaxy-absorber pairs are either significantly below or close to the cut-off point ($b\sim30$~km~s$^{-1}$) chosen by \citet{2016ApJ...833...54W} to classify between broad and narrow absorption systems.

We now consider individual systems of interest: 1) those where O~{\sc vi} is detected; 2) those where a galaxy or set of galaxies falls within $b\lesssim2R_{\rm vir}$ of the quasar sightline but no O~{\sc vi} is detected; or 3) a candidate group/cluster environment is found within the WFC3 data field of view. In this instance, we focus on galaxies within impact parameter $b/R_{\rm vir}=2$ cut prompted in part by results at low-redshift \citep{2011Sci...334..948T,2014ApJ...784..142S} but also for brevity in focusing on only the systems of most interest where we might have expected to find absorption systems based on previous studies. We provide a list of the proximate galaxies along the sightline and their properties in Table~\ref{tab:proxgals}.

\subsubsection{Absorbers at $z<0.68$}

\begin{figure*}
	\includegraphics[width=0.9\textwidth]{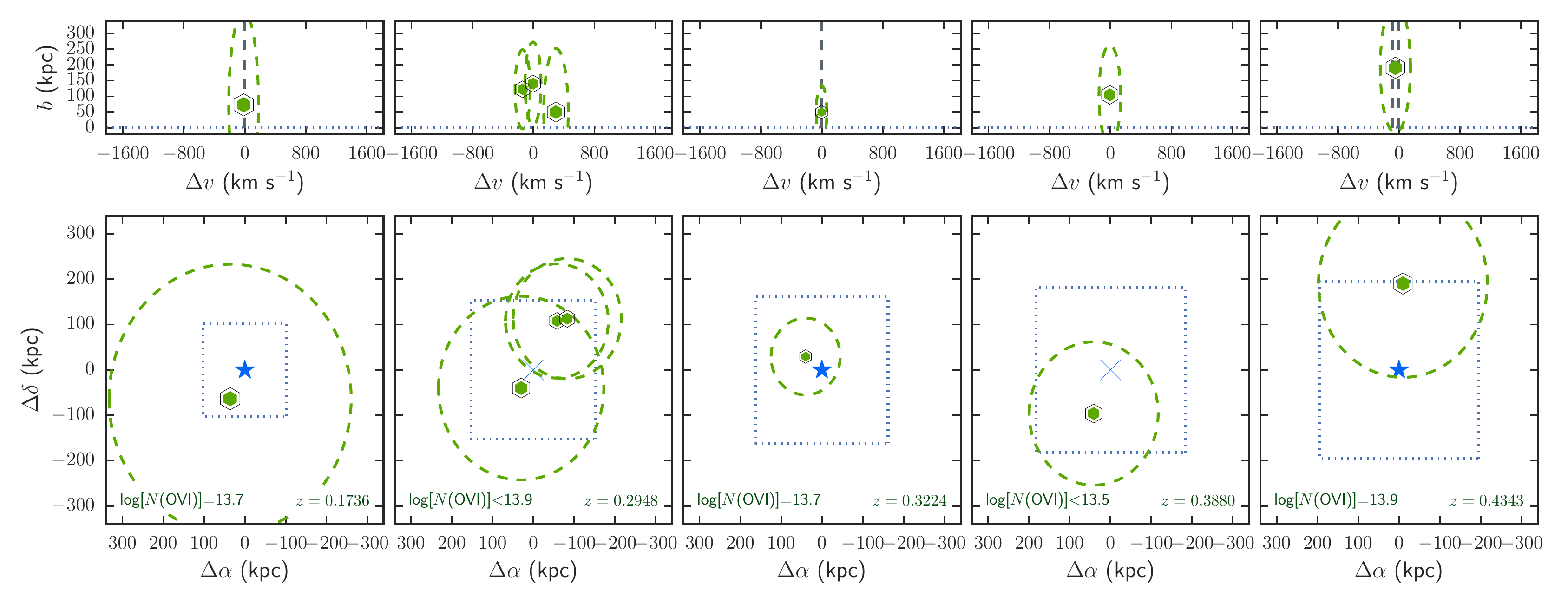}
	\caption{The velocity (top panels) and spatial (lower panels) distribution of galaxies (hexagons) around each of the $z<0.68$ systems where one or more galaxies are found within $b=2R_{\rm vir}$ of the sightline. The presence of any O~{\sc vi} absorption lines at each redshift is denoted by the vertical dashed lines in the top panels and a central blue star in the lower panels. Dashed circles centred on the galaxy positions illustrate the scales of $2R_{\rm vir}$ and $2\sigma_v$ for each galaxy. The dotted square in each spatial panel indicates the extent of the MUSE FOV at the system redshift.}
    \label{fig:loz_fovs}
\end{figure*}

Whilst the data at $z<0.68$ do not probe the large physical scales afforded by the higher redshift sample, we do detect a number of galaxies coincident with absorbers in the sightline. These are shown in Fig.~\ref{fig:loz_fovs} alongside all $z<0.68$ galaxies that lie within $b=2R_{\rm vir}$ where no O~{\sc vi} absorption is detected.

The absorbers at $z=0.1736$, $z=0.3224$, and $z=0.4343$ are all found to lie within $2R_{\rm vir}$ of galaxies detected in the MUSE data. The velocity offsets between the galaxies and absorbers in all three cases are $\lesssim50$~km~s$^{-1}$, i.e. relatively small in comparison to the nominal halo $\sigma_v$ estimates for the galaxies ($\gtrsim200$~km~s$^{-1}$).  We note that the absorbers detected at $z=0.218$, $z=0.356$, $z=0.364$, and $z=0.512$ show no detected galaxies in the deep MUSE data within a few 1000~km~s$^{-1}$ and therefore are not shown in Fig.~\ref{fig:loz_fovs}. The MUSE field of view at these redshifts corresponds to maximum probed impact parameters of $b=109$~kpc (compared to a virial radius of $R_{\rm vir,12}\approx 195$~kpc for a $M_\star\approx10^{12}$~M$_\odot$ mass galaxy), 154~kpc ($R_{\rm vir,12}\approx 176$~kpc), 156~kpc ($R_{\rm vir,12}\approx 174$~kpc) and 191~kpc ($R_{\rm vir,12}\approx 157$~kpc) respectively. Given previous studies at low redshift have found `host' galaxies of O~{\sc vi} absorption at impact parameters of up to $b\approx2R_{\rm vir}$, and some deep surveys at low redshift have found O~{\sc vi} absorbers at very large distances from the closest galaxy \citep{2006ApJ...643L..77T, 2013MNRAS.434.1765J}, it is not unreasonable that associated galaxies for these absorbers may fall outside of the field of view. Indeed, from a search of the NASA Extragalactic Database (NED) and the soon to be published CASBAH data, we find galaxies associated with these absorbers outside the MUSE field of view. In relation to the $z=0.512$ absrber, we find a galaxy at $z=0.511$ listed in NED, which is also detected in the WFC3 grism data via absorption features.

Taking the reverse, i.e. galaxies with no detectable associated O~{\sc vi} absorption in the sightline data, we find 4 such galaxies within $b\approx2R_{\rm vir}$ of the sightline at $z<0.68$. Three of these form a small grouping, lying within $\approx600$~km~s$^{-1}$ of each other at $z\approx0.295$. The upper limit on any associated O~{\sc vi} coincident with this grouping is $N_{\rm OVI}<10^{13.9}$~cm$^{-2}$ (which we note is higher than the detected systems at $z<0.68$). 

\subsubsection{Galaxy environments probed by the sightline at $z>0.68$}

\begin{figure*}
	\includegraphics[width=0.9\textwidth]{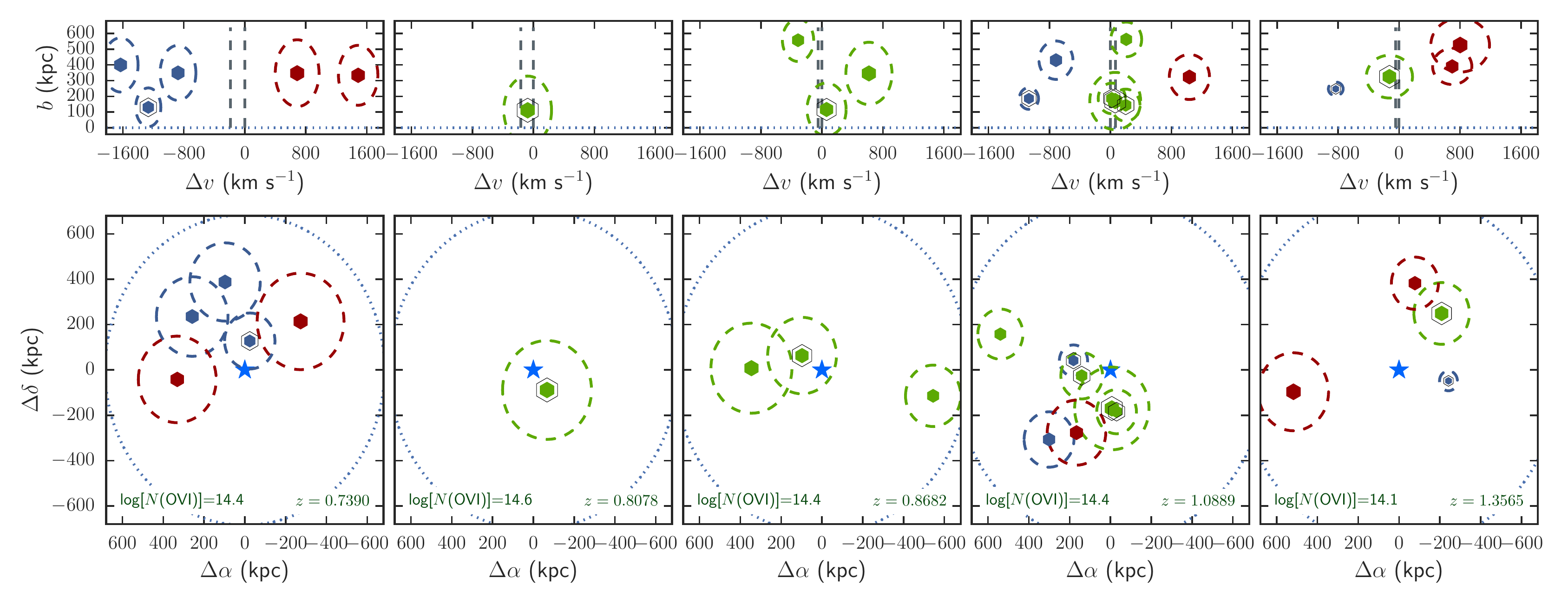}
	\caption{The velocity (top panels) and spatial (lower panels) distributions of galaxies at redshifts centred on detected O~{\sc vi} absorbers at $z>0.68$. The sightline position is marked by a star in each case, whilst the O~{\sc vi} absorption lines at each redshift are denoted by the vertical dashed lines in the top panels. Galaxies marked in green are within $\Delta v<680$~km~s$^{-1}$ of the absorber, whilst blue and red points denote galaxies blueshifted and redshifted by $\Delta v>680$~km~s$^{-1}$ respectively. Points with a black hexagonal outline have been detected in [O~{\sc ii}] emission in the MUSE datacube. We mark twice the inferred virial radii of each galaxy within $\Delta v=1800$~km~s$^{-1}$ of the detected absorber. The large dotted circle marks the extent of the WFC3 grism data field of view at the redshift of interest.}
	\label{fig:det_fov}
\end{figure*}

Fig.~\ref{fig:det_fov} shows the galaxy distributions around each of the 5 O~{\sc vi} absorption systems detected at $z>0.68$. In each case, galaxies within $\Delta  v=\pm680$~km~s$^{-1}$ are marked with green points, galaxies at $-1600$~km~s$^{-1}<\Delta  v<-680$~km~s$^{-1}$ with blue points, and galaxies at $680$~km~s$^{-1}<\Delta v<1600$~km~s$^{-1}$ with red points. Galaxies within this range that have been identified in the MUSE IFU data are given a black outline, whilst the estimated $2R_{\rm vir}$ scale for each galaxy is illustrated by the dashed ellipses. In each case $\Delta v=0$~km~s$^{-1}$ is centred on the strongest component observed in the O~{\sc vi} absorption system at each redshift. The large dotted circle in each of the spatial panels shows the approximate field of view of the WFC3 observations at each redshift. As with the low redshift sample, we find the absorbers predominantly lie within $\approx 100$~km~s$^{-1}$ of a nearby galaxy, except in the case of the absorber at $z\approx0.73$. Additionally, the impact parameters in three of these cases are at scales of $b<2R_{\rm vir}$. 

The $z\approx1.089$ strong ($N_{\rm OVI}=10^{14.43\pm0.19}~{\rm cm}^{-2}$) O~{\sc vi} absorber in the quasar spectrum coincides with a detected over-density in the galaxy population. Four of the galaxies lie within $b=200$~kpc of the quasar sightline, with three of these having velocities within a range of $\Delta v\approx250$~km~s$^{-1}$ of each other (based on the MUSE [O~{\sc ii}] measured redshifts). Of these three most closely associated galaxies, we find that the stellar mass is dominated by a $M_\star=10^{10.0\pm0.1}$~M$_\odot$ galaxy, which lies at $\Delta v=+230$~km~s$^{-1}$ and $b=170$~kpc from the absorption system. The remaining two closely associated galaxies are estimated to have masses of $M_\star\approx10^{8.5}$~M$_\odot$.

Given the spatial and velocity distribution, we assume these to be the primary members forming a triplet/low-mass group (with the remaining galaxies at larger separations spatially and in velocity potentially tracing the large scale structure environment). Given only three detected members, it is not possible to reasonably estimate a halo mass from the velocity dispersion, however from the stellar mass, we infer a halo mass for the central group galaxy of $M_{\rm halo}=10^{11.8\pm0.1}$~M$_\odot$. From this halo mass, we estimate a virial radius of $R_{\rm vir}\approx90$~kpc, meaning the dominant galaxy lies at $b=1.9R_{\rm vir}$ from the sightline.

For the $z\approx0.73$ system, no galaxy lies within $\Delta_{\rm v}<680$~km~s$^{-1}$ of the absorber in the data sample. Indeed, this system may be a prospective candidate as the sightline tracing the cosmic web outside of galaxy halos as discussed by \citet{2018MNRAS.tmp..710P}. Those authors set a limit of $\frac{\Delta v}{\sigma_v}\frac{b}{R_{\rm vir}}>2$ in their analysis to identify candidate cosmic web absorbers, which is clearly satisfied here (see Fig.~\ref{fig:det_fov} where the dashed ellipses in the top panels show the extent of $2\sigma_v$ in velocity space and $2R_{\rm vir}$ in impact parameter). We note that the redshift uncertainties on these galaxies (which lack MUSE coverage) are $\approx680$~km~s$^{-1}$ and as such the lack of alignment in velocity space could in part be due to the redshift accuracy of the galaxies.

\begin{figure*}
	\includegraphics[width=0.9\textwidth]{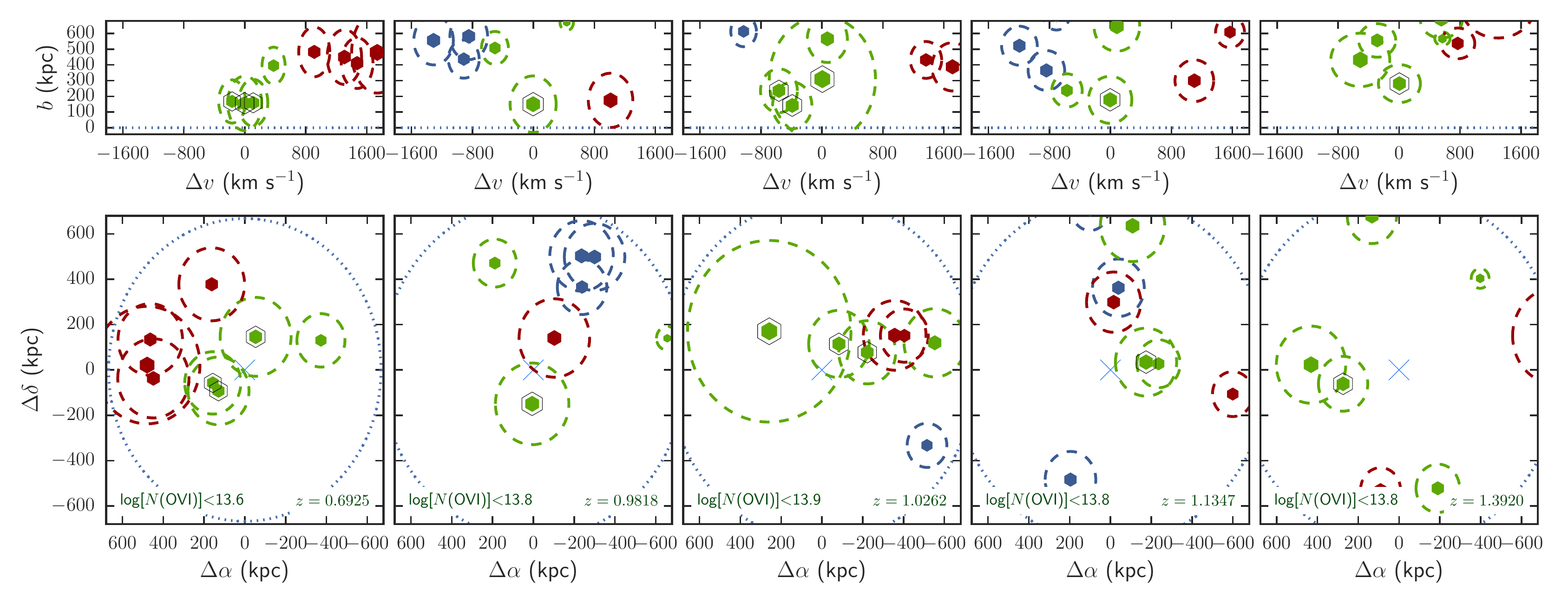}
	\caption{As in Fig.~\ref{fig:det_fov}, but for the galaxy over-density regions marked in Fig.~\ref{fig:zVimp}. The sightline position is marked by a `$\times$' symbol in each case.}
    \label{fig:hiden_fovs}
\end{figure*}

We now look to the remaining over-dense regions along the sightline as traced by the galaxy distribution. From the galaxy density, we find peaks above twice the median galaxy density at $z\approx0.698$, $z\approx0.977$, $z\approx1.026$, $z\approx1.089$, $z\approx1.131$ and $z\approx1.396$ (marked in Fig.~\ref{fig:zVimp}). For each of these we plot the galaxy distribution, both spatially (lower panels) and in velocity space (top panels), around the sightline in Fig.~\ref{fig:hiden_fovs} (except for the $z\approx1.089$ system which has already been discussed and shown in Fig.~\ref{fig:det_fov}). We centre the velocity axis in the top panels on the nearest galaxy (in units of $R_{\rm vir}$) to the sightline. The largest stellar mass galaxy detected in the foreground of the quasar lies at $z=1.0262$. It has a strong continuum and [O~{\sc ii}] line emission detection in the VLT MUSE data, as well as H$\alpha$ detection in the grism data. The sightline  lies at $\approx1.5R_{\rm vir}$ of the galaxy position. Two further lower-mass galaxies are detected in the MUSE data at impact distances of $b\approx2R_{\rm vir}$, whilst four further H$\alpha$ emitting galaxies are detected within $\Delta v<1800$~km~s$^{-1}$ in the grism data. Despite three galaxies lying at $b\lesssim2R_{\rm vir}$, we find no detectable O~{\sc vi} at the galaxy redshifts. We calculate an upper limit on the O~{\sc vi} column density of $N_{\rm OVI}<10^{13.9}$~cm$^{-2}$. For the most massive galaxy, we estimate a halo mass of $M_{\rm halo}=10^{12.8\pm0.1}$~M$_\odot$, consistent with this being a galaxy group intersecting the sightline.

For the over-density at $z\approx1.39$, we find an upper limit on the O~{\sc vi} column density of $N_{\rm OVI}<10^{13.6}$~cm$^{-2}$. This over-density is significantly less concentrated both spatially and in velocity than both the $z\approx1$ over-densities, suggesting it may be a random fluctuation as opposed to a physical system. The most massive galaxy in this redshift range ($M_\star=10^{10.22\pm0.21}$~M$_\star$) lies at $b\approx780$~kpc, whilst the nearest lies at $b\approx280$~kpc.

The over-densities at $z\approx0.69$, $z\approx0.98$ and $z\approx1.13$ all consist of multiple low mass galaxies close to the sightline and each have galaxies again lying at $b\approx2R_{\rm vir}$. Overall, we find little sign of O~{\sc vi} absorption features tracing over-densities along the sightline probed, finding only one case out of 6 in which there is an association between O~{\sc vi} and a galaxy over-density.

\begin{table*}
\caption{Properties of proximate galaxies to the quasar sightline.}
\label{tab:proxgals}
\begin{tabular}{lccccccccc}
\hline
ID & R.A. & Dec.   & $m_{\rm F140W}$ & $z$ & $b$  & log[$M_\star$] & log[$M_{\rm halo}$] & $R_{\rm vir}$ & log[SFR]\\
  & \multicolumn{2}{c}{(J2000)}&(AB)&     &(kpc) & (M$_\odot$)    & (M$_\odot$)         & (kpc)         & (M$_\odot$yr$^{-1}$)\\
\hline
\hline
QSAGE J023508.12+023508.1 &  38.78385 &  -4.04062 & $19.41\pm0.01$ & 0.1735 &   73 & $ 9.45_{-0.01}^{+0.01}$ & $11.5_{-0.1}^{+0.1}$ & $148_{-23}^{+23}$ & $-0.25_{-0.01}^{+0.01}$ \\ 
QSAGE J023506.49+023506.5 &  38.77705 &  -4.02821 & $23.79\pm0.05$ & 0.2942 &  122 & $ 7.95_{-0.19}^{+0.21}$ & $10.6_{-0.2}^{+0.2}$ & $ 63_{-38}^{+41}$ & $-1.79_{-0.08}^{+0.07}$ \\ 
QSAGE J023506.12+023506.1 &  38.77551 &  -4.02791 & $23.06\pm0.01$ & 0.2948 &  140 & $ 8.04_{-0.01}^{+0.01}$ & $10.6_{-0.1}^{+0.1}$ & $ 66_{-10}^{+10}$ & $-2.21_{-0.10}^{+0.08}$ \\ 
QSAGE J023507.78+023507.8 &  38.78240 &  -4.03727 & $22.63\pm0.02$ & 0.2961 &   50 & $ 8.76_{-0.16}^{+0.08}$ & $11.2_{-0.2}^{+0.1}$ & $101_{-48}^{+26}$ & $ 0.17_{-0.12}^{+0.09}$ \\ 
QSAGE J023507.89+023507.9 &  38.78287 &  -4.03313 & $24.73\pm0.19$ & 0.3224 &   50 & $ 7.27_{-0.28}^{+0.29}$ & $10.1_{-0.3}^{+0.3}$ & $ 42_{-40}^{+41}$ & $-2.02_{-0.23}^{+0.15}$ \\ 
QSAGE J023507.23+023507.2 &  38.78012 &  -4.02573 & $21.82\pm0.01$ & 0.4341 &  191 & $ 9.13_{-0.04}^{+0.04}$ & $11.4_{-0.1}^{+0.1}$ & $103_{-18}^{+18}$ & $-0.09_{-0.06}^{+0.05}$ \\ 
QSAGE J023508.77+023508.8 &  38.78653 &  -4.03700 & $24.09\pm0.06$ & 0.6916 &  167 & $ 8.59_{-0.07}^{+0.07}$ & $11.1_{-0.1}^{+0.1}$ & $ 69_{-16}^{+16}$ & $-0.12_{-0.03}^{+0.03}$ \\ 
QSAGE J023506.86+023506.9 &  38.77857 &  -4.02929 & $23.64\pm0.04$ & 0.6925 &  155 & $ 9.19_{-0.14}^{+0.14}$ & $11.4_{-0.1}^{+0.1}$ & $ 87_{-36}^{+35}$ & $-0.00_{-0.04}^{+0.03}$ \\ 
QSAGE J023508.51+023508.5 &  38.78546 &  -4.03833 & $25.05\pm0.09$ & 0.6931 &  159 & $ 8.81_{-0.26}^{+0.24}$ & $11.2_{-0.3}^{+0.2}$ & $ 75_{-65}^{+57}$ & $-0.63_{-0.05}^{+0.05}$ \\ 
QSAGE J023511.66+023511.7 &  38.79860 &  -4.03402 & $20.12\pm0.01$ & 0.7023 &  479 & $10.29_{-0.01}^{+0.01}$ & $11.9_{-0.1}^{+0.1}$ & $130_{-20}^{+20}$ & $ 1.41_{-0.01}^{+0.01}$ \\ 
QSAGE J023508.20+023508.2 &  38.78417 &  -4.02041 & $21.94\pm0.01$ & 0.7296 &  400 & $ 9.24_{-0.01}^{+0.01}$ & $11.4_{-0.1}^{+0.1}$ & $ 86_{-13}^{+13}$ & $ 0.60_{-0.84}^{+0.27}$ \\ 
QSAGE J023507.12+023507.1 &  38.77968 &  -4.03009 & $24.92\pm0.11$ & 0.7317 &  130 & $ 8.47_{-0.20}^{+0.19}$ & $11.0_{-0.2}^{+0.2}$ & $ 62_{-39}^{+36}$ & $-0.45_{-0.14}^{+0.11}$ \\ 
QSAGE J023509.63+023509.6 &  38.79012 &  -4.02609 & $22.59\pm0.02$ & 0.7339 &  349 & $ 9.29_{-0.06}^{+0.06}$ & $11.4_{-0.1}^{+0.1}$ & $ 88_{-19}^{+19}$ & $ 0.08_{-0.10}^{+0.08}$ \\ 
QSAGE J023504.92+023504.9 &  38.77051 &  -4.02692 & $25.11\pm0.09$ & 0.7430 &  347 & $ 9.88_{-0.27}^{+0.22}$ & $11.7_{-0.3}^{+0.2}$ & $107_{-96}^{+72}$ & $-0.17_{-0.07}^{+0.06}$ \\ 
QSAGE J023510.26+023510.3 &  38.79274 &  -4.03637 & $21.95\pm0.02$ & 0.7476 &  334 & $ 9.57_{-0.02}^{+0.02}$ & $11.6_{-0.1}^{+0.1}$ & $ 95_{-15}^{+15}$ & $-0.12_{-0.12}^{+0.10}$ \\ 
QSAGE J023506.76+023506.8 &  38.77818 &  -4.03800 & $21.81\pm0.01$ & 0.8074 &  111 & $10.04_{-0.07}^{+0.16}$ & $11.8_{-0.1}^{+0.2}$ & $109_{-25}^{+52}$ & $ 0.28_{-0.08}^{+0.07}$ \\ 
QSAGE J023508.16+023508.2 &  38.78398 &  -4.03261 & $23.34\pm0.03$ & 0.8685 &  116 & $ 9.43_{-0.13}^{+0.16}$ & $11.5_{-0.1}^{+0.2}$ & $ 84_{-32}^{+40}$ & $-0.04_{-0.10}^{+0.08}$ \\ 
QSAGE J023510.23+023510.2 &  38.79264 &  -4.03456 & $21.77\pm0.01$ & 0.8720 &  346 & $ 9.91_{-0.05}^{+0.12}$ & $11.7_{-0.1}^{+0.1}$ & $ 99_{-20}^{+35}$ & $-0.15_{-0.17}^{+0.12}$ \\ 
QSAGE J023507.39+023507.4 &  38.78078 &  -4.03990 & $23.43\pm0.04$ & 0.9818 &  150 & $ 9.81_{-0.10}^{+0.08}$ & $11.7_{-0.1}^{+0.1}$ & $ 90_{-28}^{+24}$ & $ 0.27_{-0.09}^{+0.07}$ \\ 
QSAGE J023506.51+023506.5 &  38.77711 &  -4.03006 & $24.08\pm0.07$ & 0.9885 &  174 & $ 9.70_{-0.17}^{+0.14}$ & $11.6_{-0.2}^{+0.2}$ & $ 87_{-46}^{+36}$ & $ 0.27_{-0.08}^{+0.07}$ \\ 
QSAGE J023505.55+023505.6 &  38.77314 &  -4.03222 & $25.15\pm0.11$ & 1.0224 &  235 & $ 9.14_{-0.23}^{+0.21}$ & $11.4_{-0.2}^{+0.2}$ & $ 70_{-52}^{+45}$ & $-0.04_{-0.10}^{+0.08}$ \\ 
QSAGE J023506.67+023506.7 &  38.77779 &  -4.03099 & $24.02\pm0.06$ & 1.0236 &  142 & $ 9.31_{-0.16}^{+0.16}$ & $11.5_{-0.2}^{+0.2}$ & $ 74_{-35}^{+35}$ & $ 0.23_{-0.11}^{+0.08}$ \\ 
QSAGE J023509.41+023509.4 &  38.78922 &  -4.02911 & $20.09\pm0.01$ & 1.0262 &  309 & $10.94_{-0.04}^{+0.04}$ & $12.8_{-0.1}^{+0.1}$ & $200_{-38}^{+36}$ & $ 1.00_{-0.07}^{+0.06}$ \\ 
QSAGE J023507.10+023507.1 &  38.77959 &  -4.04093 & $24.77\pm0.13$ & 1.0891 &  187 & $ 8.43_{-0.23}^{+0.21}$ & $11.0_{-0.2}^{+0.2}$ & $ 49_{-36}^{+32}$ & $ 0.07_{-0.11}^{+0.09}$ \\ 
QSAGE J023507.29+023507.3 &  38.78036 &  -4.04046 & $21.59\pm0.01$ & 1.0894 &  170 & $10.00_{-0.10}^{+0.09}$ & $11.8_{-0.1}^{+0.1}$ & $ 91_{-27}^{+25}$ & $ 1.16_{-0.02}^{+0.02}$ \\ 
QSAGE J023508.46+023508.5 &  38.78524 &  -4.03568 & $25.06\pm0.10$ & 1.0903 &  143 & $ 8.50_{-0.21}^{+0.22}$ & $11.0_{-0.2}^{+0.2}$ & $ 51_{-33}^{+35}$ & $ 0.09_{-0.10}^{+0.08}$ \\ 
QSAGE J023508.66+023508.7 &  38.78608 &  -4.04398 & $25.30\pm0.10$ & 1.0961 &  323 & $ 9.33_{-0.30}^{+0.28}$ & $11.5_{-0.3}^{+0.3}$ & $ 72_{-75}^{+68}$ & $-0.16_{-0.12}^{+0.09}$ \\ 
QSAGE J023505.48+023505.5 &  38.77284 &  -4.03393 & $25.07\pm0.09$ & 1.1306 &  237 & $ 8.59_{-0.23}^{+0.24}$ & $11.1_{-0.2}^{+0.2}$ & $ 53_{-39}^{+41}$ & $-0.06_{-0.14}^{+0.11}$ \\ 
QSAGE J023505.97+023506.0 &  38.77486 &  -4.03370 & $24.00\pm0.05$ & 1.1347 &  178 & $ 9.52_{-0.18}^{+0.12}$ & $11.5_{-0.2}^{+0.1}$ & $ 75_{-40}^{+28}$ & $ 0.28_{-0.06}^{+0.05}$ \\ 
QSAGE J023507.22+023507.2 &  38.78007 &  -4.02501 & $24.18\pm0.06$ & 1.1425 &  299 & $ 9.18_{-0.12}^{+0.18}$ & $11.4_{-0.1}^{+0.2}$ & $ 67_{-25}^{+37}$ & $ 0.18_{-0.09}^{+0.07}$ \\ 
QSAGE J023505.73+023505.7 &  38.77387 &  -4.02682 & $24.59\pm0.10$ & 1.3555 &  325 & $ 9.54_{-0.23}^{+0.22}$ & $11.6_{-0.2}^{+0.2}$ & $ 68_{-49}^{+48}$ & $ 1.50_{-0.26}^{+0.16}$ \\ 
QSAGE J023509.45+023509.4 &  38.78936 &  -4.03679 & $24.38\pm0.08$ & 1.3920 &  281 & $ 9.24_{-0.20}^{+0.18}$ & $11.4_{-0.2}^{+0.2}$ & $ 60_{-37}^{+33}$ & $ 0.26_{-0.10}^{+0.08}$ \\ 
\hline 
\hline

\end{tabular}
\end{table*}

\subsection{The properties of O~{\sc vi} associated galaxies}

\label{sec:ovi_stats}

Matter in the Universe is distributed hierarchically, such that galaxies exist within shared halos. Assuming this present paradigm of hierarchically ordered dark matter halos and sub-halos, the conditions at any point within large scale structure can be driven and influenced by a number of components of the surrounding structure, e.g. the nearest galaxy, individual galaxies in the local neighbourhood, and as the overall mass and size scale of the matter halo hosting both the gas and the galaxies. As such (and combined with the large scales over which the O~{\sc vi} ion is observed around galaxies) it is worth identifying any correlations between not just the properties of the nearest galaxy to a given absorption system, but also to the wider galaxy population and environment. As discussed, this is one of the primary advantages of the strategy invoked in this survey, that we take a blind snapshot of the galaxy population around the quasar sightline.

\begin{figure}
    \centering
	\includegraphics[width=0.9\columnwidth]{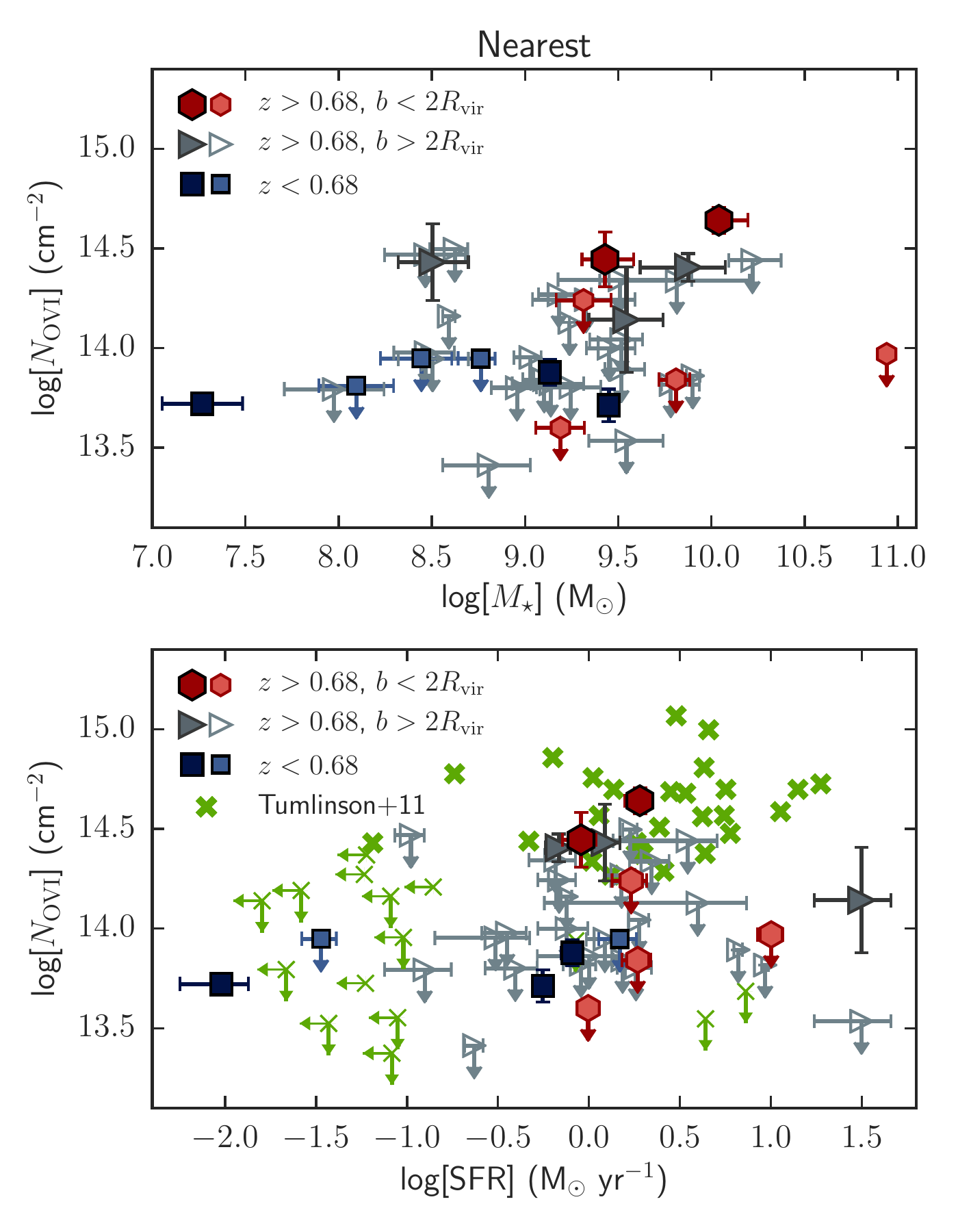}
	\caption{Sightline O{\sc vi} absorber column density versus measured stellar mass (top panel) and SFR (lower panel) of the nearest galaxy within a velocity offset of $\Delta v_{\rm max}$ (as defined in the text). Red hexagons show O~{\sc vi} absorber-galaxy alignments at $0.68<z<1.44$ within $b=2R_{\rm vir}$ of the sightline; grey triangles denote absorber-galaxy alignments at $0.68<z<1.44$ in the range $2R_{\rm vir}<b<5R_{\rm vir}$ of the sightline; and blue squares denote $z<0.68$ absorber-galaxy alignments. The paler points in each case highlight points where only upper limit measurements are available on the absorption. In the lower panel, we show the $z<0.2$ data-points of \citet{2011Sci...334..948T} as green crosses for comparison.}
    \label{fig:NOVIvNearestGals}
\end{figure}

\begin{figure}
    \centering
	\includegraphics[width=0.9\columnwidth]{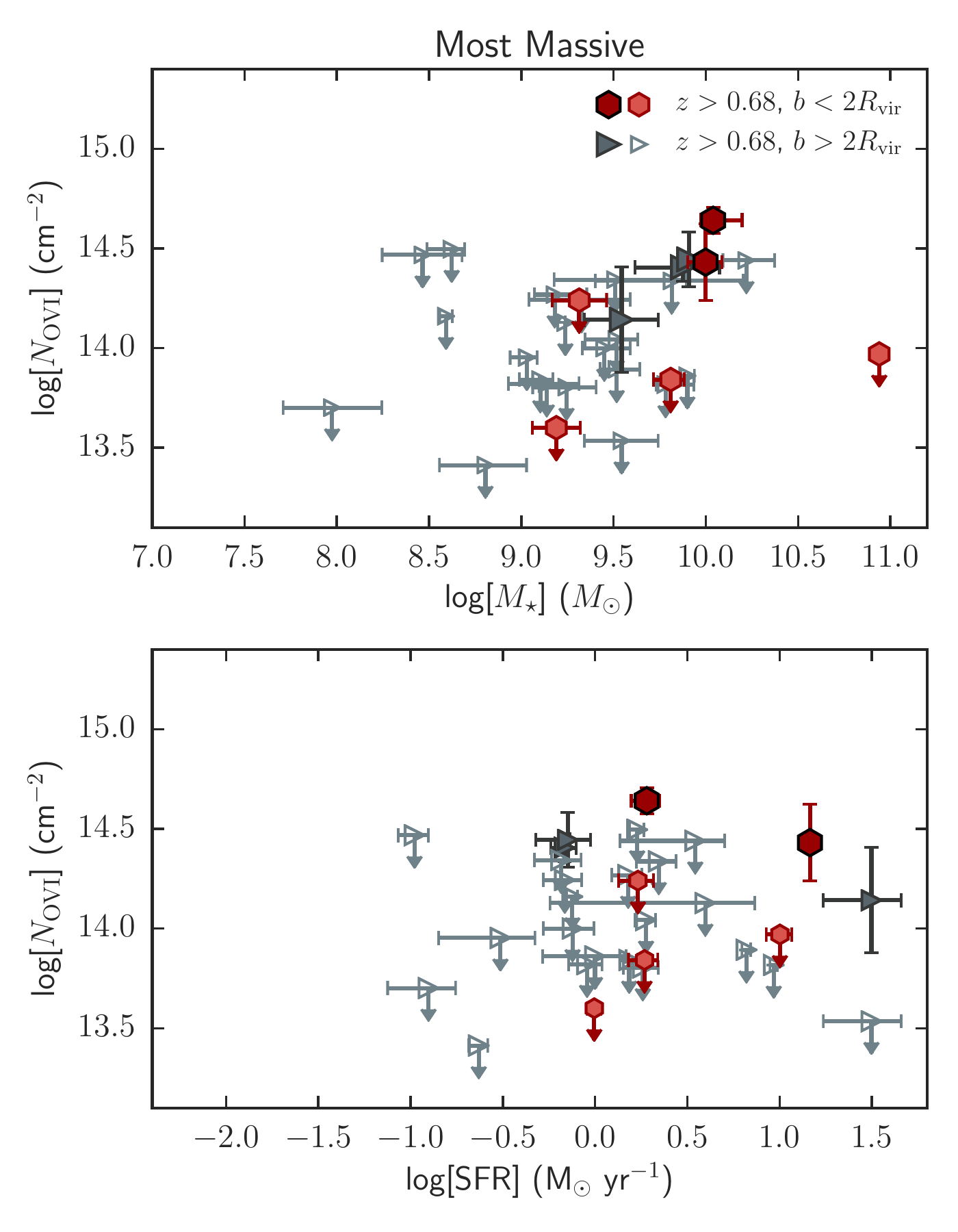}
	\caption{As in Fig.~\ref{fig:NOVIvNearestGals} but taking the stellar mass and SFR properties for the most massive proximate galaxies (i.e. up to $5R_{\rm vir}$). Only the $0.68<z<1.44$ sample is now shown, given the more limited field of view of the $z<0.68$ data.}
    \label{fig:NOVIvMassiveGals}
\end{figure}

We take two primary approaches in this section: 1. what the observed properties of the most proximate galaxies to the sightline are, and 2. what those same properties for the most massive galaxies detected at each given redshift probed are. The motivation for the latter is that, by isolating the most massive galaxies, these more closely relate the total structure mass that is being probed, thus opening the possibility of discerning any correlation between absorber properties and parent-halo properties \citep[e.g.][]{2016MNRAS.460.2157O,2018MNRAS.477..450N}.

We begin by analysing the properties of galaxies lying closest in impact parameter to the sightline, within our defined velocity window around an absorber redshift. These are shown in Fig.~\ref{fig:NOVIvNearestGals}, with the absorber column density versus nearest galaxy stellar mass in the top panel and the absorber column density versus SFR of the nearest galaxy in the lower panel. In both cases, the blue squares show the galaxy-absorber pairs at $z<0.68$, whilst the $0.68<z<1.44$ pairs are split into two groups: those with $b<2R_{\rm vir}$ (red hexagons); and those at $2R_{\rm vir}<b<5R_{\rm vir}$ (grey triangles). In each case the paler points denote upper-limits on O~{\sc vi} absorption, whilst the darker points denote significant detections.  

At $z<0.68$, we find the three significant detections of O~{\sc vi} are associated with galaxies covering a wide range in properties with stellar masses over the range $10^{7.2}\lesssim{\rm log}[M_\star/{\rm M_\odot}]\lesssim10^{9.5}$ and star-formation rates over the range $0.01{\rm M_\odot~yr^{-1}}\lesssim{\rm SFR}\lesssim2{\rm M_\odot~yr^{-1}}$. The O~{\sc vi} upper-limits found at proximate galaxy positions are comparable to the 3 significant detections leaving the possibility that these may also exhibit O~{\sc vi} absorption in the sightline if not for blended interloping absorption features. No correlations between $N_{\rm OVI}$ and either of the galaxy properties presented is evident in this small $z<0.68$ dataset. We show the $Z<0.2$ data-points of \citet{2011Sci...334..948T} (taken from the same dataset as those of \citealt{2016ApJ...833...54W} used earlier) as the green crosses in the lower panel of Fig.~\ref{fig:NOVIvNearestGals}, showing that our sample lies in the lower column density region of the COS-Halos set of star-forming galaxies. 

Focussing on the $z>0.68$ sample, at impact parameters of $b/R_{\rm vir}<2$ we find that galaxies associated with both the significant O~{\sc vi} detections and the upper limits of $N_{\rm OVI}\lesssim10^{13.5-14}$~cm$^{-2}$ have consistent stellar mass and star-formation rate properties. The same is seen for the galaxies at larger impact parameters, with no indication of any correlation. The field of view does not include any detected passive galaxies at $0.68<z<1.44$, although we would have been able to identify such galaxies in the MUSE field of view (i.e. up to $\approx300$~kpc) to a magnitude limit of $i\approx24$ if any were present. We note that the space density of passive galaxies to $i\lesssim24$ at $z\sim1$ is $n\approx0.0003h^{-1}$Mpc \citep[e.g.][]{2014A&A...568A..24B}, equating to $\approx1$ in a 1 sq. arcmin, $0.68<z<1.44$ volume on average (i.e. not taking into account the highly clustered nature of such galaxies). Our data in this field can do little then to inform both the low-SF range ($\lesssim 10^{-0.5}$~M$_\odot$~yr$^{-1}$) and by association the high mass range beyond $M_\star\gtrsim10^{11}$~M$_\odot$. However, the fact that we find such low limits on O~{\sc vi} absorption around such star-forming galaxies presents significant implications for any causal connection between ongoing star-formation in a galaxy and the detection of O~{\sc vi} ions in the galaxy's vicinity. 

We now reproduce the plot of column density versus galaxy properties from Fig.~\ref{fig:NOVIvNearestGals}, but with the stellar mass (top panel) and star-formation rates (lower panel) of the most massive galaxies lying at the redshift of each absorption system in Fig.~\ref{fig:NOVIvMassiveGals}. The symbols used are consistent with those in Fig.~\ref{fig:NOVIvNearestGals}. For clarity and given the smaller field of view probed by MUSE, we omit the $z<0.68$ O~{\sc vi} absorber-galaxy pairs.

We find a marginal preference for the $N_{\sc OVI}\approx10^{14.5}$~cm$^{-2}$ absorbers to be associated with galaxies of stellar masses $M_\star\approx10^{10}$~M$_\odot$ (for both galaxy-absorber pairs lying at $b<2R_{\rm vir}$ and $b<5R_{\rm vir}$). Indeed, the galaxies that we associated with the detections based on our matching criteria all lie within a mass range of $10^{9.5}~{\rm M_\odot}<M_\star<10^{10}~{\rm M_\odot}$, whilst the non-detections that also lie in this mass range offer only relatively poorly constrained upper limits of $N_{\rm OVI}\approx10^{14.3}$~cm$^{-2}$. Beyond stellar masses of $M_\star>10^{10}~{\rm M_\odot}$, our dataset from this first field poorly samples the galaxy population, with only a single massive galaxy in this range (which corresponds to an O~{\sc vi} limit of $N_{\rm OVI}\approx10^{13.9}$~cm$^{-2}$). Moving to the SFR distribution as shown in the lower panel, we see no indication of any correlations between SFR and column density in the available data. The distribution of points shows a large scatter in column densities across the full SFR range probed.

\subsection{Covering fraction of O~{\sc vi} around $z\approx1$ galaxies}

\begin{figure*}
 	\includegraphics[width=\textwidth]{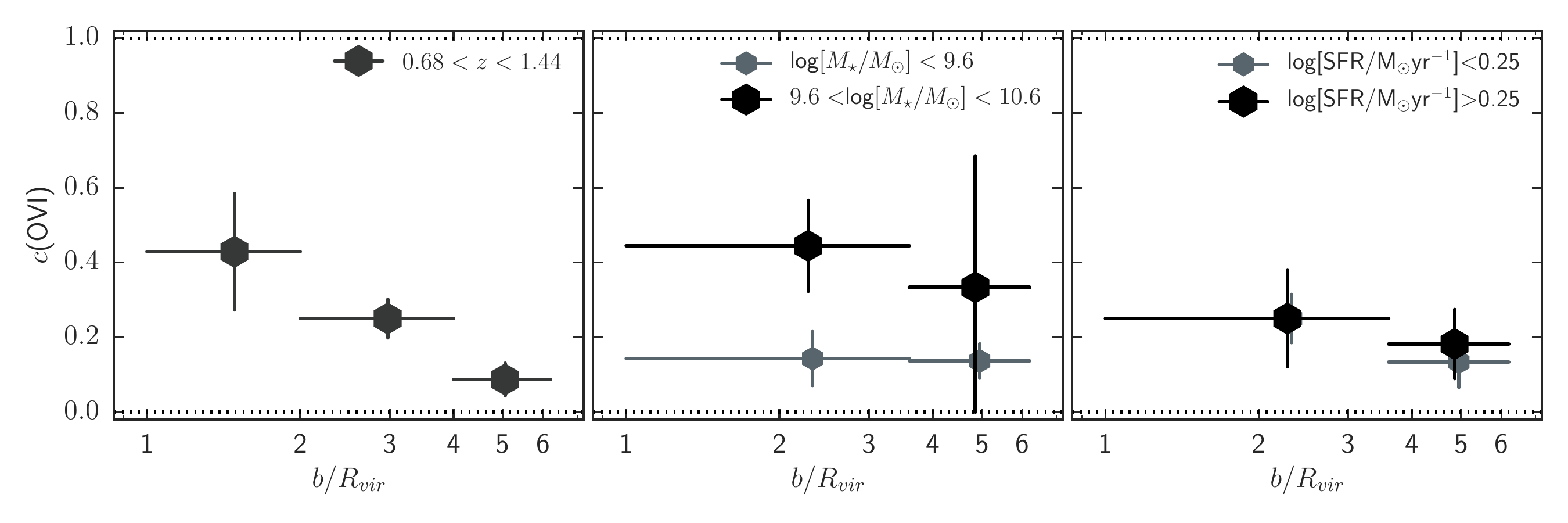}
 	\caption{The covering fraction of $N_{\rm OVI} > 10^{14}$~cm$^{-2}$ O~{\sc vi} absorbers as a function of impact parameter (left panel); separated into mass bins at $M_\star=10^{9.6}$~M$_\odot$ (centre panel); and separated into SFR bins at ${\rm SFR}=10^{0.25}$~M$_\odot$~yr$^{-1}$ (right panel).}
     \label{fig:CovFrac}
\end{figure*}

In order to quantify any correlations in a more systematic way, we now calculate the covering fraction of O~{\sc vi} around the galaxy population as a function of impact parameter, galaxy stellar mass and galaxy star-formation rate. For this, we include all galaxies around the sightline and not just the nearest or most massive. We measure the covering fraction for O~{\sc vi} column densities of $N_{\rm OVI}>10^{14}$~cm$^{-2}$, such that the covering fraction is given by $c=\frac{n_{\rm gal}(\rm N>10^{14})}{n_{\rm gal}}$. Galaxies found within $\Delta_v={\rm max}\{\sigma_v,400{\rm km~s^{-1}}\}$ of a column density upper limit of $N_{\rm OVI}>10^{14}$~cm$^{-2}$ (with no corresponding significant detection) are not included in the calculation (i.e. as these are likely the result of contamination from other absorption lines masked in the analysis).

The resulting covering fraction of O~{\sc vi} absorbers with column densities of $N>10^{14}$~cm$^{-2}$ is shown as a function of impact parameter (in units of the virial radius) in the left hand panel of Fig.~\ref{fig:CovFrac}. We find a trend for increasing covering fraction with decreasing impact parameter in the $0.68<z<1.5$ sample, with covering fractions of $c\approx0.2$ up to $b/R_{\rm vir}=4$. In the central panel of Fig.~\ref{fig:CovFrac}, we show the covering fraction for galaxies split into two mass bins: $M_\star<10^{9.6}$~M$_\odot$ and $10^{9.6}$~M$_\odot < M_\star < 10^{10.2}$~M$_\odot$ (whilst noting that we have only a single galaxy in the sampled redshift range at $M_\star>10^{10.2}$~M$_\odot$). We find a moderately significant relation between covering fraction and stellar mass at $1<b/R_{\rm vir}<4$, with $c=0.44\pm0.12$ for $9.6<{\rm log}[M_\star/M_\odot]<10.2$ and $c=0.18\pm0.06$ for ${\rm log}[M_\star/M_\odot]<9.6$.  The right hand panel shows the covering fraction split into samples based on a star-formation rate limit of 0.25~M$_\odot$~yr$^{-1}$. In this case we find no sign of a difference of the covering fraction based on the two different samples.

\section{Discussion}
\label{sec:discussion}

Our results have shown tentative evidence for O~{\sc vi} absorption preferentially being found in the vicinity of moderate mass ($M_\star\approx10^{9.5-10}$~M$_\odot$) star-forming galaxies. The star-forming properties of these $z\approx1$ galaxies are comparable (${\rm SFR}\gtrsim1$~M$_\odot$~yr$^{-1}$) to that observed for isolated $L^\star$ galaxies at $z\approx0.2$ coincident with O~{\sc vi} absorption \citep{2011Sci...334..948T,2014ApJ...792....8W}.

In terms of the scales at which O~{\sc vi} is detected around galaxies, we find absorbers lying up to $b\approx300-400$~kpc (or $b\approx4-5R_{\rm vir}$) to the most proximate detected galaxy alongside a significant measured covering fraction ($c\approx0.2$) out to these scales. Whilst \citet{2011Sci...334..948T} and \citet{2014ApJ...784..142S} detected O~{\sc vi} at impact parameters of up to $\approx150$~kpc (or $\approx2R_{\rm vir}$) at $z<0.2$ this corresponded to the maximum distances probed by their survey. Looking to a more comparable survey at low redshift, \citet{2015MNRAS.449.3263J} found O~{\sc vi} absorption (at a covering fraction of $c\approx0.15$) up to $b\approx3R_{\rm vir}$ around individual galaxies, consistent with our findings.

Two of the $z>0.68$ absorption systems are coincident with multiple galaxies in our survey (a galaxy pair at $z\approx0.868$ and the galaxy triplet group at $z\approx1.089$), whilst around 3 we detect only a single galaxy given our survey limits. We find a more massive ($M_{\rm halo}\approx10^{12.6}$~M$_\odot$) group system for which we detect no O~{\sc vi} absorption at $z=1.026$. \citet{2011ApJ...740...91P} at low-redshift similarly find a picture whereby the association between galaxies and O~{\sc vi} shows a strong dependence on galaxy luminosity. They find O~{\sc vi} absorption most commonly to be found within $b\approx200-300$~kpc of a $0.1L^\star<L<1L^\star$ (intermediate sub-$L^\star$) galaxies, and that where a dwarf galaxy is found closest to O~{\sc vi} sightline absorption, there invariably also exists an intermediate sub-$L^\star$ galaxy within $b\approx300$~kpc of that same absorber. 
 
The association of the absorbers in our sample with galaxies in a given mass range, is consistent with predictions by \citet{1996ApJ...456L...5M}, \citet{2016MNRAS.460.2157O} and \citet{2018MNRAS.477..450N}. Indeed, the EAGLE hydrodynamical simulations \citep{2015MNRAS.446..521S,2015MNRAS.450.1937C} predict that the O~{\sc vi} column density peaks ($N_{\rm OVI}\approx10^{14.2}$ cm$^{-2}$) for galaxies in halos of masses $\approx10^{11.7-12.4}$~M$_{\odot}$, with the gas within lower mass halos too cool, and the gas within higher mass halos too hot, to lead to significant O~{\sc vi} absorption in quasar sightlines \citep[albeit at $z=0.2$;][]{2016MNRAS.460.2157O}. Similarly, the IllustrisTNG predicts, at $z=0$, that O~{\sc vi} should be predominantly found in diffuse $T\approx10^{5.6\pm0.2}$~K halos of half-mass radii $\approx10^{2.1\pm0.2}$~kpc around $M_\star\approx10^{10.2\pm0.3}$~M$_\odot$ galaxies \citep{2018MNRAS.477..450N}.

\begin{figure}
	\includegraphics[width=\columnwidth]{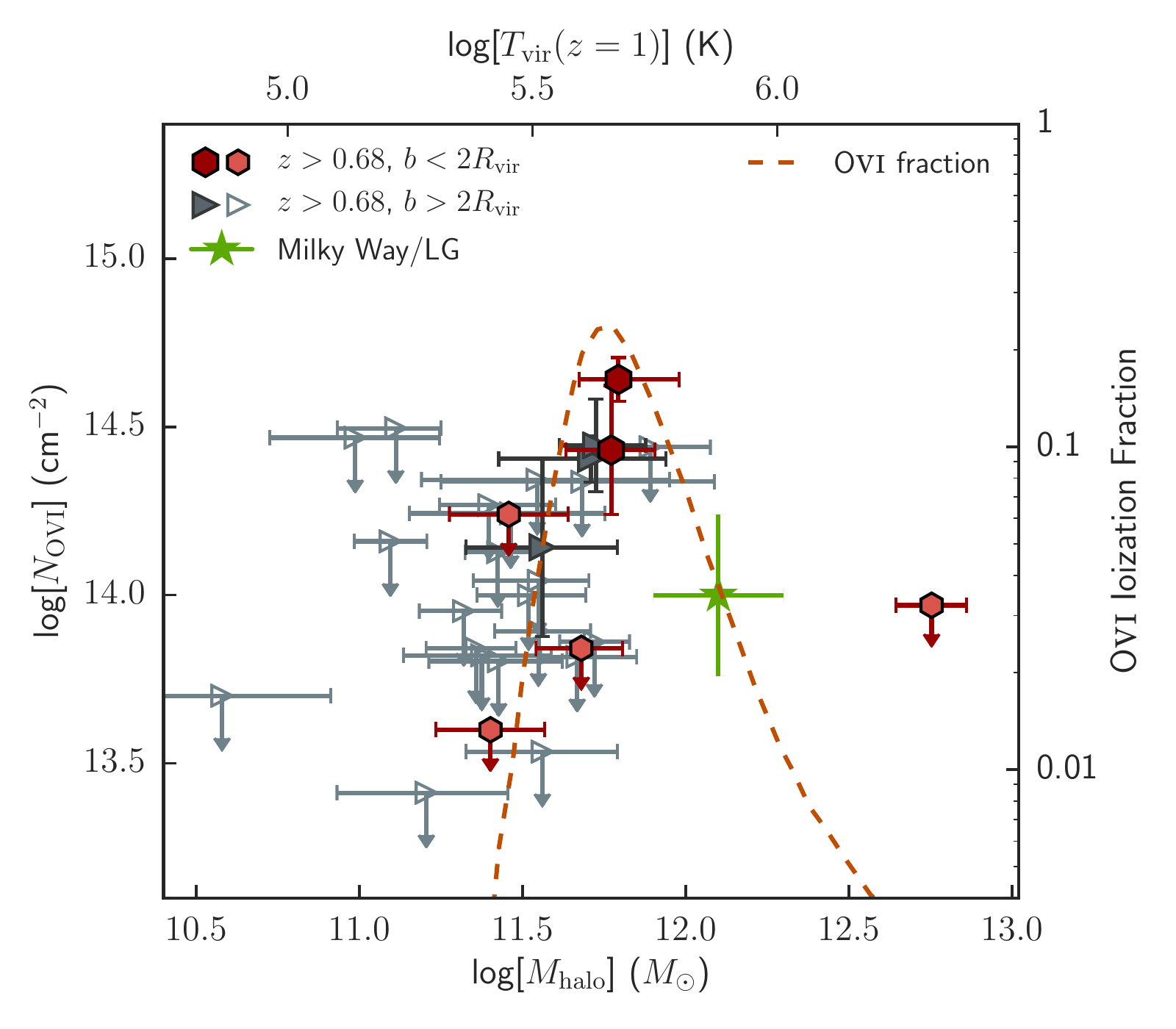}
	\caption{Column densities of O~{\sc vi} absorbers versus their estimated dark matter halo mass for the $z>0.68$ galaxy sample based on identified associations from Fig.~\ref{fig:NOVIvMassiveGals}. Hexagons denote associated galaxies within $b=2R_{\rm vir}$, whilst triangles denote associated galaxy-absorber pairs at impact parameters of $2R_{\rm vir}<b<5R_{\rm vir}$. The dashed curve shows the predicted O~{\sc vi} ionization fraction for a given halo mass, assuming the halo gas being probed to be close to the virial temperature of the halo.}
    \label{fig:OVIvGHmass}
\end{figure}

In Fig.~\ref{fig:OVIvGHmass} we show the inferred halo masses (as calculated above) assuming the absorbers to be associated with the most massive proximate galaxy (i.e. within $b=2R_{\rm vir}$; dark red hexagons). We find that the $N_{\rm OVI}\gtrsim10^{14}$~cm$^{-2}$ O~{\sc vi} absorption is almost exclusively found within in the proximity of a galaxy corresponding to a $10^{11.5}M_\odot\lesssim M_\star\lesssim10^{12}M_\odot$ halo mass. Upper limits on the O~{\sc vi} column density adjacent to other galaxies along the sightline (within $0.68<z<1.4$) preferentially show upper-limit constraints indicating lower levels of the O~{\sc vi} ion.

The dashed curve shows the predicted O~{\sc vi} ionization fraction assuming that the halo gas being probed is at temperatures of $\approx70\%$ of the virial temperature (i.e. accounting for predicted halo temperature profiles \citealt[e.g.][]{2001MNRAS.327.1353K}). This provides a simple physical picture in which the detected gas represents a diffuse warm halo component \citep[e.g.][]{2017ApJ...846L..24M}. The concept is somewhat complicated by the presence of a number of non-detections within this same mass range. Whilst for 2 of these the upper limit constraints allow for column densities comparable to the actual detections (i.e. $N_{\rm OVI}\approx10^{14.2}$~cm$^{-2}$), there are $\approx4-5$ that may be expected to have detectable O~{\sc vi} absorption under the assumed simple model. Indeed, this may be the effect of a patchy/clumpy halo medium \citep{2018Natur.554..493L}, or AGN heating of the gas \citep{2017arXiv170907577O}. 

Alternatively, the O~{\sc vi} gas may trace low-pressure gas photo-ionised by the UV background \citep[e.g.][]{2016ApJ...830...87S, 2018arXiv180305446S}. In this case clouds are hierarchically embedded in the CGM structure, allowing for a patchy medium and potentially explaining the large scatter in column densities that we measure. The halo masses that we find associated with the O~{\sc vi} absorbers are comparable to the mass of the local group. Indeed, observations of the Milky Way's halo have shown detection of O~{\sc vi} absorption ($N_{\rm OVI}\gtrsim10^{13.6}$~cm$^{-2}$) in $\approx70\%$ of sightlines through the halo \citep[e.g.][]{2003ApJS..146....1W, 2003ApJS..146..165S}. The halo mass of the Milky Way is estimated to be $M_{\rm halo}\approx10^{12}$~M$_\odot$ \citep[e.g.][]{2018arXiv180810456C}, whilst the mass of the local group as a whole is estimated to be $M_{\rm halo}\approx2\times10^{12}$~M$_\odot$ \citep[e.g.][]{1999AJ....118..337C}.

In this first QSAGE field, given the relatively small area covered, we have only a single galaxy with which to probe the higher mass galaxy environment beyond $M_{\star}\gtrsim10^{10.5}$~M$_\star$. At $M_{\star}\approx10^{10.9}$~M$_\star$, this galaxy and its associated group shows no detectable O~{\sc vi} in the sightline data. Incorporating low redshift results, \citet{2011Sci...334..948T} find 50\% of galaxies at $M_\star\gtrsim10^{10.5}$~M$_\odot$ show associated O~{\sc vi} absorption (compared to 100\% at $10^{9.5}$~M$_\odot\lesssim M_\star\lesssim10^{10.5}$~M$_\odot$). \citet{2014ApJ...791..128S} report the alignment of warm O~{\sc vi} absorbers with several galaxy groups at $z\lesssim0.2$, with group velocity dispersions of $\sigma_v\approx100-600$~km~s$^{-1}$. This equates approximately to a halo mass range of $M_{\rm halo}\approx10^{12.5-14.5}$~M$_\odot$, suggesting groups of all masses, up to low mass clusters have the potential to exhibit warm gas capable of being detected in O~{\sc vi} absorption. Conversely, \citet{2018MNRAS.475.2067B} found no O~{\sc vi} absorption within cluster halos at $M_{\rm halo}\approx10^{14-14.5}$~M$_\odot$. The caveat remains however, that it is not often clear whether the halo gas probed by a given sightline is reflecting the nature of the properties of a sub-halo or the overall group halo \citep[e.g.][]{2017ApJ...838...37S}, whilst star-forming winds remain a potentially significant influence on the presence of warm gas at small scales \citep[e.g.][]{2011Sci...334..948T,2011Sci...334..952T}.

\section{Conclusions}
\label{sec:conclusions}

We have presented methods and results from the first field in the QSAGE survey - a blind HST/WFC3 grism survey of galaxies in the region of bright $z>1.2$ quasars with archival HST/STIS and COS spectra. Our key results are:

\begin{enumerate}
\item We find O~{\sc vi} up to impact parameters of $b\approx350$~kpc from the nearest detected galaxy at $0.68<z<1.42$. Column densities of absorbers over the impact parameters probed ($100<b<400$~kpc) show a large scatter, corresponding to covering fractions of $c({\rm OVI})\lesssim0.5$ (for $N({\rm OVI})>10^{14}$~cm$^{-2}$).
\item Whilst all 5 of the detected $z\approx1$ O~{\sc vi} absorbers are found to lie within $b\approx400$~kpc of a star-forming galaxy, we also find comparably star-forming galaxies within the same range in impact parameter with no detected O~{\sc vi} absorption. Taking a limit in impact parameter of $b=2R_{\rm vir}$, we find 50\% of the sample relate to upper limits on the sightline O~{\sc vi} column density of $N_{\rm OVI}\lesssim10^{13.9}$~cm$^{-2}$.
\item We identify a low-mass galaxy group at $z=1.08$ coincident in redshift with significant O~{\sc vi} absorption in the quasar sightline, potentially probing the intra-group medium. The group consists of 3 confirmed members and estimate a group halo mass of $M_{\rm halo}\approx10^{11.8}$~M$_\odot$. We find several further galaxy over-densities close to the sightline, with estimated halo masses of up to $M_{\rm halo}\approx10^{12.8}$~M$_\odot$. None are coincident with detected O~{\sc vi} absorption to a detection limit of $N_{\rm OVI}\lesssim13.9~{\rm cm}^{-2}$.
\item Estimating the host halo masses of $N_{\rm OVI}\gtrsim10^{14}$~cm$^{-2}$ absorbers suggests the majority of such absorbers are found in the proximity of halos of mass $M_{\rm halo}\approx10^{11.8}$~M$_\odot$, consistent with diffuse gas at the virial temperature of such halos. Significantly, we find a higher covering fraction of $N_{\rm OVI}\ge10^{14}$~cm$^{-2}$ absorbers around higher mass star-forming galaxies (at $\approx2\sigma$) at impact parameters of $\lesssim4R_{\rm vir}$.  
\end{enumerate}

This first of 12 fields from our HST WFC3 Large Program, QSAGE, acts as a proof of concept of what the full survey can deliver. Whilst we have focussed on tracing the properties of the galaxy population around O~{\sc vi} absorbers here, these data will provide a comprehensive basis for studies of the CGM across a range of absorption species, allowing insights into a broad range of phases of material within the CGM.

\section*{Acknowledgements}

RMB, MF, RGB and SLM acknowledge the Science and Technology Facilities Council (STFC) through grants ST/P000541/1 and ST/L00075X/1 for support. This project has received funding from the European Research Council (ERC) under the European Union's Horizon 2020 research and innovation programme (grant agreement No 757535). TMT, JNB, and JXP received financial support for this research through NASA Grant HST-GO-11741 from the Space Telescope Science Institute, which is operated by the Association of Universities for Research in Astronomy, Inc., under NASA Contract NAS5-26555. This work is based in part on observations obtained with MegaPrime/MegaCam, a joint project of CFHT and CEA/IRFU, at the Canada-France-Hawaii Telescope (CFHT) which is operated by the National Research Council (NRC) of Canada, the Centre National de la Recherche Scientifique (CNRS) of France, and the University of Hawaii. The data used in this paper are available via the Mikulski Archive for Space Telescopes (HST COS, STIS and WFC3 data), the Canadian Astronomy Data Centre (CFHTLS data) and the ESO archive (VLT/MUSE data). The WHT/ACAM imaging data are available at \url{http://astro.dur.ac.uk/qsage/}. We have produced data products using {\sc Le Phare} and {\sc GazPar} (located at the Laboratoire d'Astrophysique de Marseille) and we thank O. Ilbert for support in using these. This research made use of Astropy, a community-developed core Python package for Astronomy \citep{2013A&A...558A..33A, 2018arXiv180102634T}. In the course of this work, we made use of A. Rohatgi's {\sc WebPlotDigitizer} (\url{http://arohatgi.info/WebPlotDigitizer/}).




\bibliographystyle{mnras}
\bibliography{rmb} 



\appendix

\bsp	
\label{lastpage}
\end{document}